\documentclass[a4paper,10pt]{elsarticle}
 \usepackage{graphics}
 \usepackage{color}
\usepackage{amssymb}
\journal{Journal of the Mechanics and Physics of Solids}

\usepackage{mathrsfs}

\newcommand{\rr}[1]{{\textrm{#1}}}
\newcommand{\bb}[1]{{\mathbb{#1}}}
\begin{document}

\begin{frontmatter}



\title{Allen--Cahn and Cahn--Hilliard--like equations for 
       \\
       \vskip 0.2 cm
       dissipative dynamics of saturated porous media}


\author[cirillo]{Emilio N.M.\ Cirillo\corref{cor}}
\ead{emilio.cirillo@uniroma1.it}
\author[ianiro]{Nicoletta Ianiro}
\ead{nicoletta.ianiro@uniroma1.it}
\author[sciarra]{Giulio Sciarra}
\ead{giulio.sciarra@uniroma1.it}

\address[cirillo]{Dipartimento di Scienze di Base e Applicate per l'Ingegneria, 
             Sapienza Universit\`a di Roma, 
             via A.\ Scarpa 16, I--00161, Roma, Italy.}
\address[ianiro]{Dipartimento di Scienze di Base e Applicate per l'Ingegneria, 
              Sapienza Universit\`a di Roma, 
              via A.\ Scarpa 16, I--00161, Roma, Italy.}
\address[sciarra]{Dipartimento di Ingegneria Chimica Materiali Ambiente,
             Sapienza Universit\`a di Roma, 
             via Eudossiana 18, 00184 Roma, Italy}

\cortext[cor]{Corresponding author}

\begin{abstract}
We consider a saturated porous medium in the  
solid--fluid segregation regime under the effect of 
an external pressure applied on the 
solid constituent.
We prove that, depending on the dissipation 
mechanism, the dynamics is described either by a 
Cahn--Hilliard or by an Allen--Cahn--like equation.
More precisely, when the dissipation is modeled via the Darcy 
law we find that, 
provided the solid 
deformation and the fluid density variations
are small,
the evolution equation is very similar 
to the Cahn--Hilliard one. 
On the other hand, when 
only the Stokes dissipation term is considered, we find that the 
evolution is governed by an Allen--Cahn--like equation.
We use this theory to describe 
the formation of interfaces inside porous media. We consider 
a recently developed model proposed to study the solid--liquid 
segregation in consolidation 
and we give a complete description of 
the formation of an interface between 
the fluid--rich and the fluid--poor phase.
\end{abstract}


\end{frontmatter}

% \linenumbers

%%%%%%%% Il lavoro
\section{Introduction}
\label{s:introduzione}
\par\noindent
Many interesting swelling and shrinking phenomena are observed 
when a porous medium deforms as a consequence 
of the variation of the fluid mass content.
The behavior of the system can be controlled 
via different external parameters, 
e.g.\ the  
fluid chemical potential~\citep{GGRS,CC02,CVBCTC}
or a mechanical pressure exerted on the solid component
\citep{CIS2011,CIS2010}.

By focusing the attention on this last class 
of phenomena we are led to 
consider the so--called consolidation problem. 
Despite the existence of 
classical and well known theories \citep{Terzaghi,biot01,Cryer63,Mandel},
the dilatant/contractant behavior of porous solids needs, however, deeper 
investigation.
We refer, in particular, to undrained conditions, i.e., to the 
case 
in which the loading rate does not allow 
pore--water pressure dissipation.
In these conditions,
the coupling between shearing and dilatancy typically induces 
the formation of strain localization bands.
In particular shear bands with transeverse size 
depending on the texture and the 
constitutive properties of the material, see e.g. \citep{Besuelle2001}, 
are seen.
For instance, in simple shear test, densely or loosely packed granular media
exhibit dilatancy or compaction behavior.
Moreover due to porosity change, the fluid can migrate through the pores and 
eventually remain segregated, possibly enhancing localised 
overpressurization and fluidization of the soil, 
see e.g. \citep{Kolymbas94,Nichols94}. 

Local dilatancy is typically modeled in the framework of 
(poro--)plasticity by assuming that irreversible deformations take 
place inside the porous medium, see e.g. the general treatises 
\citep{Coussy,Coussybook2010}. 
Special important applications to multi--phase 
porous media and geomaterials have been considered in the recent 
literature \citep{Zhang99,CollinChambon2006,Chambon2001,Andrade2011}
by approaching plasticity with strain gradient 
\citep{FleckHutchinson97,FleckHutchinson2001} or 
multi--scale analysis \citep{Christoff_NN1981,NematNasser}.

In the papers \citep{CIS2011,CIS2009,CIS2010}
we have attacked this problem from the 
point of view of bifurcation theory and we have shown that 
it is possible to describe interesting phenomena (still in the 
range of non--linear elasticity) taking place when 
the pressure exerted on the solid 
exceeds a suitable limiting value. 
More precisely, 
we have proven that
enhanced models based on gradient elasticity allow for describing the
transition between phases of the porous medium associated with different
fluid content, say a fluid--rich and a fluid--poor phase. 
In other words the theory proposed in those papers is able to 
describe 
the occurrence of segregation and local overpressurization
of the pore--water, see the above quoted references for more details.

Experiments demonstrate that the presence of a pressurized fluid--rich phase 
at the base of a layered granular material can give rise to an 
unstable (with respect to possible fluidization)
configuration of the system \citep{Kolymbas98}.
This result is achieved in laboratory, see \citep{Nichols94}, 
considering a fluidized column 
test setup where a fluid is forced to flow through a saturated sample from the
bottom. 
By tuning the velocity of the fluid, 
the drag force acting on the solid
grains, and hence the possible unbalance of gravity, is controlled
\citep{Vardoulakis04_1,Vardoulakis04_2}.
The model 
in \citep{CIS2009}
is able to explain 
the formation 
of fluid--rich 
regions inside a porous material via a sort of segregation process, 
due to 
different initial data and parametrized by the pressure applied to the solid;
note that the control parameter in the experiment quoted above is, 
instead, the velocity of the injected fluid.
Different dissipation mechanisms, 
driven by the microstructural properties of the
material, give rise to different deformation paths 
and fluid mass content evolutions.

In this paper we show that, depending 
on the dissipative mechanism which is taken into account, the evolution 
is described either by an Allen--Cahn or by a Cahn--Hilliard--like system 
of partial differential equations, which indeed play an important role in the
theory of phase ordering dynamics~\citep{Bray,Langer,Eyre}.
When a system is quenched from the homogeneous phase into a 
broken--symmetry one, think to a ferromagnet or to a gas abruptly 
cooled below their critical temperature, the two 
phases have to separate and order has to increase throughout the system via
domain coarsening. This phenomenon can be described via a 
field $u$ on the physical space $\Omega\subset\bb{R}^d$, 
with $d$ the physical dimension,
having the proper physical interpretation, for instance 
local magnetization in ferromagnets, density in liquid--vapor
systems, and concentration in alloys.
A possible model for the evolution of this field is the Allen--Cahn 
equation
\begin{equation}
\label{ac}
\frac{\partial u}{\partial t}
=
\varepsilon^2\Delta u-W'(u)
\end{equation}
where $\varepsilon$ is a positive constant and $W(u)$ a 
double well regular function.
The Allen--Cahn equation, also called 
the time--dependent Ginzburg--Landau equation, was introduced in \citep{Allen} 
to describe the motion of anti--phase boundaries in crystalline solids.
In this context $u$ represents the concentration of one of the two 
components of the alloy and the parameter $\varepsilon^2$ 
the interface width. 

The Allen--Cahn equation is not suited to describe phase separation when 
the order parameter is conserved, namely, when the integral 
of the field $u$ on the whole space $\Omega$ is constant. 
On the other hand this is possible
in the framework of the Cahn--Hilliard equation 
\begin{equation}
\label{ch}
\frac{\partial u}{\partial t}
=
-\Delta(
\varepsilon^2\Delta u-W'(u)
)
\end{equation}
by assuming 
$\partial_\nu u=\partial_\nu\Delta u=0$ on $\partial\Omega$, 
that is by requiring the derivative of $u$ and that of the 
Laplacian of $u$ orthogonal to the 
boundary to vanish on the boundary $\partial\Omega$ itself.
The Cahn--Hilliard equation was firstly introduced to describe 
spinodal decomposition in alloys~\citep{Cahn}.

A straightforward way to derive 
the Allen--Cahn and the Cahn--Hilliard equations 
is that of assuming the evolution of the 
field $u$ to be governed by a gradient equation \citep{Fife,ABF}
$\partial u/\partial t = -\delta F/\delta u$ associated with 
the Landau energy functional 
\begin{equation}
\label{grad-f}
F(u)
:=
\int_\Omega\Big[\frac{1}{2}\varepsilon^2\|\nabla u\|^2+W(u)\Big]\rr{d}x
\end{equation}
The functional above has a clear physical interpretation:
the term $W$ has two minima corresponding to the two
phases, while the gradient--squared term associates an energy cost with 
an interface between the two phases. 
If no constraint to the total value of the order parameter is imposed, 
it is possible to compute the gradient of the Landau functional 
in the Hilbert space $L^2(\Omega)$
to get the Allen--Cahn equation. When the integral of the field $u$ 
is assumed to be constant throughout the evolution, computing 
the gradient in the $L^2(\Omega)$ results into a non--local 
evolution equation, while by using the Hilbert space 
$H^{-1}(\Omega)$ the Cahn--Hilliard equation is found.

The above described approach is general and applies to any physical 
interpretation of the two equations. 
When the discussion is bounded to particular physical systems,
phenomenological approaches can also be used
requiring the a priori specification of the constitutive equations.
The related physical literature is huge and, among others, we refer to 
\citep{Gurtin}. In the context of binary fluids, 
for example, $u$ is the fluid concentration.
Assume the continuity equation in the form  
$u_t+\nabla\cdot J=0$, where $J$ is the fluid current
prescribed in terms of the chemical 
potential $\mu$ by Fick's law $J=-k\nabla\mu$, with $k$ 
a positive constant. By letting $\mu$ be the 
functional derivative of $F$, one gets
\begin{displaymath}
\frac{\partial u}{\partial t}
=
-\nabla\cdot J
=
-\nabla\cdot(-k\nabla\mu)
=
k\Delta\mu
=
k\Delta\frac{\delta F}{\delta u}
=
k\Delta(-\varepsilon^2\Delta u+W'(u))
\end{displaymath}
which is the Cahn--Hilliard equation.

In our problem the Allen--Cahn and the Cahn--Hilliard--like equations 
appear in a completely natural way by using the 
variational description of continuum mechanics.
By following~\citep{Coussy,SIIM}
we describe the porous material via two fields, the 
\textit{strain} $\varepsilon$ and the \textit{fluid density}
(measured with respect to the solid reference volume) $m$. 
We then assume that the possible motions of the system are those 
such that the variation of the action 
equals the opposite of the time integral of the virtual work of the 
dissipative forces corresponding to the variation of the fields.
Solving the variational problem we get the equations of motion. 

The explicit form of the equations of motion depends
not only on the form of the action but also on the way in which 
the dissipation phenomenon is modeled~\citep{Nield,LC}. 
Assuming the potential energy to be quadratic 
in the first derivatives of the strain and of the fluid 
density, the evolution is described by 
an Allen--Cahn--like equation for the field $m$ provided 
that dissipative forces are assumed
to be proportional to the second derivative of the 
velocity of the fluid.
We refer to this assumption as to the Stokes dissipation.
On the other hand, we find that 
the evolution is described by 
a Cahn--Hilliard--like equation for the field $m$ provided 
dissipation is modeled via the Darcy law, namely, the 
dissipative forces are proportional to the
velocity of the fluid.

It is possible to give a nice physical interpretation to 
the fact that the evolution is described by the Allen--Cahn or the 
Cahn--Hilliard--like equations depending on the way 
in which dissipation is
put into the game.
With homogenization arguments it has been proven that 
when the 
characteristic length
at the local $(\ell)$ and at the macroscopic $(L)$ scales separate, 
say $\ell/L\ll1$,
then the resistance experienced by the fluid when flowing through the
porous material can be described by the Darcy law 
\citep{Ene&SanchezP75}.
On the other hand when the separation of scales is poor, the Darcy limit
unfairly approximates the flow, since the macroscopic characteristic
length is of the same order of magnitude as the pore characteristic size 
\citep{Auriault2005}.
In other words in this case the solid grains have to be considered 
as microporous themselves.

Moreover, 
we note that 
the steady Stokes equations, 
which describe
the behavior of the fluid at the microscopic level, 
can be upscaled to a set of
equations describing a non--Stokesian flow only 
if the size of the solid obstacles to
the fluid motion have a critical size with respect to the 
inter--obstacle distance,
see for more details \citep[Theorem~3.1]{Allaire}. 
This condition provide the so--called
Brinkman limit. Conversely when the ostables are
too small the homogenized equations coincide with the Stokes equations.
Thus the Allen--Cahn--like equations, which correspond to a purely Stokesian
flow, are controlled by the  viscosity of the fluid; on
the other hand the Cahn--Hilliard--like equations, which correspond to a Darcy flow,
are controlled by the permeability of the solid skeleton. In particular 
the Cahn--Hilliard--like model reads, within the framework of 
soil consolidation theory, as a generalization to strain gradient materials
of the one--dimensional Terzaghi problem, see \citep{MIIS}.

The paper is organized as follows. We first introduce the model 
in Section~\ref{s:model}. In Section~\ref{s:lin}  
we assume small 
variations of the fields with respect to some reference values, 
and we find the Allen-Cahn and Cahn--Hilliard--like equations. 
In Section~\ref{s:risultati} we consider a special model allowing 
to describe solid--fluid segregation in consolidation~\citep{CIS2011,CIS2009,CIS2010} 
and 
we give a numerical study of the equations deduced in the previous sections.

\section{The model}
\label{s:model}
\par\noindent
In this section we introduce the one dimensional poromechanical model
whose geometrically linearized version will be studied 
in the following sections. Kinematics will be briefly resumed starting from the
general statement of the model \citep{Coussy} together with some
particular issue introduced in \citep{SIIM}. 
The equations governing 
the behavior of the porous system will then be deduced prescribing 
the conservative part of
the constitutive law through a suitable potential energy density 
$\Phi$ and the dissipative 
contributions through purely Stokes or Darcy terms. 
Special emphasis will be given to 
the boundary conditions and  
the extended definition of the essential 
and the natural ones will be discussed.

\subsection{Poromechanics setup}
\label{s:setup}
\par\noindent
Let $B_\rr{s}:=[\ell_1,\ell_2]\subset\bb{R}$, with 
$\ell_1,\ell_2\in\bb{R}$, 
and $B_\rr{f}:=\bb{R}$ be the \textit{reference}
configurations
for the solid and fluid components~\citep{Coussy}. 
The \textit{solid placement} 
$\chi_\rr{s}:B_\rr{s}\times\bb{R}\to\bb{R}$ is a $C^2$ function such that 
the map $\chi_\rr{s}(\cdot,t)$, 
associating to each $X_\rr{s}\in B_\rr{s}$
the position occupied at time $t$ by the particle labeled 
by $X_\rr{s}$ in the reference configuration $B_\rr{s}$,
is a $C^2$--diffeomorphism.
The \textit{fluid placement} map 
$\chi_\rr{f}:B_\rr{f}\times\bb{R}\to\bb{R}$
is defined analogously.
The \textit{current configuration} $B_t:=\chi_\rr{s}(B_\rr{s},t)$ at time 
$t$ is the set of positions 
of the superposed solid and fluid particles.

Consider the $C^2$ function
$\phi:B_\rr{s}\times\bb{R}\to B_\rr{f}$ 
such that 
$\phi(X_\rr{s},t)$ is 
the fluid particle that at time $t$ occupies the 
same position of the solid particle $X_\rr{s}$; 
assume, also, that $\phi(\cdot,t)$ is a $C^2$--diffeomorphism 
mapping univocally a solid particle 
into a fluid one.
The three fields $\chi_\rr{s}$, $\chi_\rr{f}$, and $\phi$ are 
not at all independent; indeed, by definition, we immediately have that 
$\chi_\rr{f}(\phi(X_\rr{s},t),t)=\chi_\rr{s}(X_\rr{s},t)$ 
for any $X_\rr{s}\in B_\rr{s}$ and $t\in\bb{R}$.

In the sequel we shall often use the inverse functions of the
space sections of the field $\chi_\rr{f}$, $\chi_\rr{s}$, and $\phi$. 
We shall misuse the notation and let 
$\phi^{-1}(\cdot,t)$ be the inverse of the map 
$X_\rr{s}\to\phi(X_\rr{s},t)$ at a given time $t$.
Similarly we shall also consider 
$\chi_\rr{s}^{-1}(\cdot,t)$ and 
$\chi_\rr{f}^{-1}(\cdot,t)$.
%We, then, let $\bar\phi:B_\rr{f}\times\bb{R}\to B_\rr{s}$ 
%be such that for any $t$ the map $\bar\phi(\cdot,t)$ is the inverse 
%function of $\phi(\cdot,t)$; we assume $\bar\phi$ to be a 
%$C^2$ function.
%The functions $\bar\chi_\rr{s}:B_t\times\bb{R}\to B_\rr{s}$ and
%$\bar\chi_\rr{f}:B_t\times\bb{R}\to B_\rr{f}$ are defined similarly.

The Lagrangian velocities are two maps associating with each time and 
each point in the solid and fluid reference space the velocities of the 
corresponding solid and fluid particles at the 
specified time. 
More precisely,
the \textit{Lagrangian velocities} are the two maps
$u_\alpha:B_\alpha\times\bb{R}\to\bb{R}$ 
defined by setting
\begin{equation}
\label{vel-lagr}
u_\alpha(X_\alpha,t):=\frac{\partial\chi_\alpha}{\partial t}(X_\alpha,t)
\end{equation}
for any $X_\alpha\in B_\alpha$, where $\alpha=\rr{s},\rr{f}$.
We also consider the \textit{Eulerian velocities} 
$v_\alpha:B_t\times\bb{R}\to\bb{R}$ associating
with each point $x\in B_t$ and for each time $t\in\bb{R}$ the velocities
of the solid and fluid particle occupying the place $x$ at time $t$;
more precisely we set 
$v_\alpha(x,t):=u_\alpha(\chi^{-1}_\alpha(x,t),t)$.

In studying the dynamics of the porous system one can arbitrarily 
choose two among the three fields 
$\chi_\rr{s}$, $\chi_\rr{f}$, and $\phi$. 
Since the reference configuration $B_\rr{s}$ of the solid component 
is know a priori, 
a good choice appears to be that of expressing all the dynamical observables 
in terms of the fields $\chi_\rr{s}$ and $\phi$ which are defined 
on $B_\rr{s}$.
Consider, for instance, the Lagrangian velocity $u_\rr{f}$ of the 
fluid component which is defined on $B_\rr{f}\times\bb{R}$; we prove 
that for any $X_\rr{s}\in B_\rr{s}$ and $t\in\bb{R}$
\begin{equation}
\label{vf}
u_\rr{f}(\phi(X_\rr{s},t),t)
=
\dot{\chi}_\rr{s}(X_\rr{s},t)
-
\frac{\dot\phi(X_\rr{s},t)}
     {\phi'(X_\rr{s},t)}
\chi_\rr{s}'(X_\rr{s},t)
\end{equation}
where the dot and the prime denote, here and in the sequel, 
the partial derivative with respect to time and to the 
space variable $X_\rr{s}$ respectively. 
The above equation 
gives the expression of the fluid Lagrangian velocity in terms of 
$\chi_\rr{s}$ and $\phi$.

In order to prove (\ref{vf}),
note that
$\chi_\rr{f}(X_\rr{f},t)=\chi_\rr{s}(\phi^{-1}(X_\rr{f},t),t)$ for 
any $X_\rr{f}$ and $t$, where we have used the 
definition of fluid placement map.
By the definition (\ref{vel-lagr}) of the Lagrangian velocity of the 
fluid, we then get 
\begin{displaymath}
u_\rr{f}(X_\rr{f},t)
=
\chi_\rr{s}'(\phi^{-1}(X_\rr{f},t),t)
\dot{\phi^{-1}}(X_\rr{f},t)
+
\dot\chi_\rr{s}(\phi^{-1}(X_\rr{f},t),t)
\end{displaymath}
Since $\phi^{-1}(\cdot,t)$ is the inverse function of $\phi(\cdot,t)$,
we have that
$\phi(\phi^{-1}(X_\rr{f},t),t)=X_\rr{f}$
for any $X_\rr{f}\in B_\rr{f}$. By deriving this equality
with respect to time we get
\begin{displaymath}
\phi'(\phi^{-1}(X_\rr{f},t),t)\dot{\phi^{-1}}(X_\rr{f},t)
+
\dot\phi(\phi^{-1}(X_\rr{f},t),t)
=0
\end{displaymath}
The last two equations yield (\ref{vf}).

\subsection{Variational principle}
\label{s:variational}
\par\noindent
In order to write the equations of motion of the system we use a 
variational approach much similar to that developed in~\citep{SIIM};
the differences between the two computations will be pointed out.
For the sake of self--sufficiency we shall
report the computation in detail for the one--dimensional case.

It is natural to assume that, if the system is acted upon only by 
conservative forces, its dynamics is described by a 
\textit{Lagrangian density} 
$\mathscr{L}$,
relative to the solid reference configuration space volume,
depending on the space variable $X_\rr{s}$ and on time 
through (in principle)
$\dot\chi_\rr{s}$, $\dot\phi$, $\chi''_\rr{s}$, $\phi''$,
$\chi'_\rr{s}$, $\phi'$, $\chi_\rr{s}$, and $\phi$.
The Lagrangian density is
equal to the 
\textit{kinetic energy density} minus
the \textit{overall potential energy density} accounting for both 
the internal and the external conservative forces.
The equations of motion for the two fields $\chi_\rr{s}$ and $\phi$ can be 
derived assuming that the possible motions of the system in an 
interval of time $(t_1,t_2)\subset\bb{R}$
are those such that the fields $\chi_\rr{s}$ and $\phi$ are 
extremals
for the \textit{action functional}
\begin{equation}
\label{azione}
 A(\chi_\rr{s},\phi)
 :=
 \int_{t_1}^{t_2}\rr{d}t
 \int_{B_\rr{s}}\rr{d}X_\rr{s}
 \,
 \mathscr{L}(
            \dot\chi_\rr{s}(X_\rr{s},t),
            \dots,
            %\dot\phi,
            %\chi''_\rr{s},
            %\phi'',
            %\chi'_\rr{s},
            %\phi',
            %\chi_\rr{s},
            \phi(X_\rr{s},t))
\end{equation}
in correspondence of the independent variations 
of the two fields
$\chi_\rr{s}$ and $\phi$ on $B_\rr{s}\times(t_1,t_2)$.
In other words any possible motion of the system in the 
considered interval is a solution of the Euler--Lagrange equations
associated to the variational principle $\delta A=0$.

A different variational principle is needed if the 
fluid component of the system is acted upon by dissipative
forces; the virtual work made by these forces
must be taken into account.
Consider the independent variations 
$\delta\chi_\rr{s}$ 
and $\delta\phi$ of the two fields $\chi_\rr{s}$ and $\phi$ and
denote by $\delta W$ the corresponding elementary 
\textit{virtual work}
made by the dissipative forces acting on the fluid component.
The possible motions of the system, see for instance 
\citep[Chapter~5]{Blanchard},
in an interval of time $(t_1,t_2)\subset\bb{R}$
are those such that the fields $\chi_\rr{s}$ and $\phi$ satisfies
the variational principle
\begin{equation}
\label{vardiss}
\delta A
=
-\int_{t_1}^{t_2}\delta W\,\rr{d}t
\end{equation}
namely, the variation of the the action integral in 
correspondence of a possible motion is equal to the 
integral over time of minus the virtual work of the dissipative 
forces corresponding to the considered variation of the fields. 

Up to now the discussion has been conducted along very general lines. 
In the following sections we shall specify first the form of the 
virtual work of the drag forces and then that of the action. 
We shall obtain, finally, an explicit 
representation of the equations of motion (see (\ref{em06}) and 
(\ref{em06bc})) with suitable boundary conditions.

\subsection{Virtual work of the dissipation forces}
\label{s:lvdis}
\par\noindent
The way in which dissipation has to be 
introduced in saturated porous media 
models is still under debate, see for instance \citep{Nield}.
In particular according to the effectiveness of the hypothesis 
of separation of scales, between the local and macroscopic level,
Darcy's or Stokes' effects are accounted for.

We first consider the so--called Darcy effect, i.e., the dissipation 
due to forces proportional to the velocity of the 
fluid component measured with respect to the solid.
By following phenomenological arguments \citep{Bear} or
by developing suitable homogenization schemes \citep{Ene&SanchezP75}
it is seen that 
the fluid flow is, in this case, controlled by the permeability of
the porous material.
% Consider, thus, the independent variations 
% $\delta\chi_\rr{s}$ 
% and $\delta\phi$ of the two fields $\chi_\rr{s}$ and $\phi$.
A natural expression for the virtual work of the dissipation forces 
acting on the fluid component and taking into account the Darcy 
effect is 
\begin{equation}
\label{lv01}
 \delta W_\rr{D}:=
 -
 \int_{B_t}
 D[v_\rr{f}(x,t)-v_\rr{s}(x,t)]
 [\delta\chi_\rr{f}(\chi^{-1}_\rr{f}(x,t),t)
 -
 \delta\chi_\rr{s}(\chi^{-1}_\rr{s}(x,t),t)]
\,\rr{d}x
\end{equation}
where $D>0$ is a constant proportional to the inverse of permeability and
$\delta\chi_\rr{f}$ is the variation of the field $\chi_\rr{f}$
induced by the independent variations $\delta\chi_\rr{s}$ and 
$\delta\phi$.
The quantity 
$-D[v_\rr{f}-v_\rr{s}]$ is the 
dissipative force density
(relative to the current configuration space volume)
which depends on the kinematic fields only through the 
velocity of the fluid component relative to the solid. 
Apparently equation (\ref{lv01}) for the virtual work of Darcy dissipative
forces is consistent with the classical expression of Darcy dissipation 
associated with the viscous flow through a porous continuum, see \citep{Coussy}.

Following the recipe described above we have to express the
virtual work in terms of the field $\chi_\rr{s}$ and $\phi$.
We first remark~\citep[Appendix C]{SIIM} that
\begin{equation}
\label{sv02}
\delta\chi_\rr{f}(\phi(X_\rr{s},t),t)
-
\delta\chi_\rr{s}(X_\rr{s},t)
=
-\frac{\chi'_\rr{s}(X_\rr{s},t)}{\phi'(X_\rr{s},t)}\,\delta\phi(X_\rr{s},t)
\end{equation}
for any $X_\rr{s}\in B_\rr{s}$ and $t\in\bb{R}$.
Moreover, noted that 
$u_\rr{s}(X_\rr{s},t)=\dot\chi_\rr{s}(X_\rr{s},t)$, 
by using the definition of the Eulerian velocities, from (\ref{vf}) 
we get 
\begin{equation}
\label{lv05}
v_\rr{f}(x,t)
-
v_\rr{s}(x,t)
=
-
\frac{\dot\phi(\chi^{-1}_\rr{s}(x,t),t)}
     {\phi'(\chi^{-1}_\rr{s}(x,t),t)}
\chi_\rr{s}'(\chi^{-1}_\rr{s}(x,t),t)
\end{equation}
In order to get an expression of the virtual work comparable to that
of the action, we have to rewrite the integral as an integral extended 
to the solid reference configuration.
Given $t$, 
by performing the change of variables $x=\chi_\rr{s}(X_\rr{s},t)$
in the integral in (\ref{lv01}) and by using equations (\ref{sv02}) 
and (\ref{lv05}) we get 
\begin{equation}
\label{lv02}
 \delta W_\rr{D}
 =
 -D
 \int_{B_\rr{s}}
 \frac{\dot\phi(X_\rr{s},t)}
      {\phi'(X_\rr{s},t)}
  \chi_\rr{s}'(X_\rr{s},t)
 \,
 \frac{\chi'_\rr{s}(X_\rr{s},t)}{\phi'(X_\rr{s},t)}
 \,\delta\phi(X_\rr{s},t)
 \,\chi'_\rr{s}(X_\rr{s},t)
 \,\rr{d}X_\rr{s}
\end{equation}

For the virtual work of the drag forces it is often considered 
also the contribution of the Stokes effect~\citep{Nield}, namely, 
the virtual work of dissipative forces controlled by the second derivative of the 
velocity of the fluid component relative to the solid. 
This contribution is typically the leading term in the case when 
separation of scales between the local and the macroscopic level is no more 
valid. In these cases 
dissipative stresses must be taken into account at 
macroscopic level too \citep{Allaire} and, thus, the second 
derivative of the fluid velocity appears in the governing equations.
% As already noted, the way in which 
% this effect is introduced in the poromechanics equation of motion
% is still under debate~\citep{Nield}.
A naive choice 
for the virtual work of the dissipation forces
would be 
\begin{equation} 
\label{lv12}
 -
 \int_{B_t}
 S
 %\Big(
 %\frac{\partial^2}{\partial x^2}
 [v_\rr{f}(x,t)-v_\rr{s}(x,t)]''
 %\Big)
 \,
 [\delta\chi_\rr{f}(\chi^{-1}_\rr{f}(x,t),t)
 -
 \delta\chi_\rr{s}(\chi^{-1}_\rr{s}(x,t),t)]
\,\rr{d}x
\end{equation}
with $S$ a positive constant; note that the second derivative 
is taken with respect to 
the space variable $x$ in the current configuration $B_t$. Following the same thermodynamic 
argument as that developed for Darcy dissipation, to the expression above of
the virtual work corresponds the effective power 
\begin{equation}
\label{lv13}
 -
 \int_{B_t}
 S 
 %\Big(
 %\frac{\partial^2}{\partial x^2}
 [v_\rr{f}(x,t)-v_\rr{s}(x,t)]''
 %\Big)
 \,
 [v_\rr{f}(x,t)-v_\rr{s}(x,t)]
\,\rr{d}x
\end{equation}
Since it is not negative definite, the proposed expression of the 
virtual work is not physically acceptable.
This problem has been nicely solved in~\citep{SIIM} where 
the following expression of the virtual work 
\begin{equation}
\label{lv10}
\delta W_\rr{S}
:=
 -
 \int_{B_t}
 S
 %\Big(
 %\frac{\partial}{\partial x}
 [v_\rr{f}(x,t)-v_\rr{s}(x,t)]'
 %\Big)
 \,
 %\frac{\partial}{\partial x}
 [\delta\chi_\rr{f}(\chi^{-1}_\rr{f}(x,t),t)
 -
 \delta\chi_\rr{s}(\chi^{-1}_\rr{s}(x,t),t)]'
\,\rr{d}x
\end{equation}
with $S>0$, has been proposed. Equation (\ref{lv10}) for the virtual
work of dissipative forces is indeed definitely consistent with
the expression of the fluid dissipation when the fluid velocity
satisfies the Navier--Stokes equations. As a matter of facts replacing 
the virtual displacement with velocities in the 
above expression yields the effective power of the Stokes 
drag forces which results to be negative definite.
Note that, by performing an integration by part of (\ref{lv10}),
a term similar to (\ref{lv12}) is found plus a contribution 
on the boundary of the actual domain. 

We rewrite, now, the expression of the Stokes virtual work 
as we have done for the Darcy contribution and we shall find an equation 
similar to (\ref{lv02}). In other words we shall express all the 
functions appearing in (\ref{lv10}) in terms of the fields $\chi_\rr{s}$ and 
$\phi$ and, via the changing of variable $x=\chi_\rr{s}(X_\rr{s},t)$ for 
any fixed $t$, extend the integral to the domain $B_\rr{s}$. 
The computation is performed in the \ref{s:bold}; the result 
is 
\begin{equation}
\label{lv16}
\delta W_\rr{S}
\!=\!
 -
 S 
 \!\!
 \int_{B_\rr{s}}
      \frac{1}{(\phi')^3}
      (\dot\phi\phi'\chi''_\rr{s}+\dot\phi'\phi'\chi'_\rr{s}
       -\dot\phi\phi''\chi'_\rr{s})
 \Big[
      \delta\phi'
  +
  \frac{1}{\chi'_\rr{s}\phi'}
      (\chi''_\rr{s}\phi'-\phi''\chi'_\rr{s})
      \delta\phi
 \Big]\,\rr{d}X_\rr{s}
\end{equation}
where we have omitted, for simplicity,
to write explicitly that 
all the functions are evaluated in $(X_\rr{s},t)$.
%In Appendix~\ref{s:dissb} we present a different derivation of (\ref{lv16})
%based on a sort of Rayleigh dissipation function.

From (\ref{lv02}) and (\ref{lv16}) we finally have the following expression
for the virtual work of the drag forces obtained by taking into 
account both the Darcy and the Stokes effects
\begin{equation}
\label{lv18}
\delta W
:=
\delta W_\rr{D}+\delta W_\rr{S}
=
-\int_{B_\rr{s}}(R\delta\phi+Q\delta\phi')
\;\rr{d}X_\rr{s}
\end{equation}
where we have set 
\begin{equation}
\label{lv19}
R:=D\frac{1}{(\phi')^2}\dot\phi(\chi'_\rr{s})^3
   +S
      \frac{1}{\chi'_\rr{s}(\phi')^4}
      (\dot\phi\phi'\chi''_\rr{s}+\dot\phi'\phi'\chi'_\rr{s}
       -\dot\phi\phi''\chi'_\rr{s})
      (\chi''_\rr{s}\phi'-\phi''\chi'_\rr{s})
\end{equation}
and
\begin{equation}
\label{lv20}
Q=S\frac{1}{(\phi')^3}
      (\dot\phi\phi'\chi''_\rr{s}+\dot\phi'\phi'\chi'_\rr{s}
       -\dot\phi\phi''\chi'_\rr{s})
\end{equation}

Since we need an expression of the virtual work in terms of the 
variations $\delta\chi_\rr{s}$ and $\delta\phi$ of the basic fields, 
we integrate (\ref{lv18}) by parts and get
\begin{equation}
\label{lv22}
\delta W
=
-
(Q\delta\phi)_{\ell_1}^{\ell_2}
-\int_{B_\rr{s}}(R-Q')\delta\phi
\;\rr{d}X_\rr{s}
\end{equation}
As already noticed only in the Brinkman limit Darcy and Stokes dissipative
effects can be accounted for together; on the other hand, in the general case,
they are treated separately. This is indeed what we shall do in 
Section \ref{s:lin}.

\subsection{Variation of the action}
\label{s:azione}
\par\noindent
In order to write explicitly the variation of the action we 
specify, now, the form of the Lagrangian density.
In the sequel we shall not consider the inertial 
effects, so that, the Lagrangian density will be the opposite of 
the potential energy 
density associated to both the internal and external 
conservative forces. 
As it has been shown in~\citep{SIIM} (see equation~(18) therein)
it is reasonable to assume that the potential energy density 
depends on the space and time variable only via two 
physically relevant functions: 
the strain of the solid and a properly normalized fluid mass 
density~\citep{CIS2011,CIS2009,CIS2010,SIIM}.

More precisely, 
consider the Jacobian $\chi'_\rr{s}(X_\rr{s},t)$ of the solid placement map,
which  measures the ratio between current and 
reference volumes of the solid component, and let 
\begin{equation}
\label{deformazione}
\varepsilon(X_\rr{s},t):=[(\chi'_\rr{s}(X_\rr{s},t))^2-1]/2
\end{equation}
be the \textit{strain field}.
Let $\varrho_{0,\rr{f}}:B_\rr{f}\to\bb{R}$ 
be the fluid reference \textit{density}; we define the 
\textit{fluid mass density} field
\begin{equation}
\label{densita}
m_\rr{f}(X_\rr{s},t)
 :=\varrho_{0,\rr{f}}(\phi(X_\rr{s},t))
   \phi'(X_\rr{s},t)
\end{equation}
Assuming that the mass is conserved, it is proven~\citep{SIIM}
that the field $m_\rr{f}$ can be interpreted as 
the fluid mass density measured with respect to the 
solid reference volume.

We assume the total potential energy density $\Phi$ to depend 
on the fields 
$m_\rr{f}$ and $\varepsilon$ and on their space
derivative
$m'_\rr{f}$ and $\varepsilon'$.
Since 
$m_\rr{f}=\varrho_{0,\rr{f}}(\phi)\phi'$,
$m'_\rr{f}=\varrho'_{0,\rr{f}}(\phi)(\phi')^2+\varrho_{0,\rr{f}}(\phi)\phi''$,
$\varepsilon=((\chi'_\rr{s})^2-1)/2$,
and $\varepsilon'=\chi'_\rr{s}\chi''_\rr{s}$, we have that 
the Lagrangian density $\mathscr{L}=-\Phi$ depends on the space and time 
variables through the fields 
$\phi$, $\phi'$, $\phi''$, $\chi'_\rr{s}$, and 
$\chi''_\rr{s}$.
We then have 
\begin{displaymath}
\delta A
=
\int_{t_1}^{t_2} \rr{d}t
\int_{B_\rr{s}} \rr{d}X_\rr{s}
\Big(
     \frac{\partial\mathscr{L}}{\partial\phi}\,\delta\phi
     +
     \frac{\partial\mathscr{L}}{\partial\phi'}\,\delta\phi'
     +
     \frac{\partial\mathscr{L}}{\partial\phi''}\,\delta\phi''
     +
     \frac{\partial\mathscr{L}}{\partial\chi'_\rr{s}}\,\delta\chi'_\rr{s}
     +
     \frac{\partial\mathscr{L}}{\partial\chi''_\rr{s}}\,\delta\chi''_\rr{s}
\Big)
\end{displaymath}
By using the above expression for the Lagrangian density we have that 
\begin{equation}
\label{va01}
\begin{array}{rl}
{\displaystyle
 \delta A
 =
 -
 \int_{t_1}^{t_2}\!\!\!\rr{d}t
 \int_{B_\rr{s}}\!\!\!\rr{d}X_\rr{s}
 \Big[
}
&
\!\!\!\!\!
{\displaystyle
      \Big(
      \frac{\partial\Phi}{\partial m_\rr{f}}
      \frac{\partial m_\rr{f}}{\partial\phi}
      +
      \frac{\partial\Phi}{\partial m'_\rr{f}}
      \frac{\partial m'_\rr{f}}{\partial\phi}
      \Big)
      \,\delta\phi
}
\vphantom{\bigg\{_\big\{}
\\
&
\!\!\!\!\!
{\displaystyle
      +
      \Big(
      \frac{\partial\Phi}{\partial m_\rr{f}}
      \frac{\partial m_\rr{f}}{\partial\phi'}
      +
      \frac{\partial\Phi}{\partial m'_\rr{f}}
      \frac{\partial m'_\rr{f}}{\partial\phi'}
      \Big)
      \,\delta\phi'
     +
     \frac{\partial\Phi}{\partial m'_\rr{f}}
     \frac{\partial m'_\rr{f}}{\partial\phi''}
     \,\delta\phi''
}
\\
&
\!\!\!\!\!
{\displaystyle
     +
      \Big(
      \frac{\partial\Phi}{\partial\varepsilon}
      \frac{\partial\varepsilon}{\partial\chi'_\rr{s}}
      +
      \frac{\partial\Phi}{\partial\varepsilon'}
      \frac{\partial\varepsilon'}{\partial\chi'_\rr{s}}
      \Big)
      \,\delta\chi'_\rr{s}
      +
      \frac{\partial\Phi}{\partial\varepsilon'}
      \frac{\partial\varepsilon'}{\partial\chi''_\rr{s}}
      \,\delta\chi''_\rr{s}
\Big]
}
\end{array}
\end{equation}
To get rid of the terms depending on the variations of the derivatives of the 
basic fields $\chi_\rr{s}$ and $\phi$ we integrate by parts. Via a standard 
computation we find
\begin{equation}
\label{va02}
\begin{array}{rl}
 \delta A
 =
&
{\displaystyle
 -
 \int_{t_1}^{t_2}\rr{d}t
 \int_{B_\rr{s}}\rr{d}X_\rr{s}
 \,
 \Big\{
}
{\displaystyle
      \Big[
      \frac{\partial\Phi}{\partial m_\rr{f}}
      \frac{\partial m_\rr{f}}{\partial\phi}
      +
      \frac{\partial\Phi}{\partial m'_\rr{f}}
      \frac{\partial m'_\rr{f}}{\partial\phi}
 \vphantom{\bigg\{_\big\{}
}
\\
&
{\displaystyle
 \phantom{
          -
          \int_{t_1}^{t_2}\rr{d}t
          \int_{B_\rr{s}}\rr{d}X_\rr{s}
          \,
          \Big\{
          \Big[
         }
      -
      \Big(
      \frac{\partial\Phi}{\partial m_\rr{f}}
      \frac{\partial m_\rr{f}}{\partial\phi'}
      +
      \frac{\partial\Phi}{\partial m'_\rr{f}}
      \frac{\partial m'_\rr{f}}{\partial\phi'}
      -
      \Big(
           \frac{\partial\Phi}{\partial m'_\rr{f}}
           \frac{\partial m'_\rr{f}}{\partial\phi''}
      \Big)'
      \Big)'
      \Big]
      \,\delta\phi
 \vphantom{\bigg\{_\big\{}
}
\\
&
{\displaystyle
 \phantom{
          -
          \int_{t_1}^{t_2}\rr{d}t
          \int_{B_\rr{s}}\rr{d}X_\rr{s}
          \,
          \Big\{
          \Big[
         }
      -
      \Big(
      \frac{\partial\Phi}{\partial\varepsilon}
      \frac{\partial\varepsilon}{\partial\chi'_\rr{s}}
      +
      \frac{\partial\Phi}{\partial\varepsilon'}
      \frac{\partial\varepsilon'}{\partial\chi'_\rr{s}}
      -
      \Big(
           \frac{\partial\Phi}{\partial\varepsilon'}
           \frac{\partial\varepsilon'}{\partial\chi''_\rr{s}}
      \Big)'
      \Big)'
      \,\delta\chi_\rr{s}
      \Big\}
 \vphantom{\bigg\{_\big\{}
}
\\
&
{\displaystyle
 -
 \int_{t_1}^{t_2}\rr{d}t
 \,
 \Big\{
      \Big(
           \frac{\partial\Phi}{\partial m'_\rr{f}}
           \frac{\partial m'_\rr{f}}{\partial\phi''}
           \delta\phi'
      \Big)_{\ell_1}^{\ell_2}
      +
      \Big(
           \frac{\partial\Phi}{\partial\varepsilon'}
           \frac{\partial\varepsilon'}{\partial\chi''_\rr{s}}
           \delta\chi'_\rr{s}
      \Big)_{\ell_1}^{\ell_2}
 \vphantom{\bigg\{_\big\{}
}
\\
&
{\displaystyle
 \phantom{
          -
          \int_{t_1}^{t_2}\rr{d}t
          \,
          \Big\{
         }
      +
      \Big[
      \Big(
      \frac{\partial\Phi}{\partial m_\rr{f}}
      \frac{\partial m_\rr{f}}{\partial\phi'}
      +
      \frac{\partial\Phi}{\partial m'_\rr{f}}
      \frac{\partial m'_\rr{f}}{\partial\phi'}
      -
      \Big(
           \frac{\partial\Phi}{\partial m'_\rr{f}}
           \frac{\partial m'_\rr{f}}{\partial\phi''}
      \Big)'
      \Big)\delta\phi
      \Big]_{\ell_1}^{\ell_2}
 \vphantom{\bigg\{_\big\{}
}
\\
&
{\displaystyle
 \phantom{
          -
          \int_{t_1}^{t_2}\rr{d}t
          \,
          \Big\{
         }
      +
      \Big[
      \Big(
      \frac{\partial\Phi}{\partial\varepsilon}
      \frac{\partial\varepsilon}{\partial\chi'_\rr{s}}
      +
      \frac{\partial\Phi}{\partial\varepsilon'}
      \frac{\partial\varepsilon'}{\partial\chi'_\rr{s}}
      -
      \Big(
           \frac{\partial\Phi}{\partial\varepsilon'}
           \frac{\partial\varepsilon'}{\partial\chi''_\rr{s}}
      \Big)'
      \Big)
      \,\delta\chi_\rr{s}
      \Big]_{\ell_1}^{\ell_2}
      \Big\}
}
\\
\end{array}
\end{equation}

\subsection{Equations of motion}
\label{s:motion}
\par\noindent
By using the variational principle (\ref{vardiss}), the equation 
(\ref{lv18}) for the virtual work of the drag forces, and 
the equation (\ref{va02}) for the variation of the action, 
we get 
the equations of motion 
\begin{equation}
\label{em04}
 \Big[\chi'_\rr{s}
 \Big(\frac{\partial\Phi}{\partial\varepsilon}
      -
      \Big(
      \frac{\partial\Phi}{\partial\varepsilon'}
      \Big)'
 \Big)
 \Big]'
 =0
\;\textrm{ and }\;
 \varrho_{0,\rr{f}}(\phi)
 \Big[\frac{\partial\Phi}{\partial m_\rr{f}}
      -
      \Big(
      \frac{\partial\Phi}{\partial m'_\rr{f}}
      \Big)'
 \Big]'
 =
 R-Q'
\end{equation}
and the boundary conditions
\begin{equation}
\label{em04bc}
\begin{array}{l}
{\displaystyle
 \Big\{
 \frac{\partial\Phi}{\partial m'_\rr{f}}
 \varrho_{0,\rr{f}}(\phi)
 \delta\phi'
 +
 \Big[
 \Big(\frac{\partial\Phi}{\partial m_\rr{f}}
      -
      \Big(
      \frac{\partial\Phi}{\partial m'_\rr{f}}
      \Big)'
 \Big)
 \varrho_{0,\rr{f}}(\phi)
 +
 \frac{\partial\Phi}{\partial m'_\rr{f}}
 \varrho'_{0,\rr{f}}(\phi)\phi'
 +Q
 \Big]
 \delta\phi
}
\vphantom{\bigg\{_\{}
\\
{\displaystyle
 \phantom{aaaaaaaaaaaaaaaaaaaaa}
 + 
 \frac{\partial\Phi}{\partial\varepsilon'}
 \chi'_\rr{s}
 \delta\chi'_\rr{s}
 +
 \Big(\frac{\partial\Phi}{\partial\varepsilon}
      -
 \Big(
      \frac{\partial\Phi}{\partial\varepsilon'}
 \Big)'
 \Big)
 \chi'_\rr{s}
 \delta\chi_\rr{s}
 \Big\}_{\ell_1}^{\ell_2}
 =0
}
\end{array}
\end{equation}
We remark that, though they are partially written in terms of 
the fields $m_\rr{f}$ 
and $\varepsilon$, the equations (\ref{em04}) are evolution 
equations for the two fields $\chi_\rr{s}$ and $\phi$. Only under a
brute approximation, see the geometrical linearization discussed in 
Section~\ref{s:lin},
they will reduce to a set of evolutionary equations 
for the fields $m_\rr{f}$ and $\varepsilon$. 

Notice that those equations are not simply the 
one dimensional particularization of the equations (30)--(32) in the 
paper~\citep{SIIM}, since there 
the reference mass density $\varrho_{0,\rr{f}}$ was assumed to be constant.
When infinite systems are considered not constant reference densities can 
be useful in order to have finite total fluid mass integral.
In the sequel we shall always consider finite systems so that 
it will be natural to assume the fluid reference density 
$\varrho_{0,\rr{f}}:B_\rr{f}\to\bb{R}$ to be constant. 
Under this hypothesis
we get the following simplified expression
\begin{equation}
\label{em06}
 \Big[\chi'_\rr{s}
 \Big(\frac{\partial\Phi}{\partial\varepsilon}
      -
      \Big(
      \frac{\partial\Phi}{\partial\varepsilon'}
      \Big)'
 \Big)
 \Big]'
 =0
\;\textrm{ and }\;
 \varrho_{0,\rr{f}}
 \Big[
 \frac{\partial\Phi}{\partial m_\rr{f}}
      -
      \Big(
      \frac{\partial\Phi}{\partial m'_\rr{f}}
      \Big)'
 \Big]'
 =
 R-Q'
\end{equation}
for the equations of motion and 
\begin{equation}
\label{em06bc}
\begin{array}{l}
{\displaystyle
 \Big\{
 \frac{\partial\Phi}{\partial m'_\rr{f}}
 \varrho_{0,\rr{f}}
 \delta\phi'
 +
 \Big[
 \Big(\frac{\partial\Phi}{\partial m_\rr{f}}
      -
      \Big(
      \frac{\partial\Phi}{\partial m'_\rr{f}}
      \Big)'
 \Big)
 \varrho_{0,\rr{f}}
 +Q
 \Big]
 \delta\phi
}
\vphantom{\bigg\{_\{}
\\
{\displaystyle
 \phantom{aaaaaaaaaaaaaaaa}
 + 
 \frac{\partial\Phi}{\partial\varepsilon'}
 \chi'_\rr{s}
 \delta\chi'_\rr{s}
 +
 \Big(\frac{\partial\Phi}{\partial\varepsilon}
      -
 \Big(
      \frac{\partial\Phi}{\partial\varepsilon'}
 \Big)'
 \Big)
 \chi'_\rr{s}
 \delta\chi_\rr{s}
 \Big\}_{\ell_1}^{\ell_2}
 =0
}
\end{array}
\end{equation}
for the boundary conditions. 
We stress that from now on the symbol $\varrho_{0,\rr{f}}$ denotes 
a real constant.

\section{Allen--Cahn and Cahn--Hilliard--like equations}
\label{s:lin}
\par\noindent
In this section we rewrite the equations of motion of the 
poromechanical system under the so called 
\textit{geometrical linearization} assumption, namely, 
in the case 
when only small deformations are present in the system.
We first introduce the \textit{displacement fields} 
$u(X_\rr{s},t)$ and $w(X_\rr{s},t)$ by setting
\begin{equation}
\label{gd00}
\chi_\rr{s}(X_\rr{s},t)=X_\rr{s}+u(X_\rr{s},t)
\;\textrm{ and }\;
\phi(X_\rr{s},t)=X_\rr{s}+w(X_\rr{s},t)
\end{equation}
for any $X_\rr{s}\in B_\rr{s}$ and $t\in\bb{R}$. We then 
assume that $u$ and $w$ are small, together with 
their space and time derivatives, and write the equations of motion 
up to the first order in $u$, $w$, and derivatives.
By using 
(\ref{deformazione}), (\ref{densita}), (\ref{lv19}), and (\ref{lv20})
we get   
\begin{equation}
\label{gd01}
m_\rr{f}=\varrho_{0,\rr{f}}(1+w'),\;
m:=m_\rr{f}-\varrho_{0,\rr{f}}=\varrho_{0,\rr{f}}w',\;
\varepsilon\approx u',\;
R\approx D\dot{w},
\textrm{ and }
Q\approx S\dot{w}'
\end{equation}
where $\approx$ means that all the terms of order larger than one 
have been neglected.

We have introduced above the field $m$. In the following we shall imagine 
$\Phi$ as a function of $m$ and $m'$ and the equations of motion and 
the boundary conditions will be written in terms of this field. 
This can be done easily; indeed, since $\varrho_{0,\rr{f}}$ is constant, 
we have that $\partial/\partial m_\rr{f}=\partial/\partial m$
and $m'_\rr{f}=m'$.
From (\ref{em06}), (\ref{em06bc}), and (\ref{gd01}), we get the 
equations of motion 
\begin{equation}
\label{gd02}
 \Big[
 (1+\varepsilon) 
 \Big(
 \frac{\partial\Phi}{\partial\varepsilon}
 -
 \Big(
 \frac{\partial\Phi}{\partial\varepsilon'}
 \Big)'
 \Big)
 \Big]'
 =0
\;\;\textrm{ and }\;\;
 \varrho_{0,\rr{f}}
 \Big[
 \frac{\partial\Phi}{\partial m}
      -
      \Big(
      \frac{\partial\Phi}{\partial m'}
      \Big)'
 \Big]'
 =
 D\dot{w}
 -S\dot{w}''
\end{equation}
and the associated boundary conditions
\begin{equation}
\label{gd02bc}
 \Big\{
 \frac{\partial\Phi}{\partial\varepsilon'}
 \delta\varepsilon
 +
 \frac{\partial\Phi}{\partial m'}
 \delta m
 +
 \Big[
 \Big(\frac{\partial\Phi}{\partial m}
      -
      \Big(
      \frac{\partial\Phi}{\partial m'}
      \Big)'
 \Big)
 \varrho_{0,\rr{f}}
 +\frac{S}{\varrho_{0,\rr{f}}}\dot{m}
 \Big]
 \delta w
 +
 \Big(\frac{\partial\Phi}{\partial\varepsilon}
      -
 \Big(
      \frac{\partial\Phi}{\partial\varepsilon'}
 \Big)'
 \Big)
 \delta u
 \Big\}_{\ell_1}^{\ell_2}
 \!\!\!\!
 =0
\end{equation}
A set of boundary conditions implying that (\ref{gd02bc}) 
are satisfied is
\begin{equation}
\label{gd03bc}
\Big(
 \frac{\partial\Phi}{\partial\varepsilon'}
 \delta\varepsilon
 +
 \frac{\partial\Phi}{\partial m'}
 \delta m
\Big)_{\ell_1,\ell_2}
 \!\!\!\!
=
 \Big[\frac{\partial\Phi}{\partial m}
      -
      \Big(
      \frac{\partial\Phi}{\partial m'}
      \Big)'
 +\frac{S}{\varrho^2_{0,\rr{f}}}\dot{m}
 \Big]_{\ell_1,\ell_2}
 \!\!\!\!
=
 \Big[\frac{\partial\Phi}{\partial\varepsilon}
      -
 \Big(
      \frac{\partial\Phi}{\partial\varepsilon'}
 \Big)'
 \Big]_{\ell_1,\ell_2}
 \!\!\!\!
=0
\end{equation}
where the notation above means that the functions in brackets are 
evaluated both in $\ell_1$ and $\ell_2$.
With this choice it is possible to fix the boundary conditions 
directly on fields $m$ and $\varepsilon$ (and derivatives).
The equations (\ref{gd02}) are evolution equations for the 
original kinematic fields, but, by treating separately the Darcy and the
Stokes effects, we can deduce a system of equations 
for the fields $m$ and $\varepsilon$.

The equations (\ref{gd02bc}) are an extension to the case of 
second gradient elasticity of classical natural and essential 
boundary conditions \citep{ISV}.
The first equation (\ref{gd03bc}) is the additional 
boundary condition due to the presence of the gradient terms in the 
potential energy density $\Phi$. This equation specifies essential boundary 
conditions on the derivatives of the displacement fields or natural 
boundary conditions on the so called double forces, see \citep{germain73}.
The generalized essential boundary conditions can be read as a 
prescription on the derivative of the independent fields $\chi_\rr{s}$ 
and $\phi$, see equations (\ref{deformazione}) and (\ref{densita}); whilst 
the extended natural boundary conditions prescribe, on one hand, 
the additional forces which the solid continuum is able to balance at
the boundary and, on the other, the wetting properties of the fluid
which fills the pores \citep{Seppecher89}.

The second and the third equations (\ref{gd03bc}) provide 
natural boundary conditions prescribing 
the chemical potential of the fluid
and the traction exerted on the overall porous solid, 
respectively.
These conditions extend the classical ones (which can be easily found in 
the literature) for a porous solid suffering internal stresses due to 
applied tractions and for a saturating fluid with fixed 
chemical potential; see \citep{baek} for more details. 
In this case we use the words generalized traction and chemical potential 
because of the additional contribution to stress 
($\partial \Phi/\partial\varepsilon$) 
and chemical potential ($\partial\Phi/\partial m$) provided by the spatial 
derivatives of the corresponding 
hyperstress fields, say $(\partial \Phi/\partial\varepsilon')'$
and $(\partial \Phi/\partial m')'$.

\textit{Pure Darcy dissipation.\/}
We consider the system of equations (\ref{gd02}) for $S=0$.
Recalling that $m=\varrho_{0,\rr{f}}w'$, 
by deriving the second between the equations
of motion with respect to $X_\rr{s}$, we get 
\begin{equation}
\label{gd03}
 \Big[
 (1+\varepsilon) 
 \Big(
 \frac{\partial\Phi}{\partial\varepsilon}
 -
 \Big(
 \frac{\partial\Phi}{\partial\varepsilon'}
 \Big)'
 \Big)
 \Big]'
 =0
\;\;\textrm{ and }\;\;
 \varrho_{0,\rr{f}}^2
 \Big[
 \Big(\frac{\partial\Phi}{\partial m}
      -
      \Big(
      \frac{\partial\Phi}{\partial m'}
      \Big)'
 \Big)
 \Big]''
 =
 D\dot{m}
\end{equation}
Those equations, together with the boundary conditions (\ref{gd03bc}),
are a partial differential equation problem for the 
two fields $m$ and $\varepsilon$. 
Note that, by exploiting the third boundary 
condition (\ref{gd03bc}) in $\ell_1$, we get the PDE problem
\begin{equation}
\label{problema-d}
\left\{
\begin{array}{l}
{\displaystyle
 \frac{\partial\Phi}{\partial\varepsilon}
 -
 \Big(
 \frac{\partial\Phi}{\partial\varepsilon'}
 \Big)'
 =0
\;\;\textrm{ and }\;\;
 \varrho_{0,\rr{f}}^2
 \Big[
      \frac{\partial\Phi}{\partial m}
      -
      \Big(
      \frac{\partial\Phi}{\partial m'}
      \Big)'
 \Big]''
 =
 D\dot{m}
\vphantom{\bigg\{_\big\{}
}
\\
{\displaystyle
\Big(
 \frac{\partial\Phi}{\partial\varepsilon'}
 \delta\varepsilon
 +
 \frac{\partial\Phi}{\partial m'}
 \delta m
\Big)_{\ell_1,\ell_2}
=
 \Big[\frac{\partial\Phi}{\partial m}
      -
      \Big(
      \frac{\partial\Phi}{\partial m'}
      \Big)'
 \Big]_{\ell_1,\ell_2}
=0
}
\\
\end{array}
\right.
\end{equation}

\textit{Pure Stokes dissipation.\/}
We consider the system of equations (\ref{gd02}) for $D=0$.
Recalling that $m=\varrho_{0,\rr{f}}w'$, 
the equations of motion (\ref{gd02}) 
become
\begin{equation}
\label{gd04}
 \Big[
 (1+\varepsilon) 
 \Big(
 \frac{\partial\Phi}{\partial\varepsilon}
 -
 \Big(
 \frac{\partial\Phi}{\partial\varepsilon'}
 \Big)'
 \Big)
 \Big]'
 =0
\;\;\textrm{ and }\;\;
 \Big[
      \frac{\partial\Phi}{\partial m}
      -
      \Big(
      \frac{\partial\Phi}{\partial m'}
      \Big)'
 +\frac{S}{\varrho_{0,\rr{f}}^2}\dot{m}
 \Big]'
 =
 0
\end{equation}
Those equations, together with the boundary conditions (\ref{gd03bc}),
are a partial differential equation problem for the 
two fields $m$ and $\varepsilon$. 
Note that, by exploiting the second and the third boundary 
condition (\ref{gd03bc}) in $\ell_1$, we get the PDE problem
\begin{equation}
\label{problema-b}
\left\{
\begin{array}{l}
{\displaystyle
 \frac{\partial\Phi}{\partial\varepsilon}
 -
 \Big(
 \frac{\partial\Phi}{\partial\varepsilon'}
 \Big)'
 =0
\;\;\textrm{ and }\;\;
      \frac{\partial\Phi}{\partial m}
      -
      \Big(
      \frac{\partial\Phi}{\partial m'}
      \Big)'
 +\frac{S}{\varrho_{0,\rr{f}}^2}\dot{m}
 =
 0
\vphantom{\bigg\{_\big\{}
}
\\
{\displaystyle
\Big(
 \frac{\partial\Phi}{\partial\varepsilon'}
 \delta\varepsilon
 +
 \frac{\partial\Phi}{\partial m'}
 \delta m
\Big)_{\ell_1,\ell_2}
=0
}
\\
\end{array}
\right.
\end{equation}

It is very interesting to note that, provided the potential energy density 
$\Phi$ depends on the space derivatives $m'$ and $\varepsilon'$ of the 
fields $m$ and $\varepsilon$ as a quadratic form with constant coefficients, 
it follows that the second between the equations of motion (\ref{problema-d}) 
becomes a Cahn--Hilliard--like equation with driving field still depending 
parametrically on $\varepsilon$, while 
the second between the equations of motion (\ref{problema-b}) 
becomes an Allen--Cahn--like equation with driving field depending 
parametrically on $\varepsilon$.
More precisely we specialize the model we are studying by choosing the 
second gradient part of the dimensionless potential energy, that is we assume 
\begin{equation}
\label{sec010}
\Phi(m',\varepsilon',m,\varepsilon)
 :=
 \frac{1}{2}[k_1(\varepsilon')^2+2k_2\varepsilon' m'+k_3(m')^2]
 +
 \Psi(m,\varepsilon)
\end{equation}
with $k_1,k_3>0$, $k_2\in\bb{R}$ such that $k_1k_3-k_2^2\ge0$.
These parameters provide energy penalties for 
the formation of interfaces; they have the physical dimensions of
squared lengths and, according with the above mentioned 
conditions, provide a well-grounded identification of the intrinsic
characteristic lengths of the one-dimensional solid and the fluid.

With this choice of the second gradient potential energy, 
in the pure Darcy case the PDE problem (\ref{problema-d}) becomes
\begin{equation}
\label{problema-d-ch}
\left\{
\begin{array}{l}
{\displaystyle
 \frac{\partial\Psi}{\partial\varepsilon}
 -
 (k_1\varepsilon''+k_2m'')
 =0
\;\;\textrm{ and }\;\;
 D\dot{m}
 =
 -\varrho_{0,\rr{f}}^2
 \Big(
      k_2\varepsilon''+k_3m''
      -
      \frac{\partial\Psi}{\partial m}
 \Big)''
\vphantom{\bigg\{_\big\{}
}
\\
{\displaystyle
\Big(
 (k_1\varepsilon'+k_2m')
 \delta\varepsilon
 +
 (k_2\varepsilon'+k_3m')
 \delta m
\Big)_{\ell_1,\ell_2}
\!\!\!\!
=
 \Big(
      k_2\varepsilon''+k_3m''
      -
      \frac{\partial\Psi}{\partial m}
 \Big)_{\ell_1,\ell_2}
\!\!\!\!
=0
}
\\
\end{array}
\right.
\end{equation}
while in the pure Stokes case the PDE problem (\ref{problema-b}) reads
\begin{equation}
\label{problema-s-ac}
\left\{
\begin{array}{l}
{\displaystyle
 \frac{\partial\Psi}{\partial\varepsilon}
 -
 (k_1\varepsilon''+k_2m'')
 =0
\;\;\textrm{ and }\;\;
 \frac{S}{\varrho_{0,\rr{f}}^2}\dot{m}
 =
  k_2\varepsilon''+k_3m''
   -
  \frac{\partial\Psi}{\partial m}
\vphantom{\bigg\{_\Big\{}
}
\\
{\displaystyle
\Big(
 (k_1\varepsilon'+k_2m')
 \delta\varepsilon
 +
 (k_2\varepsilon'+k_3m')
 \delta m
\Big)_{\ell_1,\ell_2}
=0
}
\\
\end{array}
\right.
\end{equation}

We discuss, now, some general properties of the PDE problems 
(\ref{problema-d-ch}) and (\ref{problema-s-ac}).
First of all we recall that an harmonic 
function of a single real variable equal to zero at the 
extremes of an interval is necessarily equal to zero in the whole 
interval. 
We then have that for both the two problems 
(\ref{problema-d-ch}) and (\ref{problema-s-ac})
the stationary solutions are the solutions of 
the \textit{stationary} problem
\begin{equation}
\label{problema-staz}
\left\{
\begin{array}{l}
{\displaystyle
 \frac{\partial\Psi}{\partial\varepsilon}
 -
 (k_1\varepsilon''+k_2m'')
 =0
\;\;\textrm{ and }\;\;
  k_2\varepsilon''+k_3m''
   -
  \frac{\partial\Psi}{\partial m}
  =
  0
\vphantom{\bigg\{_\Big\{}
}
\\
{\displaystyle
\Big(
 (k_1\varepsilon'+k_2m')
 \delta\varepsilon
 +
 (k_2\varepsilon'+k_3m')
 \delta m
\Big)_{\ell_1,\ell_2}
=0
}
\\
\end{array}
\right.
\end{equation}
widely 
studied in \citep{CIS2011,CIS2009,CIS2010}.

The dissipative character of the physical problem we are studying 
reflects in the existence of the not increasing energy functional 
\begin{equation}
\label{liapunov}
\begin{array}{l}
{\displaystyle
F(\chi_\rr{s},\phi)
:=
\!\!
\int_{\ell_1}^{\ell_2}
\!\!
\Phi(m',\varepsilon',m,\varepsilon)\,\rr{d}X_\rr{s}
\vphantom{\bigg\{_\big\}}
}
\\
{\displaystyle
 \phantom{aaaaaaaaa}
=
\!\!
\int_{\ell_1}^{\ell_2}
\Big[
 \frac{1}{2}[k_1(\varepsilon')^2+2k_2\varepsilon' m'+k_3(m')^2]
 +
 \Psi(m,\varepsilon)
\Big]
\rr{d}X_\rr{s}
}
\end{array}
\end{equation}
for both the systems
(\ref{problema-d-ch}) and (\ref{problema-s-ac}).
To prove this we compute the time derivative  of the functional $F$ 
evaluated on the fields 
$\chi_\rr{s}(X_\rr{s},t)$ and $\phi(X_\rr{s},t)$.
We first get 
\begin{displaymath}
\frac{\rr{d}F}{\rr{d}t}
=
\int_{\ell_1}^{\ell_2}
\Big[
     (k_1\varepsilon'+k_2m')\dot{\varepsilon}'
     +
     (k_2\varepsilon'+k_3m')\dot{m}'
     +\frac{\partial\Psi}{\partial m}\dot{m}
     +\frac{\partial\Psi}{\partial \varepsilon}\dot{\varepsilon}
\Big]
\rr{d}X_\rr{s}
\end{displaymath}
and, by integrating by parts, we obtain 
\begin{equation}
\label{derF}
\begin{array}{rcl}
{\displaystyle
 \frac{\rr{d}F}{\rr{d}t}
}
&\!\!=&\!\!
{\displaystyle
     \big[(k_1\varepsilon'+k_2m')\dot{\varepsilon}
     + (k_2\varepsilon'+k_3m')\dot{m}\big]_{\ell_1}^{\ell_2}
}
\vphantom{\bigg\{}
\\
&&\!\!
{\displaystyle
 +\int_{\ell_1}^{\ell_2}
 \Big[
     -(k_1\varepsilon''+k_2m'')\dot{\varepsilon}
     -(k_2\varepsilon''+k_3m'')\dot{m}
     +\frac{\partial\Psi}{\partial m}\dot{m}
     +\frac{\partial\Psi}{\partial \varepsilon}\dot{\varepsilon}
 \Big]
 \rr{d}X_\rr{s}
}
\\
\end{array}
\end{equation}

Assume, first, that the fields $m$ and $\varepsilon$ are solution 
of the Allen--Cahn--like evolution equations (\ref{problema-s-ac}), i.e.,
the PDE problems describing the evolution of the system when 
Stokes dissipation is considered. Moreover, assume that the boundary 
conditions have been chosen on the fields or on their first space 
derivatives so that those in (\ref{problema-s-ac}) are fulfilled, 
we have that 
\begin{equation}
\label{acdecr}
 \frac{\rr{d}F}{\rr{d}t}
 =
 -\frac{S}{\varrho_{0,\rr{f}}^2}
  \int_{\ell_1}^{\ell_2}
  \dot{m}^2
  \rr{d}X_\rr{s}
\end{equation}
which proves that the functional $F$ in not increasing along 
the motions of the Allen--Cahn--like system
(\ref{problema-s-ac}).

Assume, now, that the fields $m$ and $\varepsilon$ are solution 
of the Cahn--Hilliard--like evolution equations (\ref{problema-d-ch}), i.e.,
the PDE problems describing the evolution of the system when 
Darcy dissipation is considered. Moreover, assume that the boundary 
conditions have been chosen on the fields or on their first space 
derivatives so that those in (\ref{problema-d-ch}) are fulfilled, 
from (\ref{derF}) 
we have that 
\begin{displaymath}
 \frac{\rr{d}F}{\rr{d}t}
 =
 \frac{\varrho_{0,\rr{f}}^2}{D}
 \int_{\ell_1}^{\ell_2}
 \Big[
     -(k_2\varepsilon''+k_3m'')
     +\frac{\partial\Psi}{\partial m}
 \Big]
 \Big[
     -(k_2\varepsilon''+k_3m'')
     +\frac{\partial\Psi}{\partial m}
 \Big]''
  \rr{d}X_\rr{s}
\end{displaymath}
By integrating by parts and exploiting the second boundary condition 
in (\ref{problema-d-ch}) we get 
\begin{equation}
\label{chdecr}
 \frac{\rr{d}F}{\rr{d}t}
 =
 -
 \frac{\varrho_{0,\rr{f}}^2}{D}
 \int_{\ell_1}^{\ell_2}
 \Big\{
 \Big[
     -(k_2\varepsilon''+k_3m'')
     +\frac{\partial\Psi}{\partial m}
 \Big]'
 \Big\}^2
  \rr{d}X_\rr{s}
\end{equation}
which proves that the functional $F$ in not increasing along 
the motions of the Cahn--Hilliard--like system
(\ref{problema-d-ch}).

In conclusion, in this section, we have written the equations of motions 
(\ref{em06}) and the associated boundary conditions (\ref{em06bc}) 
under the geometrical linearization assumption.
We have proven that, provided the Darcy and the Stokes 
effects are treated separately, those equations reduce respectively 
to the Cahn--Hilliard--like (\ref{problema-d-ch}) and to the 
Allen--Cahn--like (\ref{problema-s-ac}) PDE problems. 
Moreover, 
we have remarked that the stationary solutions 
of those problems are given by the same system of equations 
(\ref{problema-staz}). 
Finally, we have proven that the energy functional (\ref{liapunov}) 
does not increase 
along the solutions of both the two systems 
(\ref{problema-d-ch}) and (\ref{problema-s-ac}), which suggests 
that in both cases the motions will tend asymptotically to the 
stationary profiles. 

\section{Pressure driven phase transition}
\label{s:risultati}
\par\noindent
We apply, now, the theory developed above to the special model 
introduced in~\citep{CIS2011,CIS2009,CIS2010} and whose stationary 
behavior has been widely discussed in those papers.
We consider the following expression for the total potential 
energy density in the perspective of describing the transition between
a fluid--poor and a fluid--rich phase 
\begin{equation}
\label{sec015}
\Psi(m,\varepsilon)
 :=
 \frac{\alpha}{12}m^2(3m^2\!-8b\varepsilon m+6b^2\varepsilon^2)
 +
 \Psi_\rr{B}(m,\varepsilon)
\end{equation}
where 
\begin{equation}
\label{sec020}
\Psi_\rr{B}(m,\varepsilon):=
 p\varepsilon+\frac{1}{2}\varepsilon^2+\frac{1}{2}a(m-b\varepsilon)^2
\end{equation}
is the Biot potential energy density~\citep{biot01},
$a>0$ is the ratio between the fluid and the solid rigidity, 
$b>0$ is a coupling between the fluid and the solid component, 
$p>0$ is the external pressure,
and
$\alpha>0$ is a material parameter responsible for the showing 
up of an additional equilibrium.

In the papers~\citep{CIS2011,CIS2009,CIS2010} we have studied the stationary 
solutions of the equations (\ref{problema-d-ch}) and (\ref{problema-s-ac}), 
that is to say we have studied the problem (\ref{problema-staz}). 
In this section we 
give a brief account of those results and 
then we will study numerically the not stationary solutions
and, in particular, discuss how the stationary ones are approached.
The two cases, pure Darcy and pure Stokes dissipation, will
be discussed separately.

First of all we write explicitly the stationary problem 
corresponding to the potential energy (\ref{sec015}). By (\ref{problema-staz})
we get 
\begin{equation}
\label{problema-staz00}
\left\{
\begin{array}{l}
-(2/3)\alpha b m^3+\alpha b^2m^2\varepsilon+p+\varepsilon-ab(m-b\varepsilon)
-k_1\varepsilon''-k_2m''=0
\\
 \alpha m^3-2\alpha b m^2 \varepsilon+\alpha b^2m\varepsilon^2
 +a(m-b\varepsilon)
 -k_2\varepsilon''-k_3m'' 
=0
\\
{\displaystyle
\Big(
 (k_1\varepsilon'+k_2m')
 \delta\varepsilon
 +
 (k_2\varepsilon'+k_3m')
 \delta m
\Big)_{\ell_1,\ell_2}
=0
}
\\
\end{array}
\right.
\end{equation}
where the last line is the boundary condition.

\subsection{Phases: constant stationary solutions}
\label{s:fasi}
\par\noindent
In~\citep{CIS2011,CIS2009}
we have studied the constant solutions of (\ref{problema-staz00}) 
which are called 
\textit{phases} of the system. We have 
proven that there exists a pressure $p_\rr{c}$, called 
\textit{critical pressure}, such that for any $p\in[0,p_\rr{c})$ there 
exists a single phase 
$(m_\rr{s}(p),\varepsilon_\rr{s}(p))$, 
called 
the \textit{standard phase}, which is very similar to the usual 
solution of the Biot model. For $p> p_\rr{c}$ a second phase 
$(m_\rr{f}(p),\varepsilon_\rr{f}(p))$, richer in fluid with respect to the 
standard phase and hence called \textit{fluid--rich} phase, appears.

We have shown that the standard phase 
$(m_\rr{s}(p),\varepsilon_\rr{s}(p))$ 
is the solution of the two equations 
$m=b\varepsilon$ and 
$p=f_1(\varepsilon)$, for any $p>0$, where 
$f_1(\varepsilon):=-\varepsilon-\alpha b^4\varepsilon^3/3$.
On the other hand 
the fluid--rich phase 
$(m_\rr{f}(p),\varepsilon_\rr{f}(p))$ 
is the solution, with the smallest value of 
$\varepsilon$ (recall $\varepsilon\in(-1/2,0)$, so that the smallest 
value has indeed largest modulus),
of the two equations 
$m=m_+(\varepsilon)$ and 
$p=f_+(\varepsilon)$, where 
\begin{displaymath}
m_+(\varepsilon)=
\frac{b}{2}
\Big[
     \varepsilon+\sqrt{\varepsilon^2-\frac{4a}{\alpha b^2}}
\Big]
\end{displaymath}
and
\begin{displaymath}
f_+(\varepsilon)
\!:=\!
-\varepsilon+ab[m_+(\varepsilon)-b\varepsilon]
  -\alpha b^2\varepsilon m_+^2(\varepsilon)
  \vphantom{\bigg\{}
  +\frac{2}{3}\alpha bm_+^3(\varepsilon)
\end{displaymath}
For $\varepsilon\le-2/(b\sqrt{\alpha/a})$ 
the function $f_+(\varepsilon)$ is positive, diverging to $+\infty$ 
for $\varepsilon\to-\infty$, and has a minimum at $\varepsilon_\rr{c}$ 
such that $f_+(\varepsilon_\rr{c})=p_\rr{c}$; this explains why the fluid--rich
phase is seen only for $p>p_\rr{c}$.

Moreover it has been shown that for any $p>0$ the point 
$(m_\rr{s}(p),\varepsilon_\rr{s}(p))$ 
is a minimum of the two variable potential 
energy $\Psi(m,\varepsilon)$ with $p$ fixed, while
$(m_\rr{f}(p),\varepsilon_\rr{f}(p))$ 
is a minimum for $p>p_\rr{c}$ and 
a saddle point for $p=p_\rr{c}$.
For more details we refer to \citep{CIS2011}.

\subsection{Profiles: not constant stationary solutions}
\label{s:profili}
\par\noindent
In \citep{CIS2010} 
it has been proven that there exists a unique value $p_\rr{co}$ 
of the pressure, called \textit{coexistence pressure}, 
such that the potential energy of the two phases is equal.
More precisely, 
it has been proven that the equation 
$\Psi(m_\rr{s}(p),\varepsilon_\rr{s}(p))
 =\Psi(m_\rr{f}(p),\varepsilon_\rr{f}(p))$ has the single 
solution $p_\rr{co}$. 

The behavior of the system at the coexistence pressure 
is particularly interesting; from now on we shall always 
consider $p=p_\rr{co}$ and, for this reason, we shall drop $p$ 
from the notation. 
When the external pressure is equal to $p_\rr{co}$,
none of the two above phases is 
favored and we ask if profiles connecting one phase to the other exist.
More precisely, in \citep{CIS2010} we have addressed the 
problem of the existence of a \textit{connection} between the two phases, 
that is, a solution of the stationary problem (\ref{problema-staz00}) 
on $\bb{R}$ tending to the standard phase 
for $X_\rr{s}\to-\infty$ 
and to the fluid--rich one for $X_\rr{s}\to+\infty$.
Using results in \citep{AF}  
we have proven that such a connection does exist provided
$k_1k_3-k_2^2>0$.

We have also shown that, for $k_1k_3-k_2^2=0$ 
(\textit{degenerate} case), the 
problem of finding a solution of the stationary problem 
can be reduced to the computation of a definite integral.
Indeed, in such a case one performs the rotation
of the Cartesian reference system 
\begin{equation}
\label{secondo07}
x:=\frac{m+k\varepsilon}{\sqrt{1+k^2}}
\;\;\;\textrm{ and }\;\;\; 
y:=\frac{-km+\varepsilon}{\sqrt{1+k^2}}
\end{equation}
in the plane $m$--$\varepsilon$, 
where $k:=k_2/k_3=\pm\sqrt{k_1/k_3}$,
and defines 
\begin{equation}
\label{secondo08}
V(x,y)
=
-\Psi(m(x,y),\varepsilon(x,y))
\end{equation}
Then one shows that the two fields 
$m(X_\rr{s})$ and $\varepsilon(X_\rr{s})$ are solutions of the two equations 
(\ref{problema-staz00}) if and only if the corresponding fields 
$x(X_\rr{s})$ and $y(X_\rr{s})$ satisfy
\begin{equation}
\label{secondo12}
k_3(1+k^2)
x''=-\frac{\partial V}{\partial x}(x,y)
\;\;\;\textrm{ and }\;\;\;
\frac{\partial V}{\partial y}(x,y)=0
\end{equation}
The root locus of 
the \textit{constraint curve}
$\partial V(x,y)/\partial y=0$ is made of 
a certain number of maximal components 
such that each of them is the graph of a 
function $x\in\bb{R}\to y(x)\in\bb{R}$;
for each of them the first between the two equations
(\ref{secondo12}) becomes
a standard one dimensional conservative mechanical system 
with $X_\rr{s}$ interpreted as time, 
kinetic energy $(1+k^2)k_3(x')^2/2$,
and
potential energy $V(x,y(x))$.

Since
the function $V$ has been obtained by flipping the sign of the 
function $\Psi$ and rotating the coordinate axes,
then at the coexistence pressure it 
has the two absolute maximum points 
$(x_\rr{s},y_\rr{s})$
and 
$(x_\rr{f},y_\rr{f})$
corresponding, respectively, to the standard and to the fluid--rich phases. 
Since $(m_\rr{s},\varepsilon_\rr{s})$
and $(m_\rr{f},\varepsilon_\rr{f})$
satisfy the equations $\Psi_m(m,\varepsilon)=0$ and 
$\Psi_\varepsilon(m,\varepsilon)=0$, we have that the two points
$(x_\rr{s},y_\rr{s})$ and 
$(x_\rr{f},y_\rr{f})$ are solutions of the constraint equation 
$\partial V(x,y)/\partial y=0$
and hence they belong to the constraint curve. 

In \citep{CIS2010} we have seen that there exist
values of the second gradient parameters $k_1$, $k_2$, and $k_3$
such that the two points above 
fall on the same maximal component of the constraint equation.
Since, in this case, the function $V$ has two isolated absolute
maximum points which, by hypothesis, belong to the 
same maximal component of the 
constraint curve, we have that the function $V(x,y(x))$ of 
$x$ has two absolute isolated maxima in 
$x_\rr{s}$ and $x_\rr{f}$. 
The motion of the equivalent one dimensional conservative 
mechanical system corresponding to the energy level 
$V_{\rr{max}}:=V(x_\rr{s},y_\rr{s})$
yields the heteroclinic orbit, that is to say it yields the 
above mentioned connection on $\bb{R}$ between the 
standard and the fluid--rich phases.

\begin{figure}[t]
\begin{picture}(200,290)(60,0)
\put(0,0)
{
\resizebox{8.18cm}{!}{\rotatebox{0}{\includegraphics{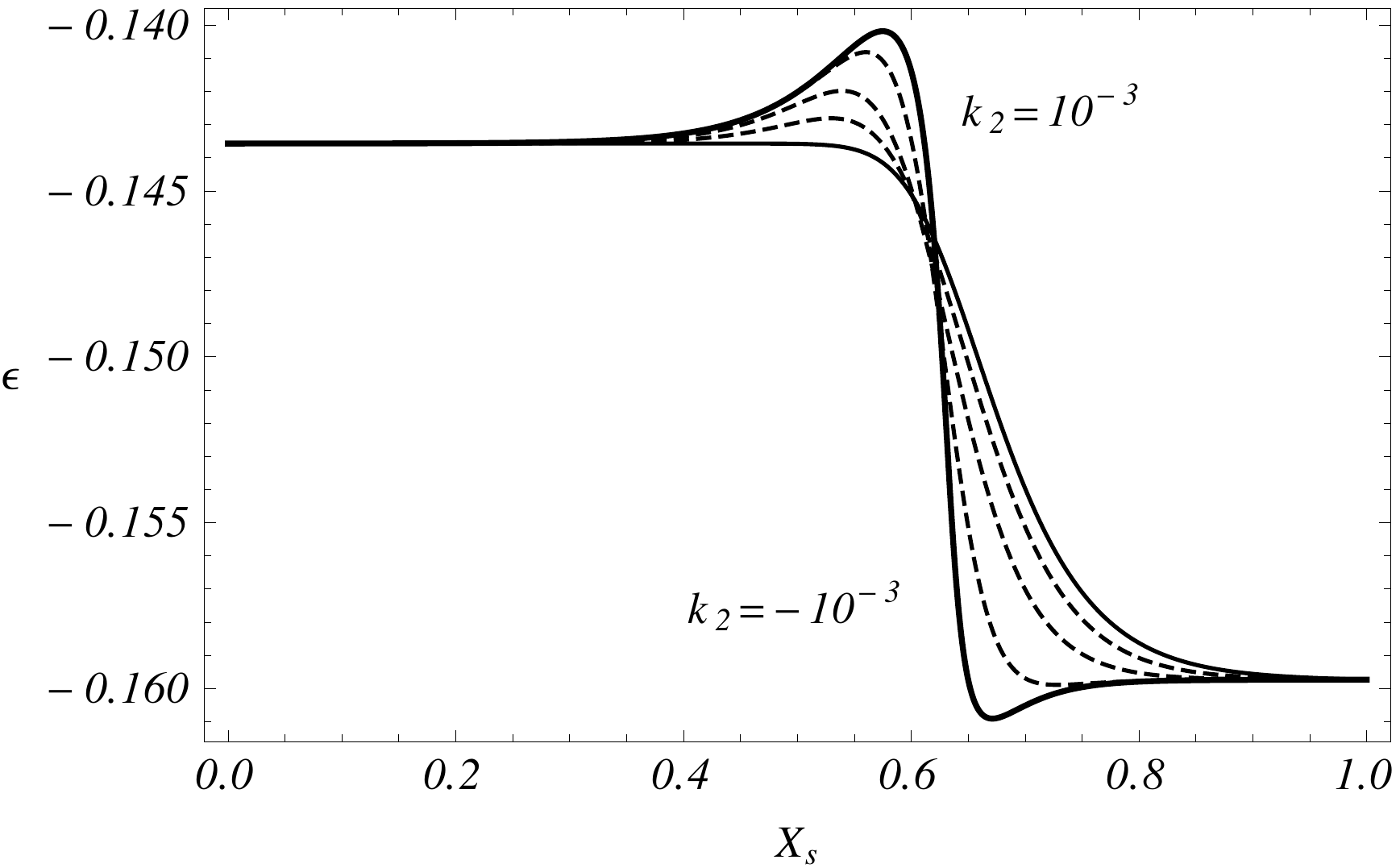}}} 
}
\put(235,0)
{
\resizebox{8.18cm}{!}{\rotatebox{0}{\includegraphics{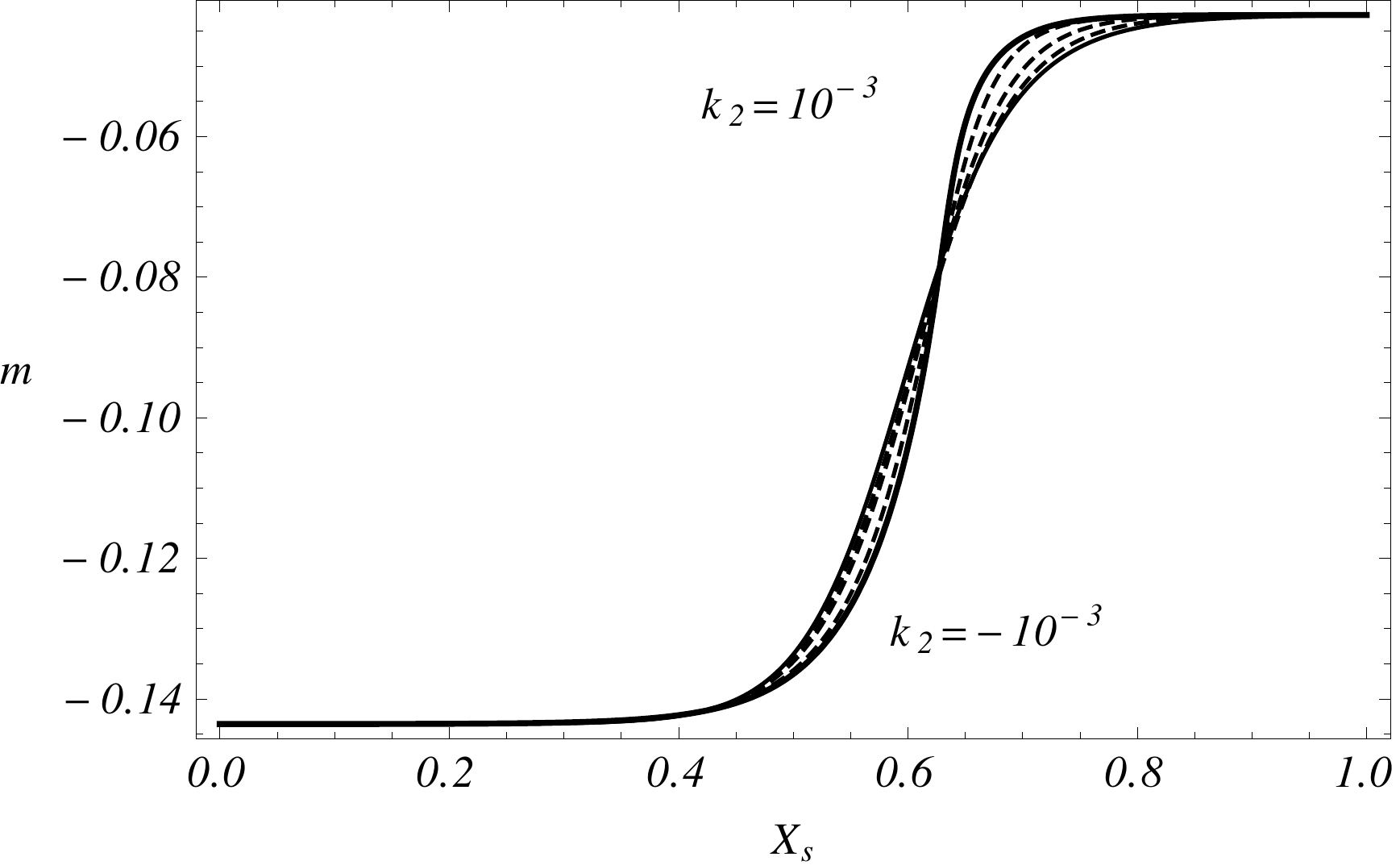}}} 
}
\put(0,150)
{
\resizebox{8cm}{!}{\rotatebox{0}{\includegraphics{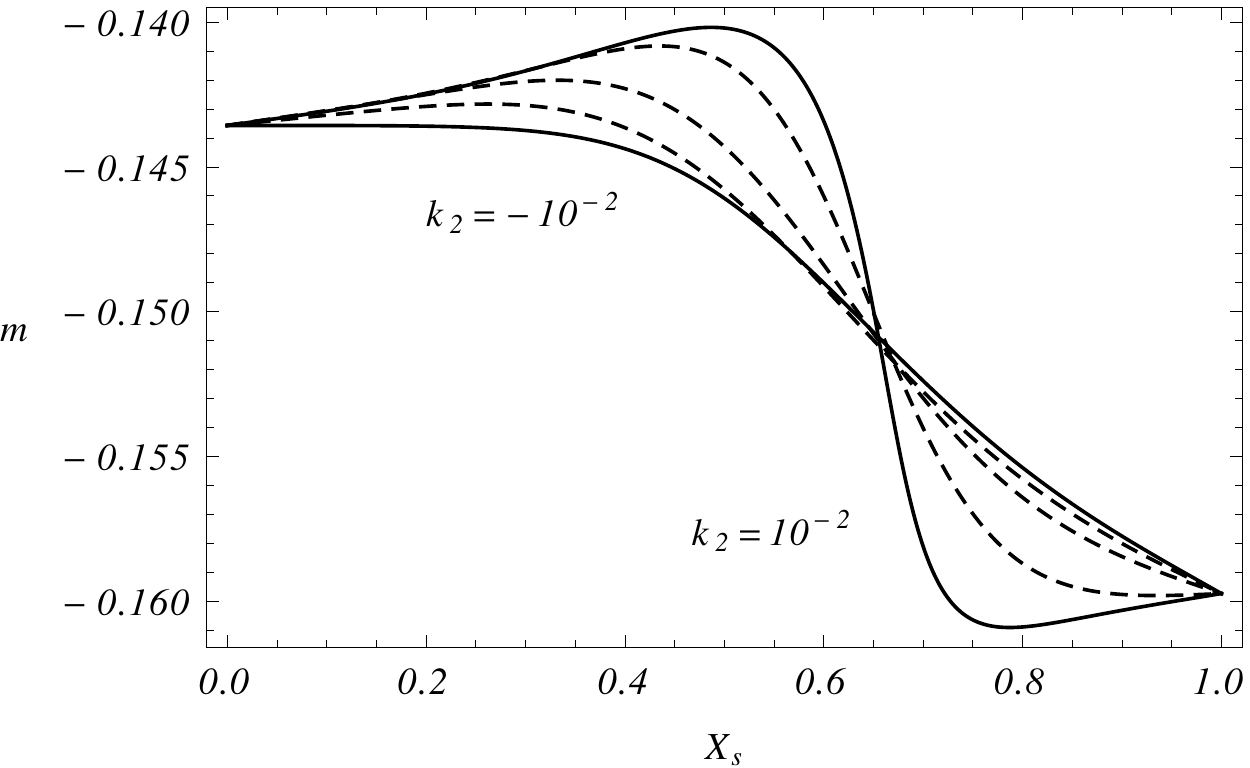}}} 
}
\put(235,150)
{
\resizebox{8cm}{!}{\rotatebox{0}{\includegraphics{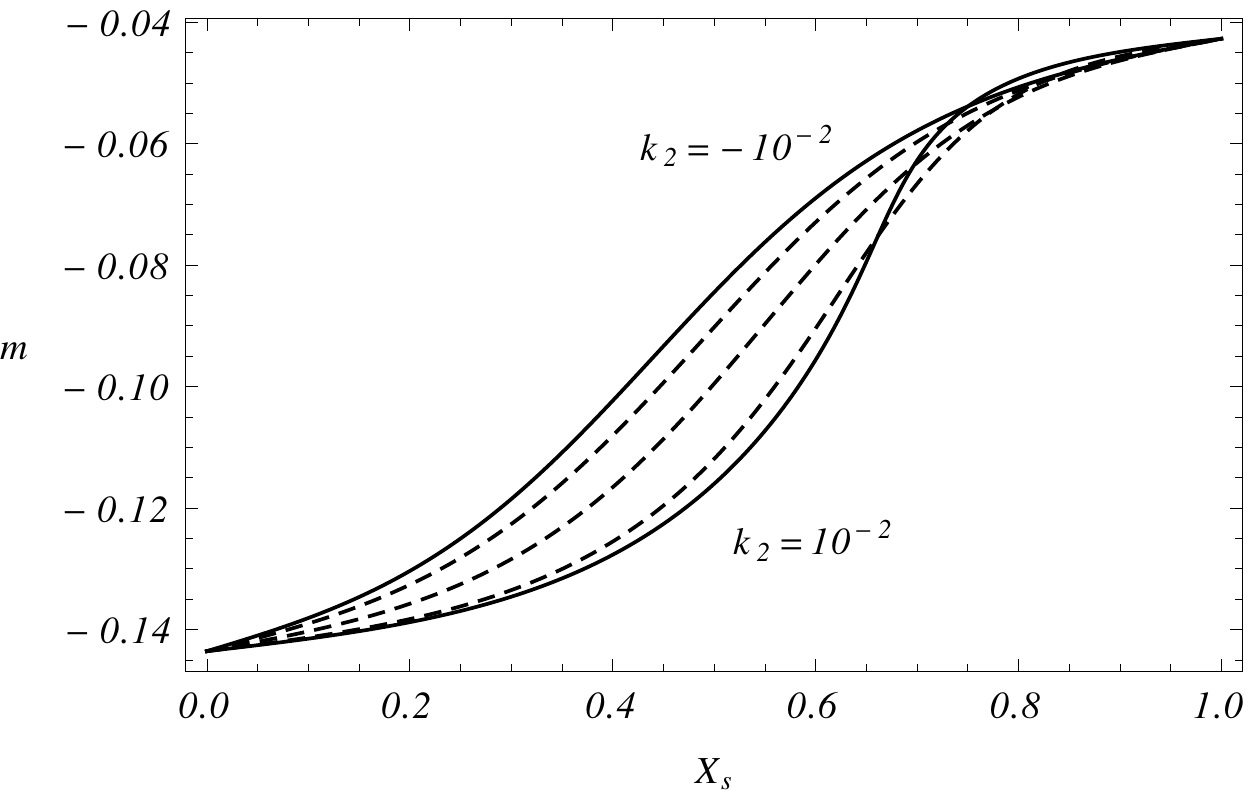}}} 
}
\end{picture}  
%\vskip 2 cm 
\caption{Bottom: solutions 
($\varepsilon(X_\rr{s})$ on the left and $m(X_\rr{s})$ on the right)
of the stationary problem (\ref{problema-staz00}) 
with the Dirichelet boundary conditions 
$m(0)=m_\rr{s}$, $\varepsilon(0)=\varepsilon_\rr{s}$,
$m(1)=m_\rr{f}$, and $\varepsilon(1)=\varepsilon_\rr{f}$
on the finite interval $[0,1]$, 
at the coexistence pressure
for 
$a=0.5$, $b=1$, $\alpha=100$,
$k_1=10^{-3}$, $k_3=10^{-3}$, 
$k_2=\pm10^{-3}$ (solid lines), and 
$k_2=-0.4\times10^{-3}, 0.2\times10^{-3}, 0.8\times10^{-3}$ (dotted lines).
Top: the same for 
$k_1=10^{-2}$, $k_3=10^{-2}$, 
$k_2=\pm10^{-2}$ (solid lines), and 
$k_2=-0.4\times10^{-2}, 0.2\times10^{-2}, 0.8\times10^{-2}$ (dotted lines).
}
\label{f:staz01} 
\end{figure}

\begin{figure}[h]
\begin{picture}(0,230)(40,0)
\put(80,0)
{
\resizebox{9cm}{!}{\rotatebox{0}{\includegraphics{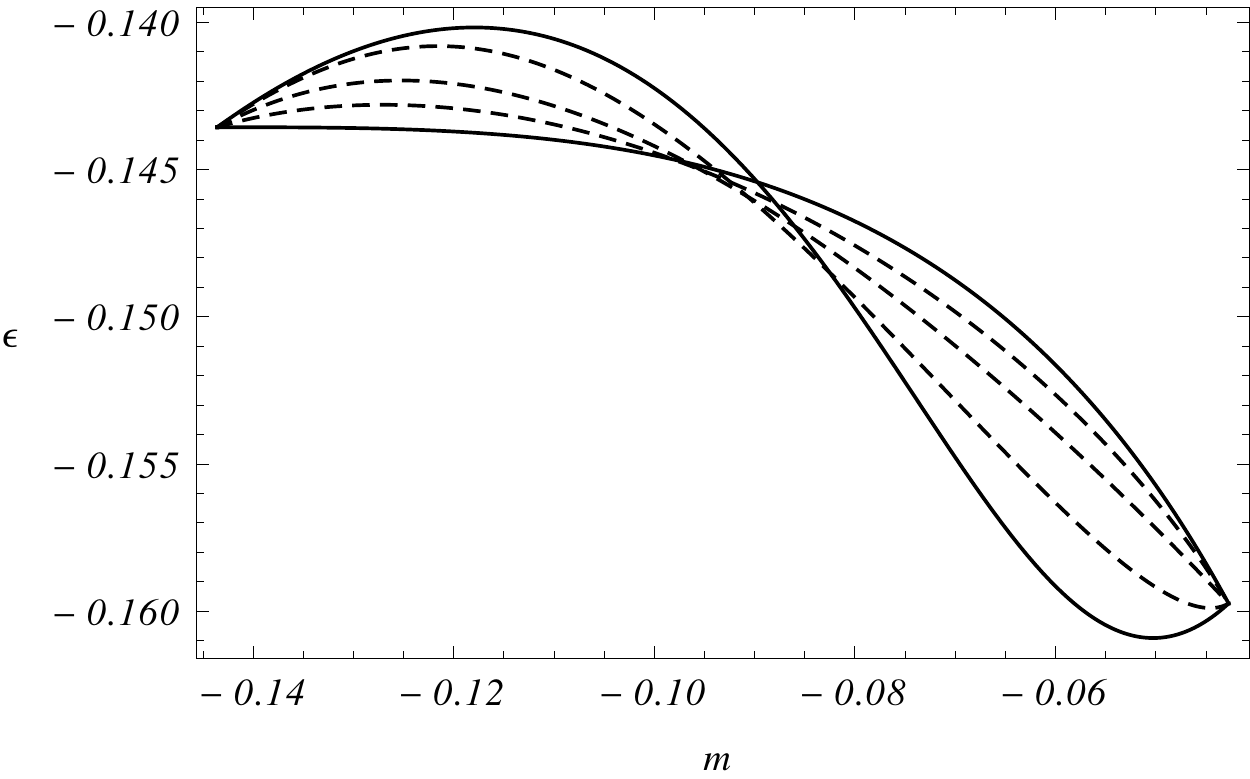}}} 
}
%\put(235,0)
%{
%\resizebox{7cm}{!}{\rotatebox{0}{\includegraphics{fig02b-porosi004.pdf}}} 
%}
\end{picture}  
%\vskip 2 cm 
\caption{Plot on the $m$--$\varepsilon$ plane of the 
solutions of the stationary problem (\ref{problema-staz00}) 
already depicted in the bottom row of the figure~\ref{f:staz01}.
}
\label{f:staz02} 
\end{figure}

With a similar approach one can find solutions of the system 
(\ref{problema-staz00}) with Dirichlet boundary conditions 
on a finite interval, say $\ell_1=0$ and $\ell_2=1$.
For instance, 
assume $x_\rr{s}<x_\rr{f}$
and consider the boundary condition 
$x(0)=x_\rr{s}$ and $x(1)=x_\rr{f}$.
Motions of the equivalent mechanical system started in $x_\rr{s}$ 
at time $0$ with 
positive velocity $x'(0)=\bar{v}>0$ (note that if it were 
$x_\rr{s}>x_\rr{f}$ one should consider a negative velocity at time 
$0$) 
will reach $x_\rr{f}$ in a finite time, which is a decreasing 
function of the initial velocity. It is then possible to choose 
properly the initial velocity so that   
\begin{displaymath}
1=\int_{x_\rr{s}}^{x_\rr{f}}
    \sqrt{
    \frac{k_3(1+k^2)}
         {2[\bar{E}-V(x,y(x))]}
         }
    \,\rr{d}x
\end{displaymath}
where $\bar{E}=(1+k^2)k_3\bar{v}^2/2+V_\rr{max}$ is the total energy of the 
motion of the equivalent one--dimensional conservative system.
Once $\bar{E}$ has been found, the solution of the stationary 
problem (\ref{problema-staz00}) 
with Dirichlet boundary conditions 
$m(0)=m_\rr{s}$, $\varepsilon(0)=\varepsilon_\rr{s}$,
$m(1)=m_\rr{f}$, and $\varepsilon(1)=\varepsilon_\rr{f}$
is implicitly given, in terms of the variable $x$, by the 
definite integral
\begin{displaymath}
X_\rr{s}=\int_{x_\rr{s}}^x
    \sqrt{
    \frac{k_3(1+k^2)}
         {2[\bar{E}-V(x,y(x))]}
         }
    \,\rr{d}x
\end{displaymath}
The solutions found in such a way for 
$k_1=10^{-3}$, $k_3=10^{-3}$, $k_2=\pm10^{-3}$, 
$a=0.5$, $b=1$, $\alpha=100$, and at coexistence pressure, have been 
depicted in figures~\ref{f:staz01} and \ref{f:staz02}.

As we have noted above, in the not degenerate case, i.e., 
$k_1k_3-k_2^2>0$, 
in \citep{CIS2010} we could prove the existence of a connection, but we were not 
able to find it explicitly.
Studying in more detail the behavior of the solutions of the 
system (\ref{problema-staz00}) is a difficult task since, in this 
case, the problem cannot be rewritten as a one dimensional 
conservative mechanical system.
We shall then study the behavior of the solutions close to the 
two phases by linearizing the stationary equations, while far from the 
points representing the phases we shall use a finite difference numerical 
approach. 

The first question that naturally arises looking at the solutions 
obtained in the degenerate case, see figure~\ref{f:staz01}, is 
the nature of the bump. 
In the degenerate case the bump is 
due to the fact that the solution of the stationary problem, 
once projected onto the $m$--$\varepsilon$ plane, has to 
lie on the constraint curve, so that its shape depends on that of 
the constraint itself. 
In the not degenerate case no constraint curve exists, in other words 
it does not exist any 
a priori
privileged manifold on the plane $m$--$\varepsilon$.
It is then possible (in principle) that the bump is associated to solutions 
oscillating close to the phases.

In order to discuss the existence of such oscillatory solutions 
we linearize the problem close to the phases.
Let $(\bar{m},\bar{\varepsilon})$ be one of the two points 
$(m_\rr{s},\varepsilon_\rr{s})$ and $(m_\rr{f},\varepsilon_\rr{f})$
representing, respectively, the standard and the fluid--rich phase.
By expanding the first gradient part of the stationary 
equations (\ref{problema-staz}) around $(\bar{m},\bar{\varepsilon})$
to the first order, we get the linearized equations
\begin{displaymath}
\left\{
\begin{array}{l}
k_1\varepsilon''+k_2m''
-\Psi_{\varepsilon m}(\bar{m},\bar{\varepsilon})(m-\bar{m})
-\Psi_{\varepsilon\varepsilon}(\bar{m},\bar{\varepsilon})
                               (\varepsilon-\bar{\varepsilon})
=0
\\
k_2\varepsilon''+k_3m''
-\Psi_{mm}(\bar{m},\bar{\varepsilon})(m-\bar{m})
-\Psi_{m\varepsilon}(\bar{m},\bar{\varepsilon})(\varepsilon-\bar{\varepsilon})
=0
\end{array}
\right.
\end{displaymath}
If we let 
$q_1=\varepsilon-\bar{\varepsilon}$ and $q_2=m-\bar{m}$, the 
above equations become 
\begin{equation}
\label{lin01}
\mathbf{T}q'' - \mathbf{\Psi} q =0
\end{equation}
where we consider the column vector $q\in\bb{R}^2$ and 
set 
\begin{displaymath}
\mathbf{T}
=
\left(\begin{array}{cc}k_1&k_2\\k_2&k_3\\\end{array}\right)
\;\;\textrm{ and }\;\;
\mathbf{\Psi}
=
\left(\begin{array}{cc}
\Psi_{\varepsilon\varepsilon}(\bar{m},\bar{\varepsilon})
&
\Psi_{\varepsilon m}(\bar{m},\bar{\varepsilon})
\\
\Psi_{m\varepsilon}(\bar{m},\bar{\varepsilon})
&
\Psi_{mm}(\bar{m},\bar{\varepsilon})
\\
\end{array}\right)
\end{displaymath}
Note that the matrix $\mathbf{T}$ is trivially real and symmetric; moreover, 
in the not degenerate case $k_1k_3-k_2^2>0$ it 
is also positive definite. 
The matrix $\mathbf{\Psi}$, on the other hand, is real and, by Schwartz 
theorem, symmetric. Moreover, as we proved in \citep{CIS2011}, it is 
positive definite at any pressure greater than the critical one $p_\rr{c}$; 
in particular it is positive definite at the coexistence pressure. 

Looking for solutions of (\ref{lin01}) in the form 
$q=c\exp\{\lambda X_\rr{s}\}$, with $c\in\bb{R}^2$ and $\lambda\in\bb{C}$, 
we get that the constant vector $c$ has to be a solution of 
$(\lambda^2\mathbf{T}-\mathbf{\Psi})c=0$.
Set $\mu=\lambda^2$;
in order to have not vanishing $c$ solving the above equation, 
$\mu$ has to be 
a solution of the secular equation 
$\det(\mu\mathbf{T}-\mathbf{\Psi})=0$.
Since $\mathbf{T}$ is real, symmetric, and positive definite and 
$\mathbf{\Psi}$ is real and symmetric, it follows that the solutions 
$\mu_1$ and $\mu_2$ of
the secular equations are real. Moreover, since $\mathbf{\Psi}$ is 
positive definite we have that those solutions are positive themselves
(see, for instance, \citep[Section~40, Chapter~6]{Gantmacher}). 
Thus the secular equations have 
four real pairwise opposite eigenvalues. 
This implies that the two fixed points are 
unstable, as it also follows from the existence of the connection, and 
that no oscillation around the fixed points is expected. 

We come, finally, to the numerical study of the
stationary problem (\ref{problema-staz00}) in the not degenerate case
on a finite 
interval (we shall always use the interval $[0,1]$). We 
use the finite difference method powered with the Newton--Raphson algorithm.
The substitution rules we adopted are standard and 
have been reported in appendix~\ref{s:numerico}.
We solved the stationary problem (\ref{problema-staz00}) 
with the Dirichlet boundary conditions 
$m(0)=m_\rr{s}$, 
$\varepsilon(0)=\varepsilon_\rr{s}$, 
$m(1)=m_\rr{f}$, 
$\varepsilon(1)=\varepsilon_\rr{f}$, 
at the coexistence pressure
for 
$a=0.5$, $b=1$, and $\alpha=100$.
For the second gradient coefficient we have considered two sets; 
the first set is  
$k_1=10^{-3}$, $k_3=10^{-3}$, 
and 
$k_2=-0.4\times10^{-3}, 0.2\times10^{-3}, 0.8\times10^{-3}$
and the related numerical results have been depicted in 
the bottom row of 
figure~\ref{f:staz01} and in figure~\ref{f:staz02}.
The second set is 
$k_1=10^{-2}$, $k_3=10^{-2}$, 
and 
$k_2=-0.4\times10^{-2}, 0.2\times10^{-2}, 0.8\times10^{-2}$
and the related numerical results have been reported in 
the top row of 
figure~\ref{f:staz01}.

On the bottom (resp.\ top) row of figure~\ref{f:staz01} we 
have depicted the $m$ and $\varepsilon$--profiles
as functions of the space variable $X_\rr{s}$;
solid lines refer to the degenerate case 
$k_2=\pm10^{-3}$ (resp.\ $k_2=\pm10^{-2}$), 
while dotted lines refer to the not degenerate cases 
$k_2=-0.4\times10^{-3}, 0.2\times10^{-3}, 0.8\times10^{-3}$
(resp.\ $k_2=-0.4\times10^{-2}, 0.2\times10^{-2}, 0.8\times10^{-2}$).
In figure~\ref{f:staz02} the same solutions have been 
depicted on the plane $m$--$\varepsilon$.
The behavior of the solutions of the stationary problem in the 
degenerate and in the 
not degenerate case is similar.
The existence of the bump in the $m$ and $\varepsilon$--profiles is 
essentially due to the shape of the manifold of the $m$--$\varepsilon$ 
plane on which the solutions lie.

It is worth noting that, even in the degenerate case, due to the asymmetry of 
the potential energy $V$, the position of the interface in the 
stationary profile depends on the value of the second gradient 
constants $k_1$, $k_2$, and $k_3$. However in the limit for small values
of $k_1$, $k_2$, and $k_3$ it is observed that the interface tends to a 
definite position \citep{CIS2012}.

\subsection{Evolution under pure Stokes dissipation: 
            Allen--Cahn--like equations}
\label{s:bri-results}
\par\noindent
The equations of motion describing the evolution of the system 
subjected to a pure Darcy drag force are the equations 
(\ref{problema-s-ac}) which, for the special model
(\ref{sec015}), read
\begin{equation}
\label{num06}
\left\{
\begin{array}{l}
-(2/3)\alpha b m^3+\alpha b^2m^2\varepsilon+p+\varepsilon-ab(m-b\varepsilon)
-k_1\varepsilon''-k_2m''=0
\\
-\varrho^2_{0,\rr{f}}[
 \alpha m^3-2\alpha b m^2 \varepsilon+\alpha b^2m\varepsilon^2
 +a(m-b\varepsilon)
 -k_2\varepsilon''-k_3m'' 
]
=S\dot{m}
\\
\end{array}
\right.
\end{equation}
with the boundary conditions
\begin{equation}
\label{num02}
((k_2m'+k_1\varepsilon')\delta\varepsilon
+
(k_3m'+k_2\varepsilon')\delta m
)_{\ell_1,\ell_2}
=0
\end{equation}

We have studied the above PDE problem with the same numerical 
approach used for the stationary problem and discussed in 
Section~\ref{s:profili}. That is we have used a finite difference 
method powered with the Newton--Raphson algorithm and with 
the standard substitution rules reported in \ref{s:numerico}.

In the figures~\ref{f:acdin01}--\ref{f:acdin03} we have 
depicted the solutions of the problem 
(\ref{num06}) 
with Dirichlet boundary conditions 
$m(0)=m_\rr{s}$, $\varepsilon(0)=\varepsilon_\rr{s}$,
$m(1)=m_\rr{f}$, and $\varepsilon(1)=\varepsilon_\rr{f}$
on the finite interval $[0,1]$, 
at the coexistence pressure
for 
$a=0.5$, $b=1$, $\alpha=100$,
$k_1=k_2=k_3=10^{-3}$. 
All the graphs refer to the same spatially random initial condition 
on the interval $[0,1]$.
Details on the depicted times can be found in the caption 
of the related figures.

Figures~\ref{f:acdin01} and \ref{f:acdin02} 
show how the limiting stationary profile is 
formed. 
At times of order one 
a deviation from the initial random profile is noticed and 
a bumped profile with 
roughness due to the randomness of the initial condition is formed. 
The profile is smeared out in a time of order $10^1$ and an interface, 
far from the position of the stationary profile, is seen. 
The interface then undergoes a metastable, purely dissipative, 
evolution approaching the stationary state in times greater than $10^4$.
This coarsening process has been widely studied for the 
standard Allen--Cahn equation, see for instance
\citep{CP,FH,W94}.

Figure~\ref{f:acdin03} shows that the solution 
is soon attracted by the manifold in the plane 
$m$--$\varepsilon$ defined by the stationary solution. 
Afterwards the projection of the profiles 
onto the $m$--$\varepsilon$ plane keeps close to such a 
curve.
We recall that, when the parameters $k_1$, $k_2$, and $k_3$, as 
in the cases illustrated in the figures~\ref{f:acdin01}--\ref{f:acdin03},
are chosen so that $k_1k_3-k_2^2=0$, the algebraic equation 
of the limiting manifold is the second above the two equations 
(\ref{secondo12}).

We have measured the way in which the stationary solution is approached
considering the difference between the 
value of the energy functional (\ref{liapunov}) at time $t$ and that 
of the corresponding stationary solution. 
Results are reported in figure~\ref{f:acdin04}.
We note that the energy functional is a decreasing function 
of time as proved in Section~\ref{s:lin} (see equation (\ref{acdecr})).

\begin{figure}
\begin{picture}(200,460)(50,0)
\put(0,0)
{
\resizebox{7.8cm}{!}{\rotatebox{0}{\includegraphics{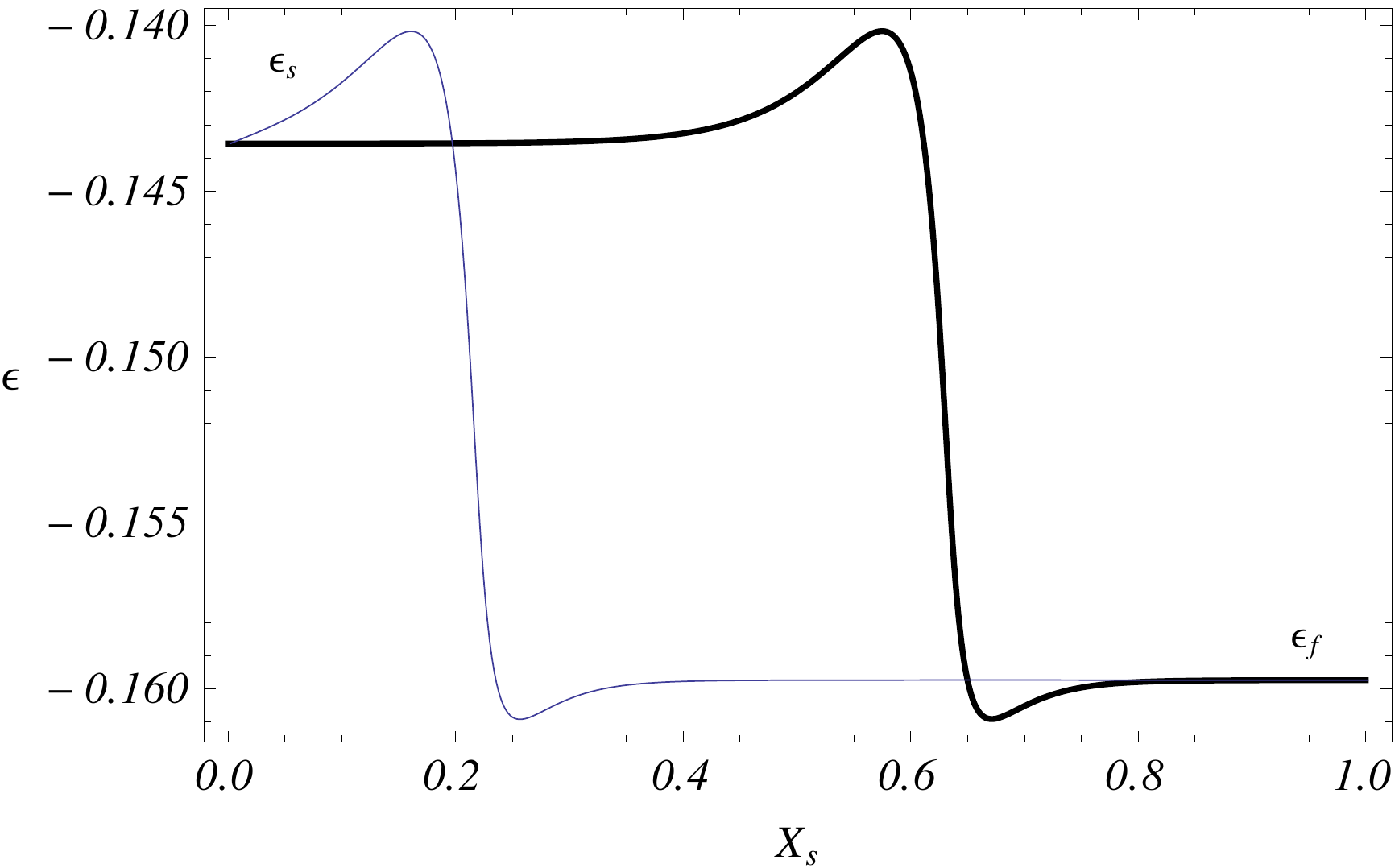}}} 
}
\put(235,0)
{
\resizebox{7.8cm}{!}{\rotatebox{0}{\includegraphics{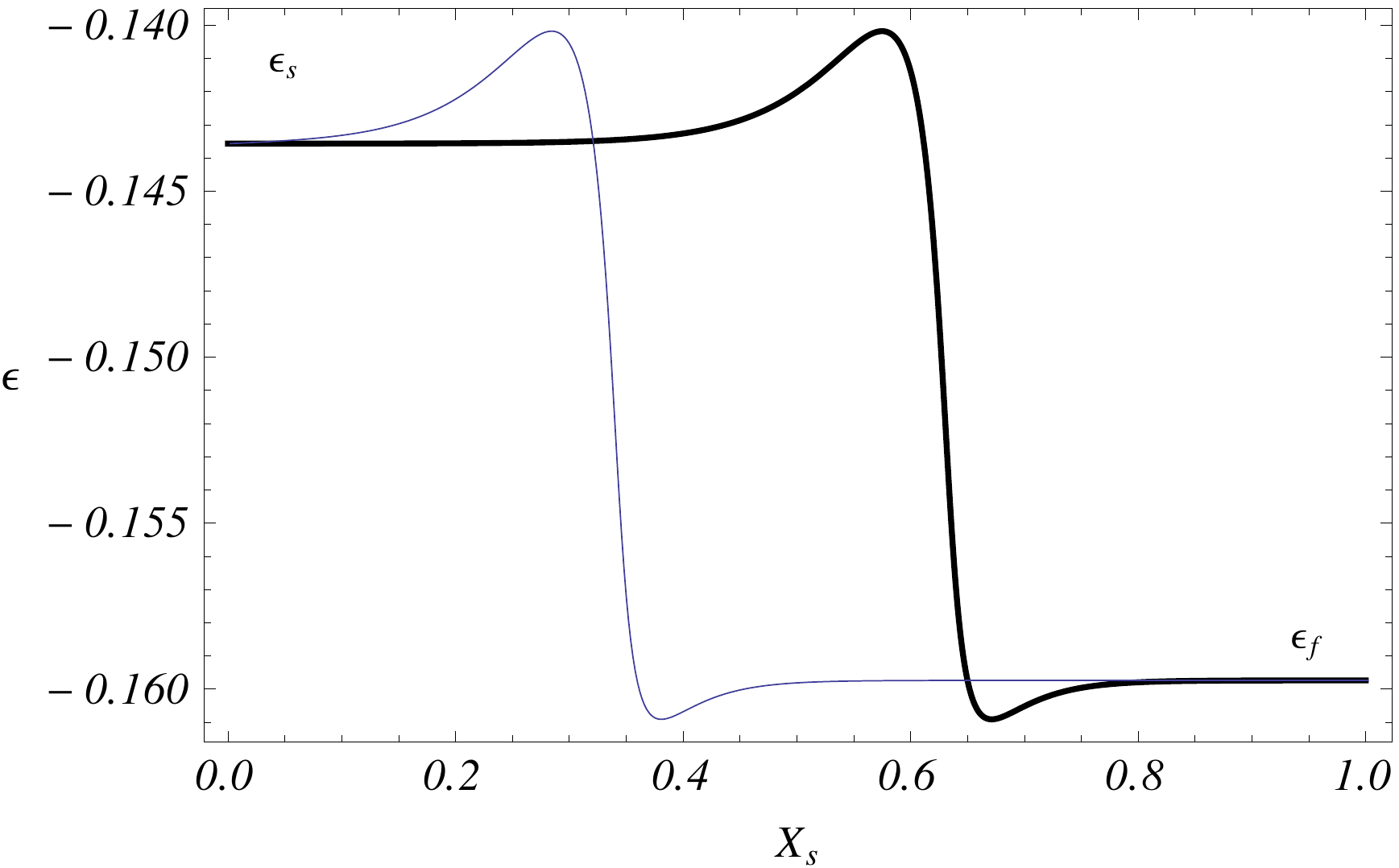}}} 
}
\put(0,160)
{
\resizebox{7.8cm}{!}{\rotatebox{0}{\includegraphics{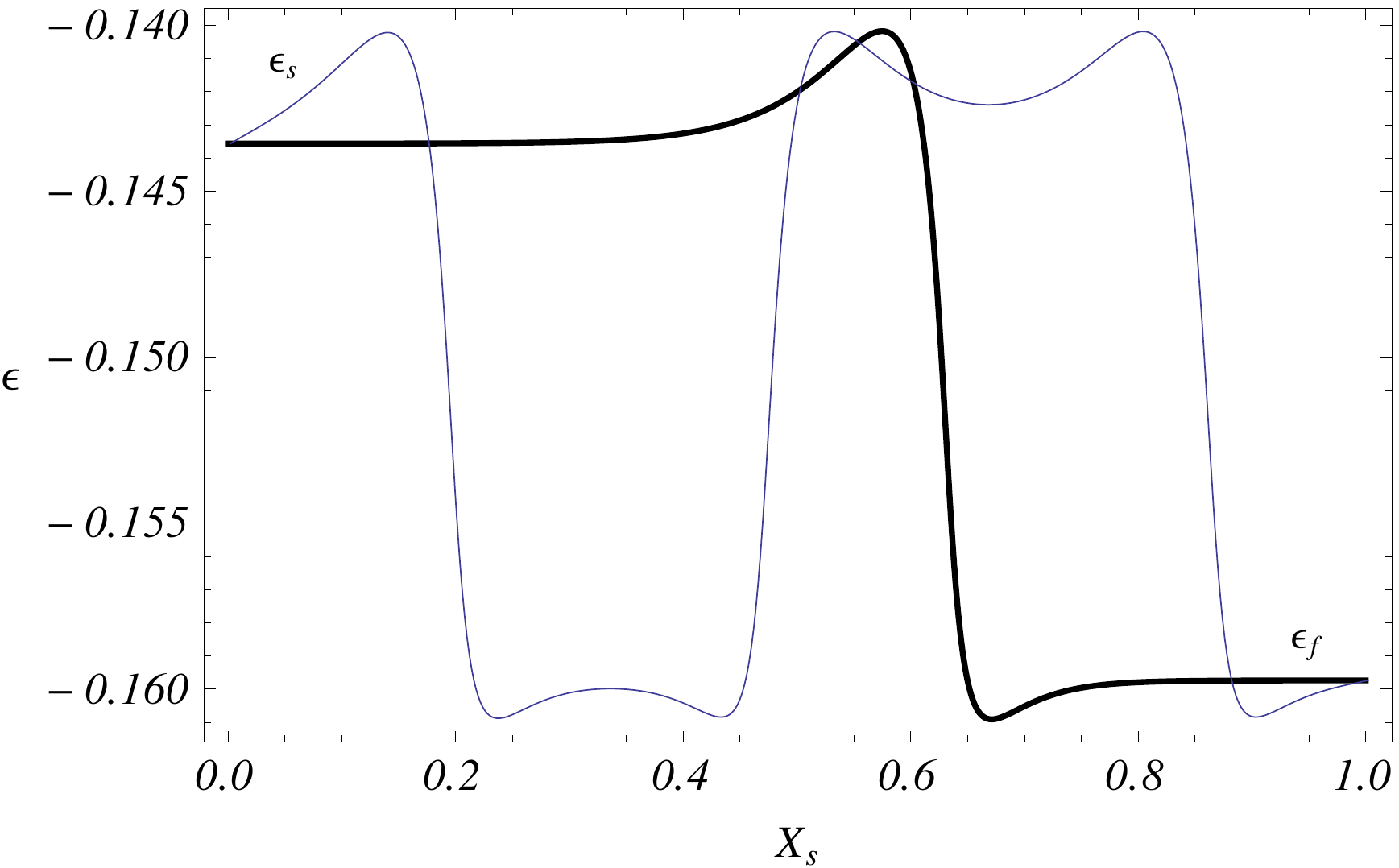}}} 
}
\put(235,160)
{
\resizebox{7.8cm}{!}{\rotatebox{0}{\includegraphics{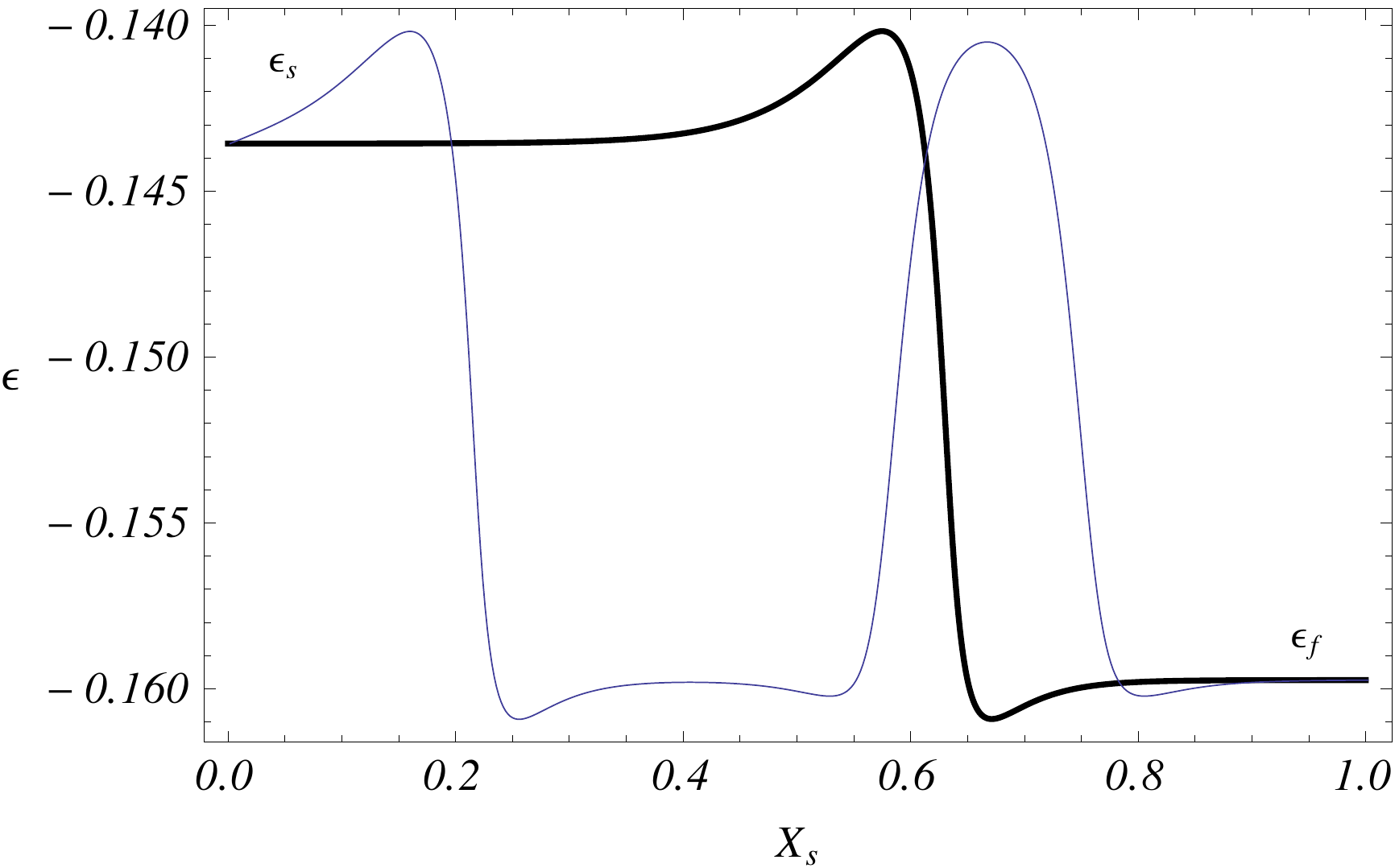}}} 
}
\put(0,320)
{
\resizebox{7.8cm}{!}{\rotatebox{0}{\includegraphics{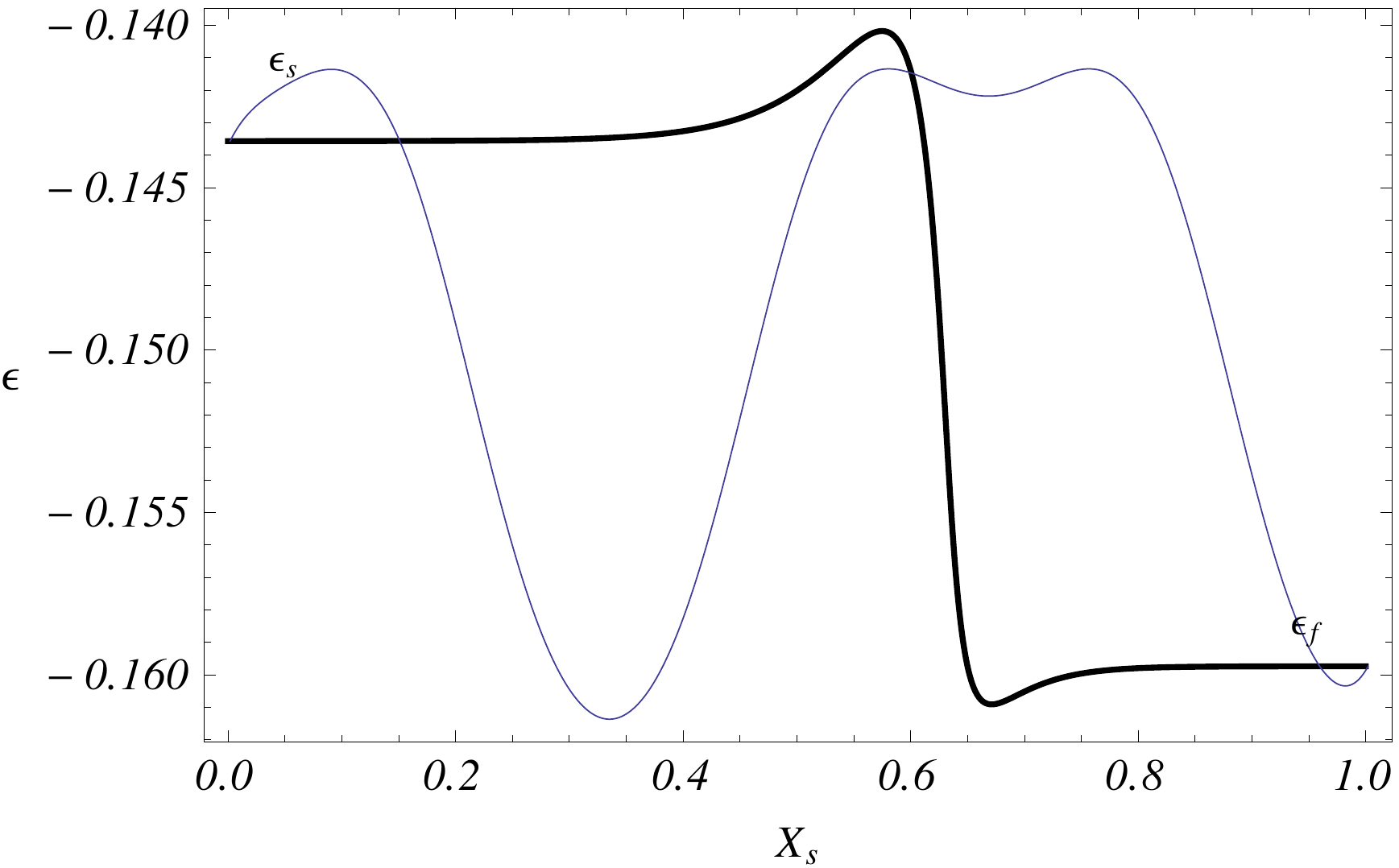}}} 
}
\put(235,320)
{
\resizebox{7.8cm}{!}{\rotatebox{0}{\includegraphics{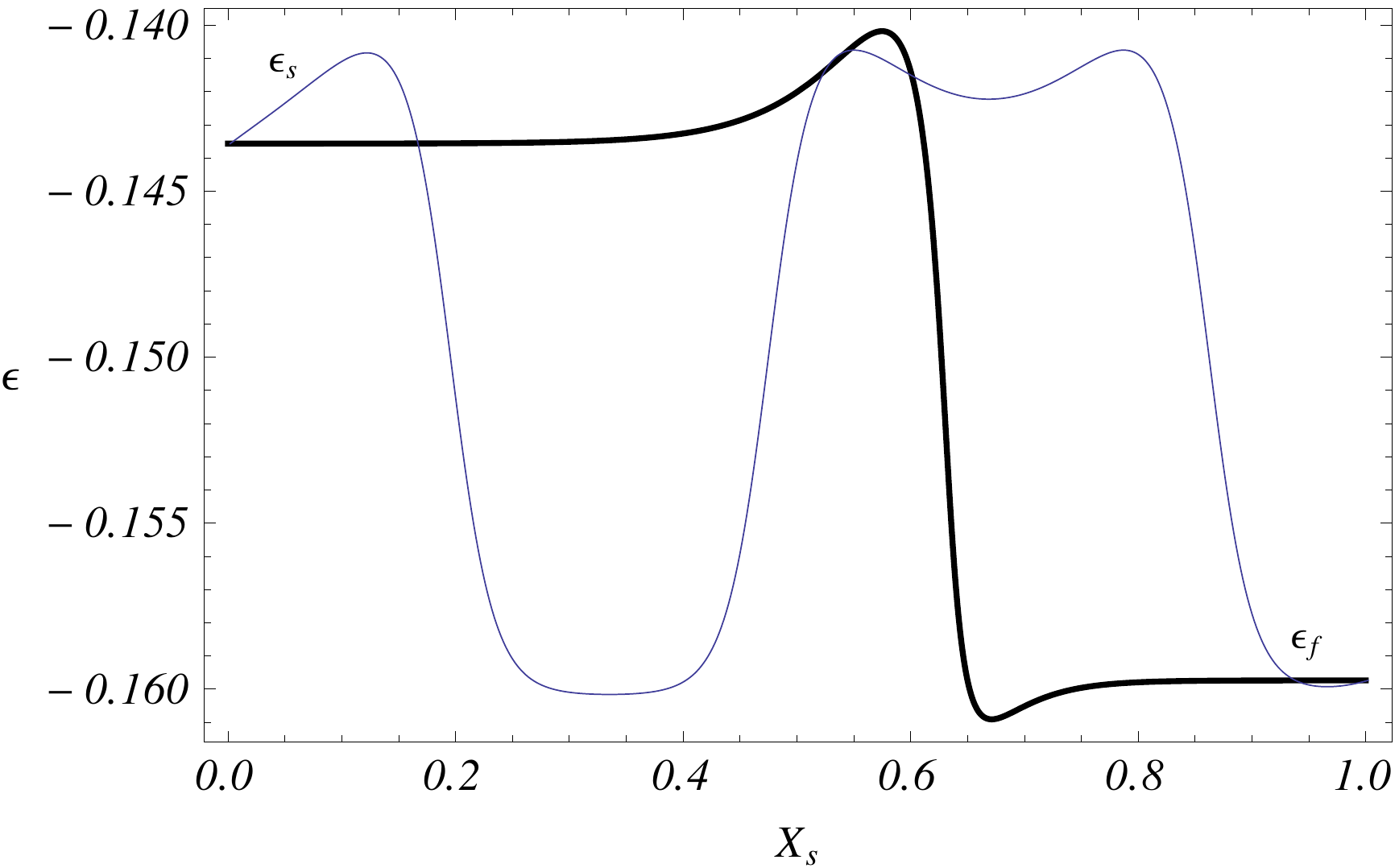}}} 
}
\end{picture}  
%\vskip 2 cm 
\caption{Profiles (solid thin lines)
$\varepsilon(X_\rr{s},t)$ 
obtained by solving the Allen--Cahn--like system (\ref{num06}) 
with a deterministic stratified initial state and 
Dirichelet boundary conditions 
$m(0)=m_\rr{s}$, $\varepsilon(0)=\varepsilon_\rr{s}$,
$m(1)=m_\rr{f}$, and $\varepsilon(1)=\varepsilon_\rr{f}$
on the finite interval $[0,1]$, 
at the coexistence pressure
for 
$a=0.5$, $b=1$, $\alpha=100$,
$k_1=k_2=k_3=10^{-3}$. 
The solid thick line is the corresponding stationary profile.
Profiles at times
$t=0.07,2,10,320,340,50000$
are depicted in lexicographic order.
}
\label{f:acdin05} 
\end{figure}

Some of the features described above are peculiar of the initial 
random condition. We have also studied the same Allen-Cahn problem, 
with the same parameters and the same boundary conditions, with 
a deterministic stratified initial state. 
Results for the $\varepsilon$--profile are depicted 
at times
%$t=0.07,2,10,100,225,232$
in the figure~\ref{f:acdin05}; we have not reported 
the $m$--profiles and the $m$--$\varepsilon$--plane view since they 
do not add any more information.

At times of order $10^{-1}$ a deviation from the initial stratified 
state is detected. The profile deforms into a new one with a 
droplet--like shape, at times of order $10^1$, 
finally an interface profile nucleates at times of order $10^2$, as
in the previous case corresponding to spatially random initial data.
Similarly to this case the stationary interface is approached 
with the same velocity.
Even in this case we have checked the way in which the solution tends 
to the stationary state via the energy functional (\ref{liapunov}). 
Results are reported in figure~\ref{f:acdin06}.

\subsection{Evolution under pure Darcy dissipation: 
            Cahn-Hilliard--like equations}
\label{s:dar-results}
\par\noindent
The equations of motion describing the evolution of the system 
subjected to a pure Darcy drag force are the equations 
(\ref{problema-d-ch}) which, for the special model
(\ref{sec015}), read
\begin{equation}
\label{num05}
\left\{
\begin{array}{l}
-(2/3)\alpha b m^3+\alpha b^2m^2\varepsilon+p+\varepsilon-ab(m-b\varepsilon)
-k_1\varepsilon''-k_2m''=0
\\
 \varrho^2_{0,\rr{f}}
 ( 
  \alpha m^3-2\alpha b m^2 \varepsilon+\alpha b^2m\varepsilon^2
  +a(m-b\varepsilon)
  -k_2\varepsilon''-k_3m'' 
 )''
 =D\dot{m}
\\
\end{array}
\right.
\end{equation}
with the boundary conditions
\begin{equation}
\label{num01}
\left\{
\begin{array}{l}
((k_2m'+k_1\varepsilon')\delta\varepsilon
+
(k_3m'+k_2\varepsilon')\delta m
)_{\ell_1,\ell_2}
=0
\\
( 
 \alpha m^3-2\alpha b m^2 \varepsilon+\alpha b^2m\varepsilon^2
 +a(m-b\varepsilon)
 -k_2\varepsilon''-k_3m'' 
)_{\ell_1,\ell_2}
=0
\end{array}
\right.
\end{equation}
We have studied the above PDE problem with the same numerical 
approach used for the stationary problem and for the Allen--Cahn--like 
system. We have repeated the same sequence of numerical 
computations, that is we have used the same parameters 
and the same initial and boundary conditions as for the 
above Allen--Cahn--like case. 

In the figures~\ref{f:chdin01}--\ref{f:chdin03} we have 
depicted the solutions of the problem 
(\ref{num05}) 
with Dirichlet boundary conditions 
$m(0)=m_\rr{s}$, $\varepsilon(0)=\varepsilon_\rr{s}$,
$m(1)=m_\rr{f}$, and $\varepsilon(1)=\varepsilon_\rr{f}$
on the finite interval $[0,1]$,
at the coexistence pressure
for 
$a=0.5$, $b=1$, $\alpha=100$,
$k_1=k_2=k_3=10^{-3}$. 
All the graphs refer to the same initial condition chosen randomly
on the interval $[0,1]$ as that considered for the solution of the
Allen--Cahn--like equations.
Details on the depicted times can be found in the caption of 
the related figures. 

Figures~\ref{f:chdin01} and \ref{f:chdin02} 
show that, starting from a random spatial distribution of strain and fluid 
mass, micro--structured paths appear at times of order $10^{-2}$. 
They progressively evolve towards a droplet--like profile which 
is formed at times of order $10^{-1}$. The droplet undergoes a metastable 
evolution at times of order $10^1$ up to the formation of 
an interface.
This coarsening process has been well studied for the 
standard Cahn--Hilliard equation, see for instance
\citep{Ward,ABF,BX1,BX2}.
The last part of the evolution is a 
the motion of the interface approaching the stationary 
state at times of order $10^2$.

Figure~\ref{f:chdin03} shows that even the solution of 
the Cahn--Hilliard--like system (\ref{num05}), as that of the Allen--Cahn-like 
system (\ref{num06}), 
is attracted immediately by the manifold in the plane 
$m$--$\varepsilon$ defined by the stationary solution
and, 
afterwards, the projection of the profiles 
onto the $m$--$\varepsilon$ plane keeps close to such a 
manifold.
Note that the profile temporarily leaves the stationary manifold
at time $3.7$ in correspondence of the shrinking of the droplet 
present in the $\varepsilon$--profile. 

As for the Allen--Cahn--like case, also in the Cahn--Hilliard--like case 
we have studied the way in which the stationary state is approached 
by using the energy--like functional (\ref{liapunov}) (see figure~\ref{f:chdin04}). 

\begin{figure}
\begin{picture}(200,460)(50,0)
\put(0,0)
{
\resizebox{7.8cm}{!}{\rotatebox{0}{\includegraphics{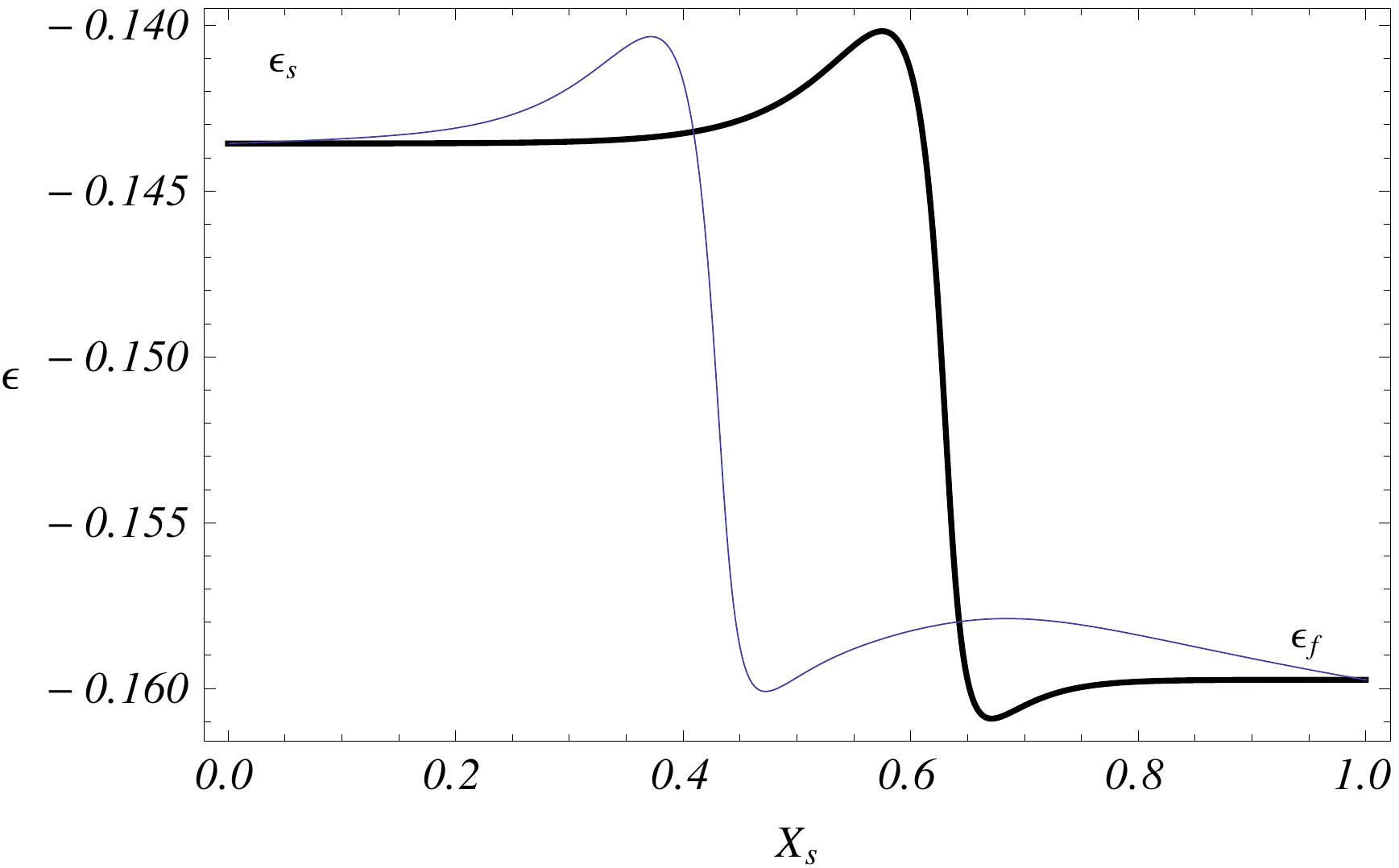}}} 
}
\put(235,0)
{
\resizebox{7.8cm}{!}{\rotatebox{0}{\includegraphics{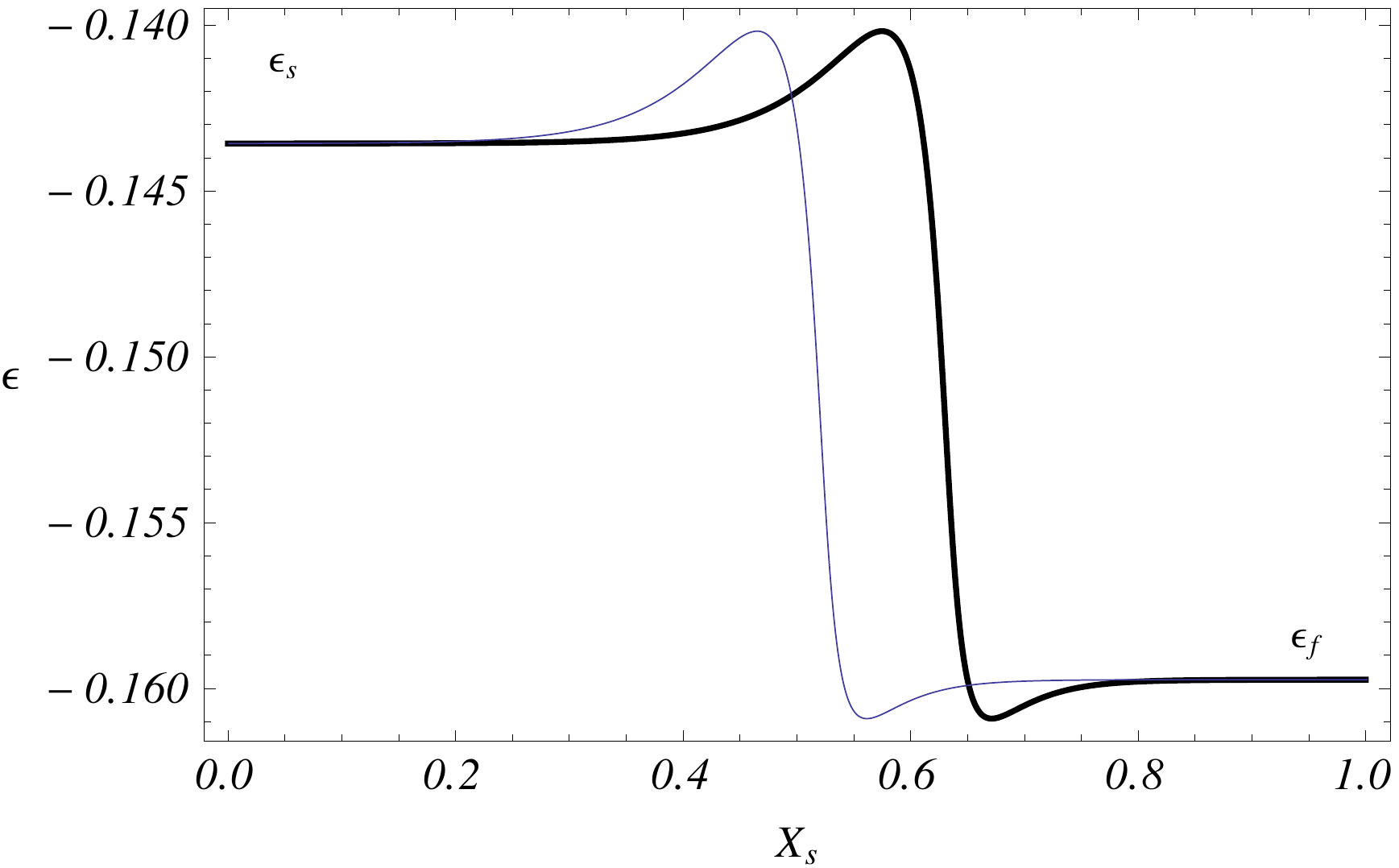}}} 
}
\put(0,160)
{
\resizebox{7.8cm}{!}{\rotatebox{0}{\includegraphics{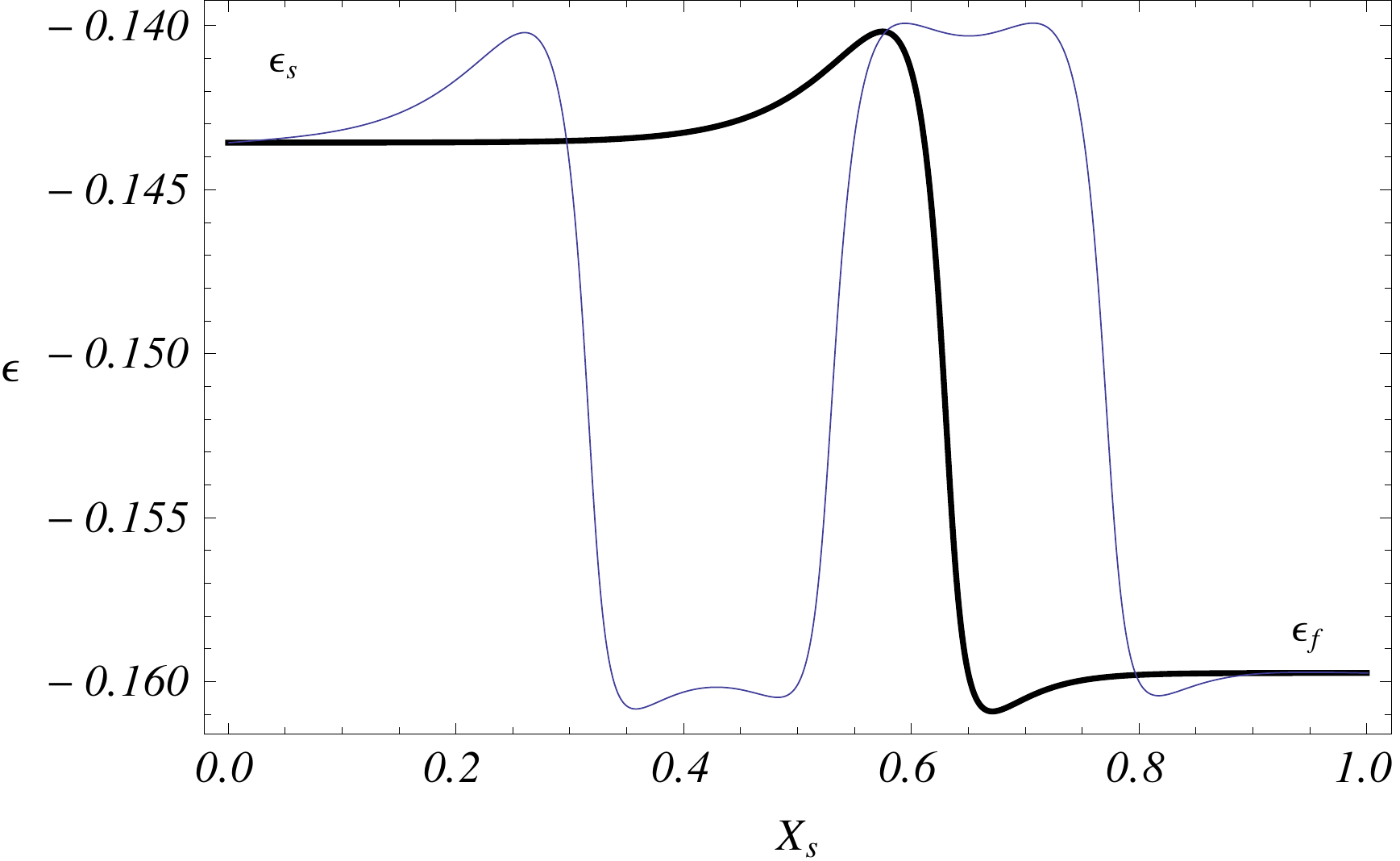}}} 
}
\put(235,160)
{
\resizebox{7.8cm}{!}{\rotatebox{0}{\includegraphics{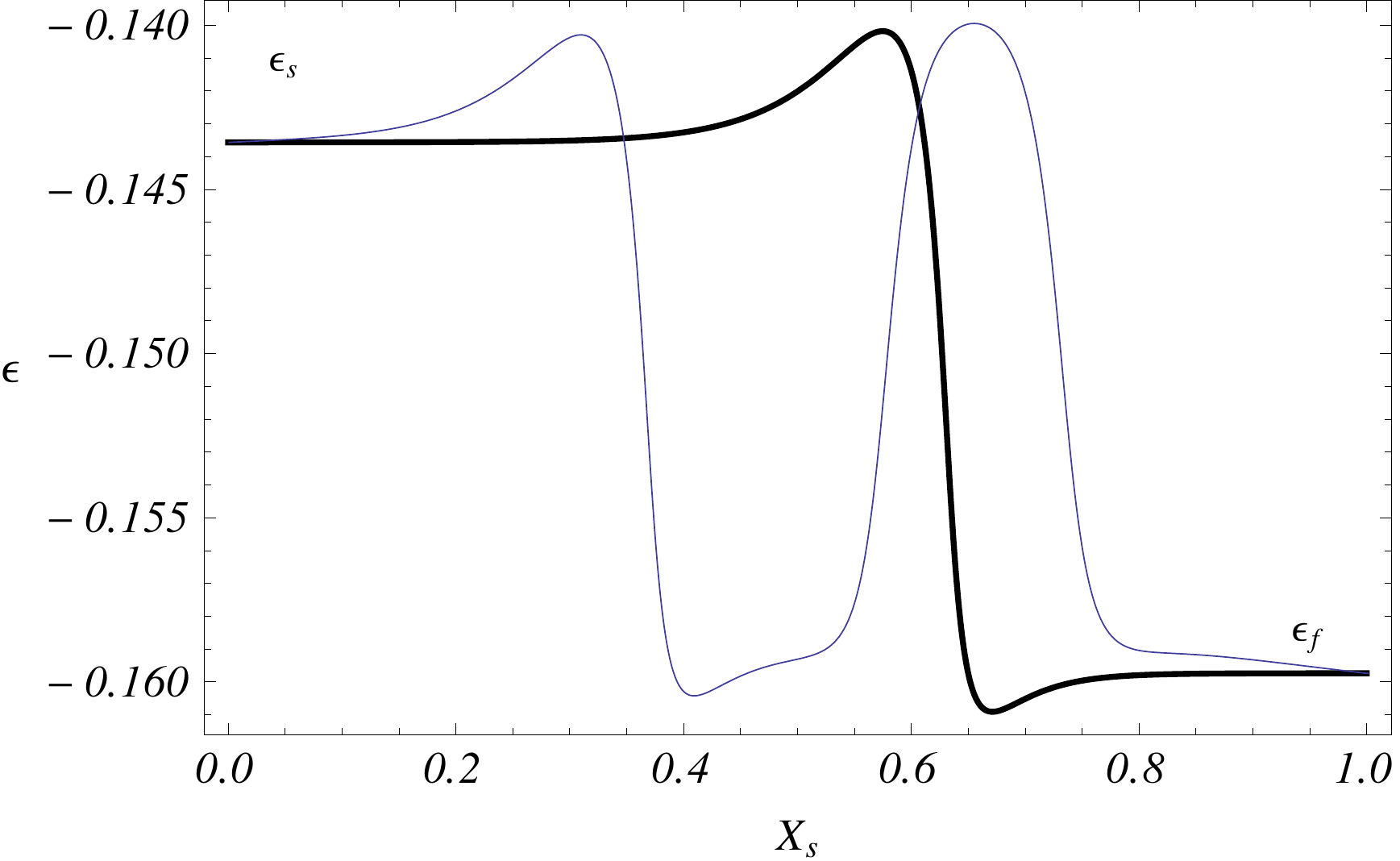}}} 
}
\put(0,320)
{
\resizebox{7.8cm}{!}{\rotatebox{0}{\includegraphics{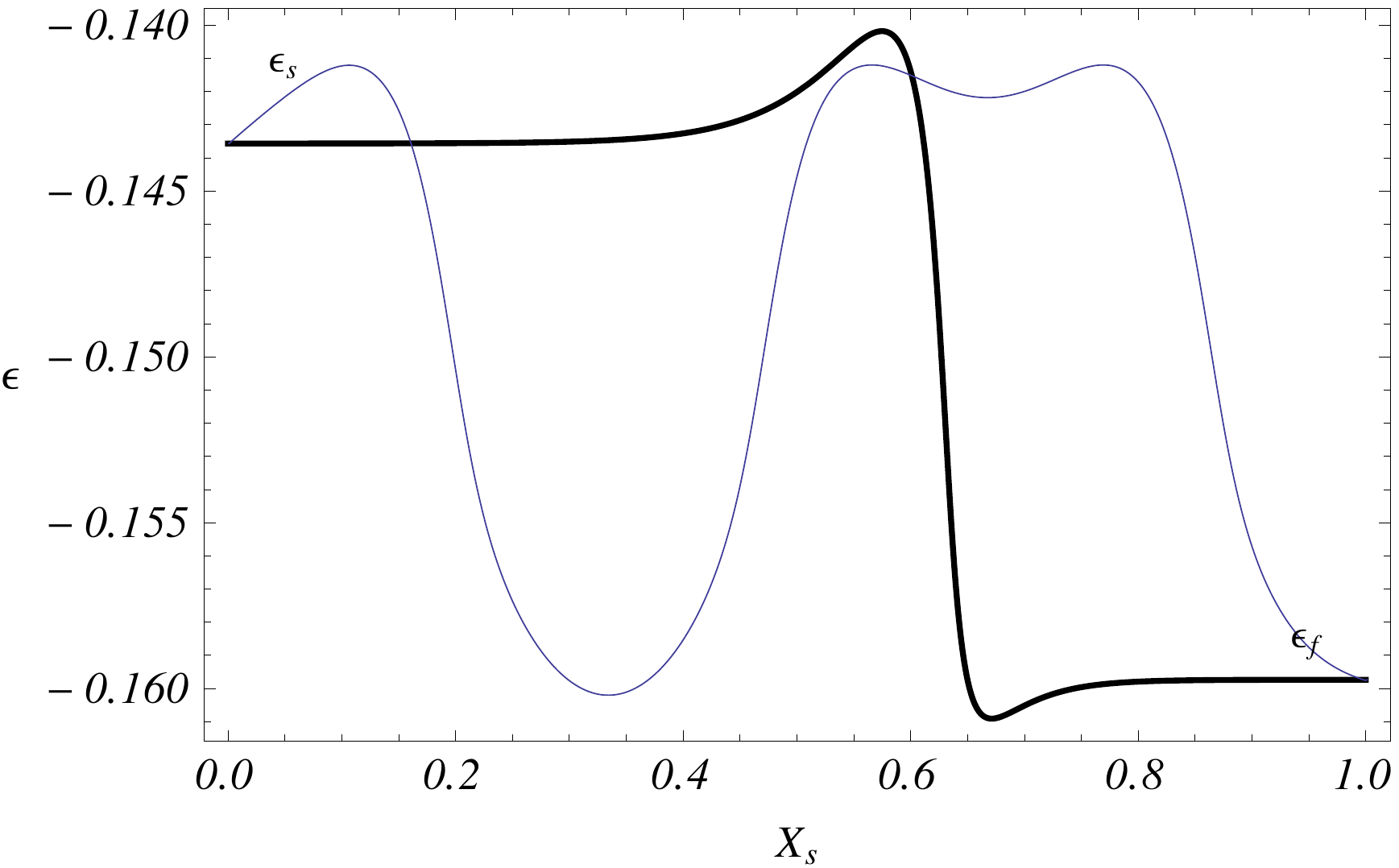}}} 
}
\put(235,320)
{
\resizebox{7.8cm}{!}{\rotatebox{0}{\includegraphics{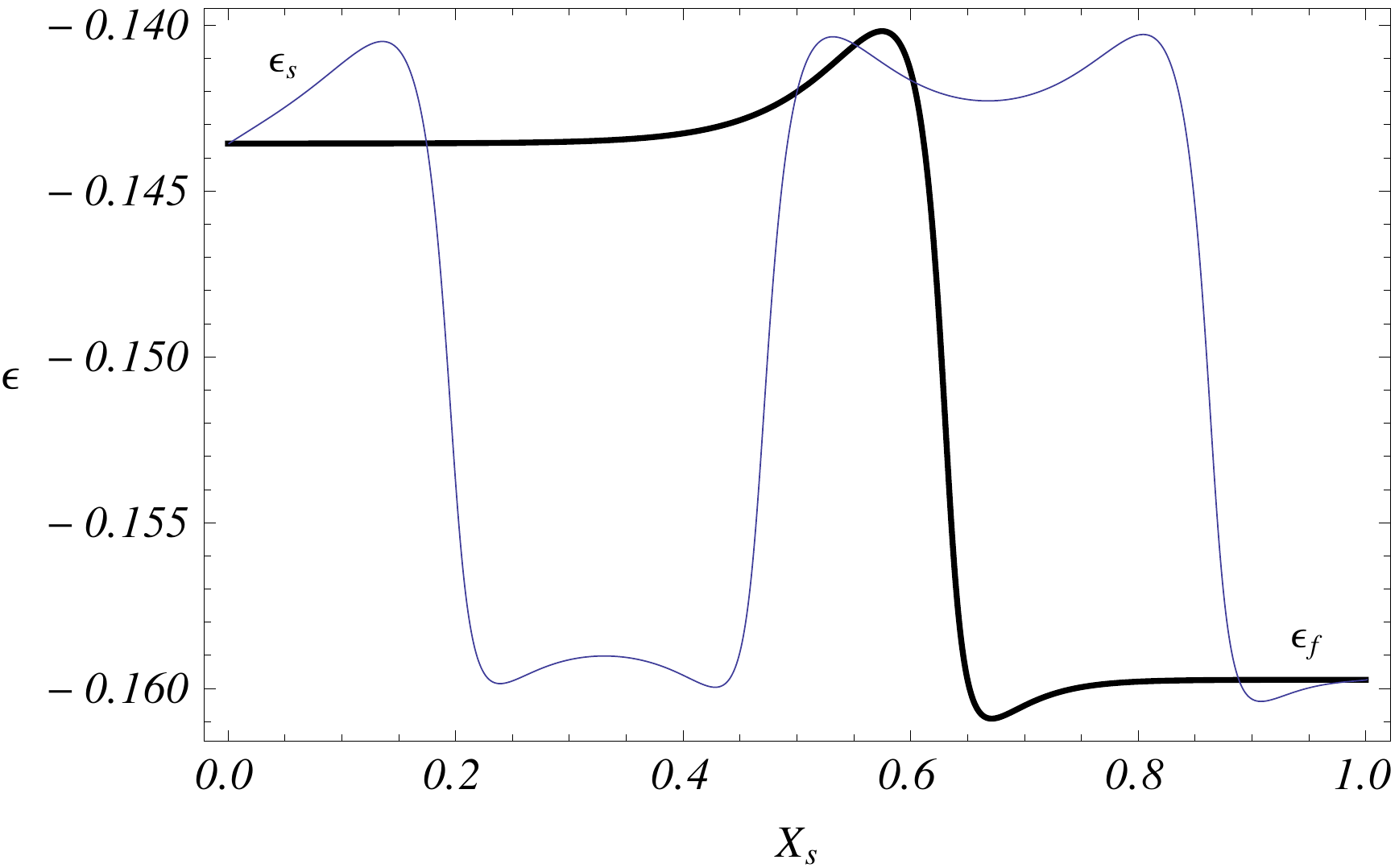}}} 
}
\end{picture}  
%\vskip 2 cm 
\caption{Profiles (solid thin lines)
$\varepsilon(X_\rr{s},t)$ 
obtained by solving the Cahn--Hilliard--like system (\ref{num05}) 
with a deterministic stratified initial state (the same used in 
figure~\ref{f:acdin05}) and 
Dirichelet boundary conditions 
$m(0)=m_\rr{s}$, $\varepsilon(0)=\varepsilon_\rr{s}$,
$m(1)=m_\rr{f}$, and $\varepsilon(1)=\varepsilon_\rr{f}$
on the finite interval $[0,1]$, 
at the coexistence pressure
for 
$a=0.5$, $b=1$, $\alpha=100$,
$k_1=k_2=k_3=10^{-3}$. 
The solid thick line is the corresponding stationary profile.
Profiles at times
$t=0.001,0.01,10.3,11,11.1,89$
are depicted in lexicographic order.
}
\label{f:chdin05} 
\end{figure}

Finally we have also studied the evolution of the system 
starting from a deterministic stratified initial datum; also in this case 
we used the same initial condition as the one used for the (\ref{num06})
problem, see figure~\ref{f:acdin05}. Results have been plotted in 
figures~\ref{f:chdin05} and \ref{f:chdin06}.

\subsection{Discussion of the numerical results}
\label{s:disc-results}
\par\noindent
Comparing the profiles of the strain and fluid mass density due to 
a pure Stokes (Cahn--Hilliard evolution) and a pure Darcy 
(Allen--Cahn evolution) dissipation, for spatially random or stratified 
initial conditions, some remarks naturally arise concerning 
short--time and long--time behavior. 

Monitoring the evolution process described by the Cahn--Hilliard partial 
differential equations and considering spatially random initial data, 
short--time micro--scale oscillations can be detected at times of order 
$10^{-2}$; 
for these characteristic times the profile is smeared out 
and progressively deforms firstly into a metastable drop--like shape, 
which is approached at times of order $10^{-1}$, and lastly into an 
interface--like graph which is reached at times of order $10^1$. 
Analogously the short--time evolution process, described by the Allen--Cahn 
differential equations and stemming from the same randomly distributed 
initial condition, still exhibits a kind of intermediate feature and finally
provides an interface--like graph which propagates towards the stationary
state. However this last is approached 
at times several order of magnitude greater than those characteristic of the 
Cahn--Hilliard evolution process (in particular $\gg 10^{4}$). 

\begin{figure}[t]
\begin{picture}(200,460)(50,0)
\put(0,0)
{
\resizebox{7.8cm}{!}{\rotatebox{0}{\includegraphics{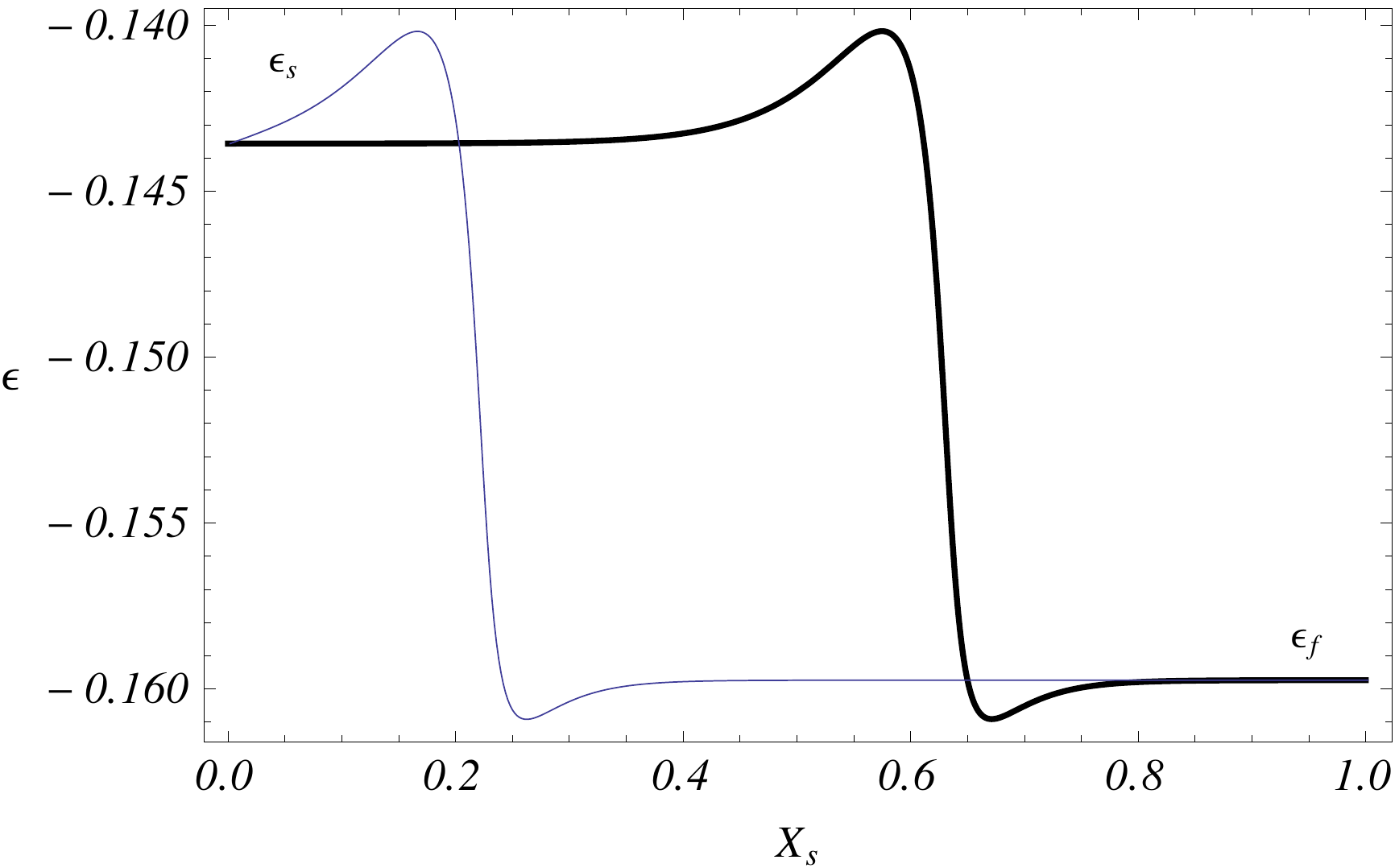}}} 
}
\put(235,0)
{
\resizebox{7.8cm}{!}{\rotatebox{0}{\includegraphics{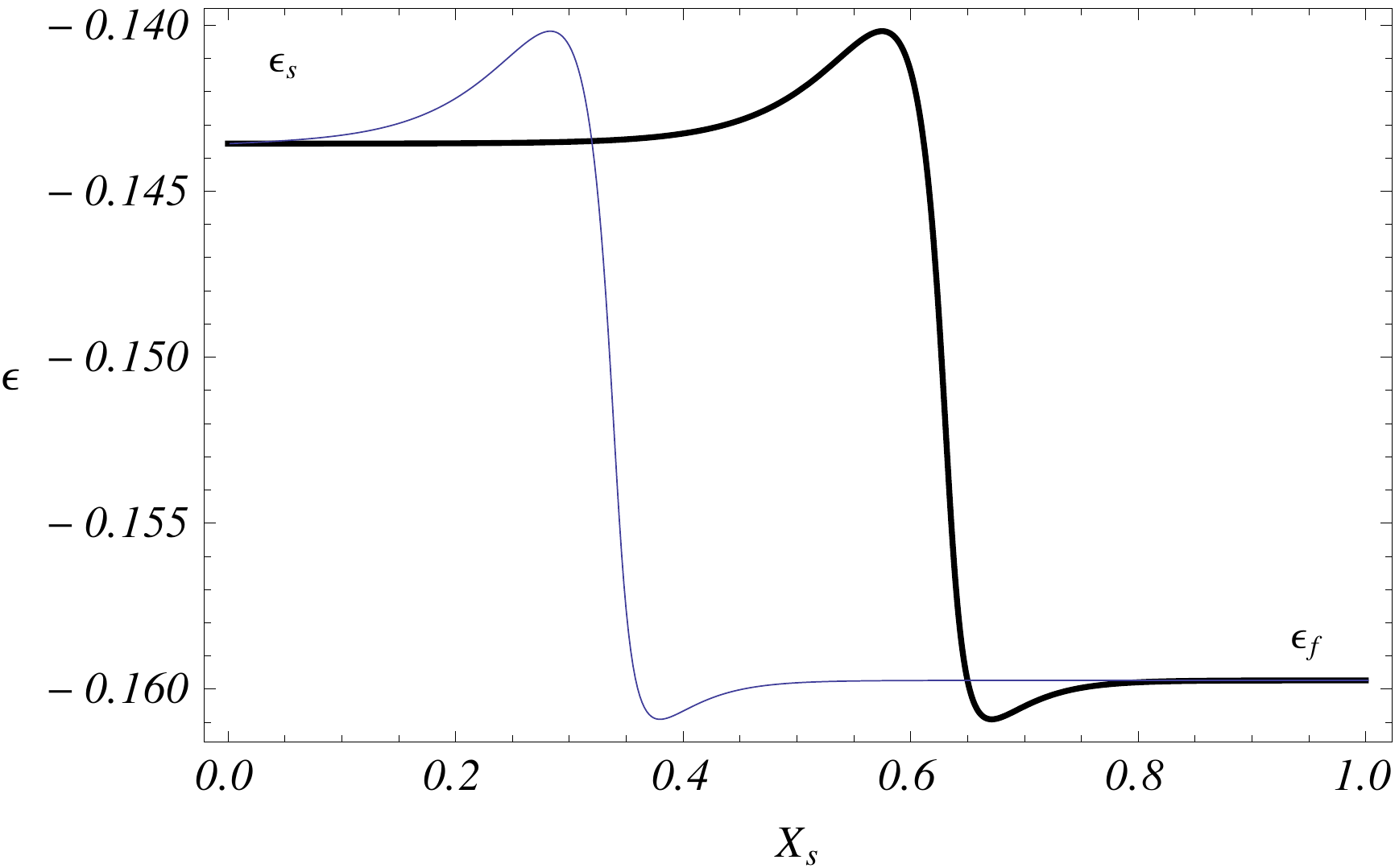}}} 
}
\put(0,160)
{
\resizebox{7.8cm}{!}{\rotatebox{0}{\includegraphics{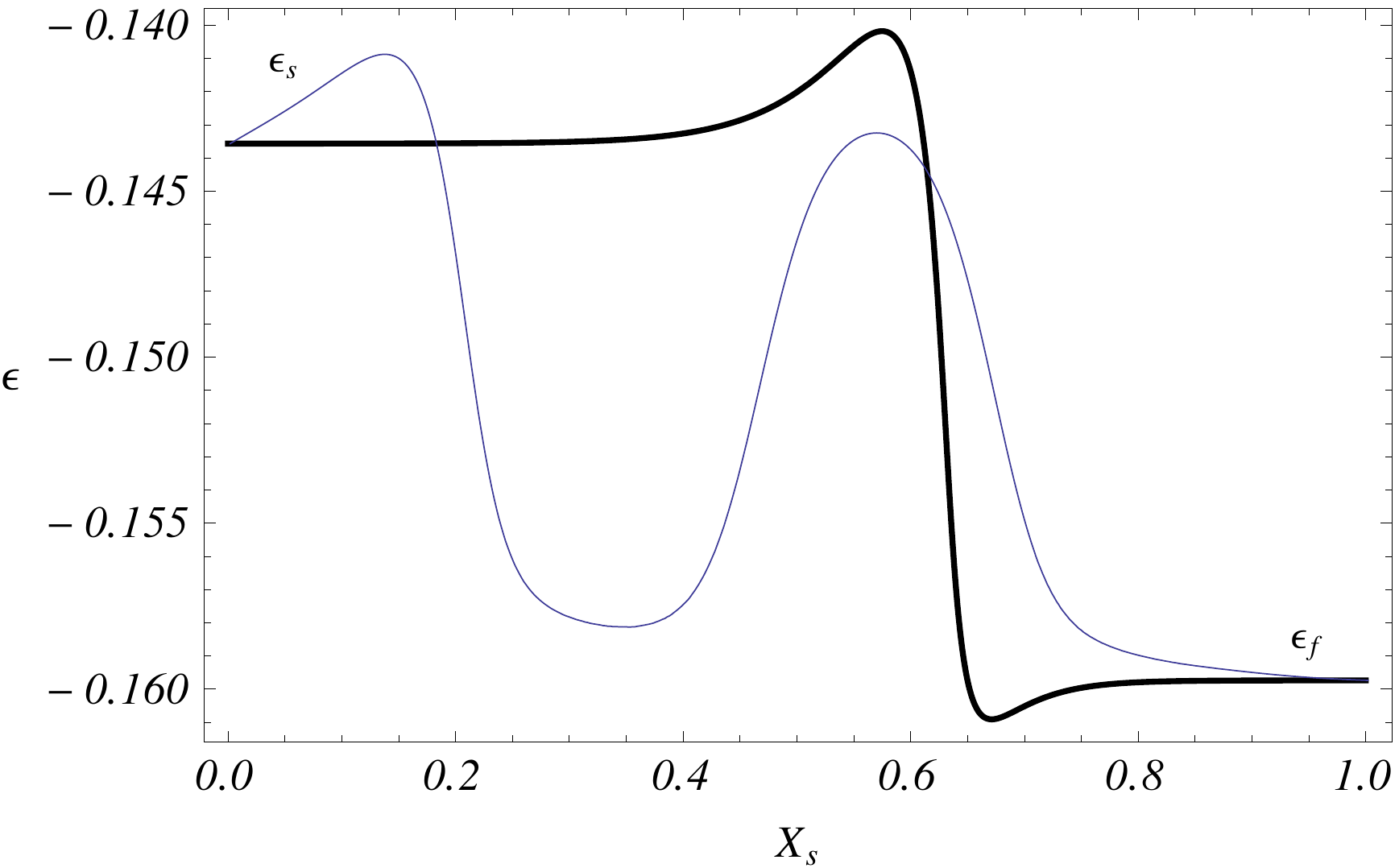}}} 
}
\put(235,160)
{
\resizebox{7.8cm}{!}{\rotatebox{0}{\includegraphics{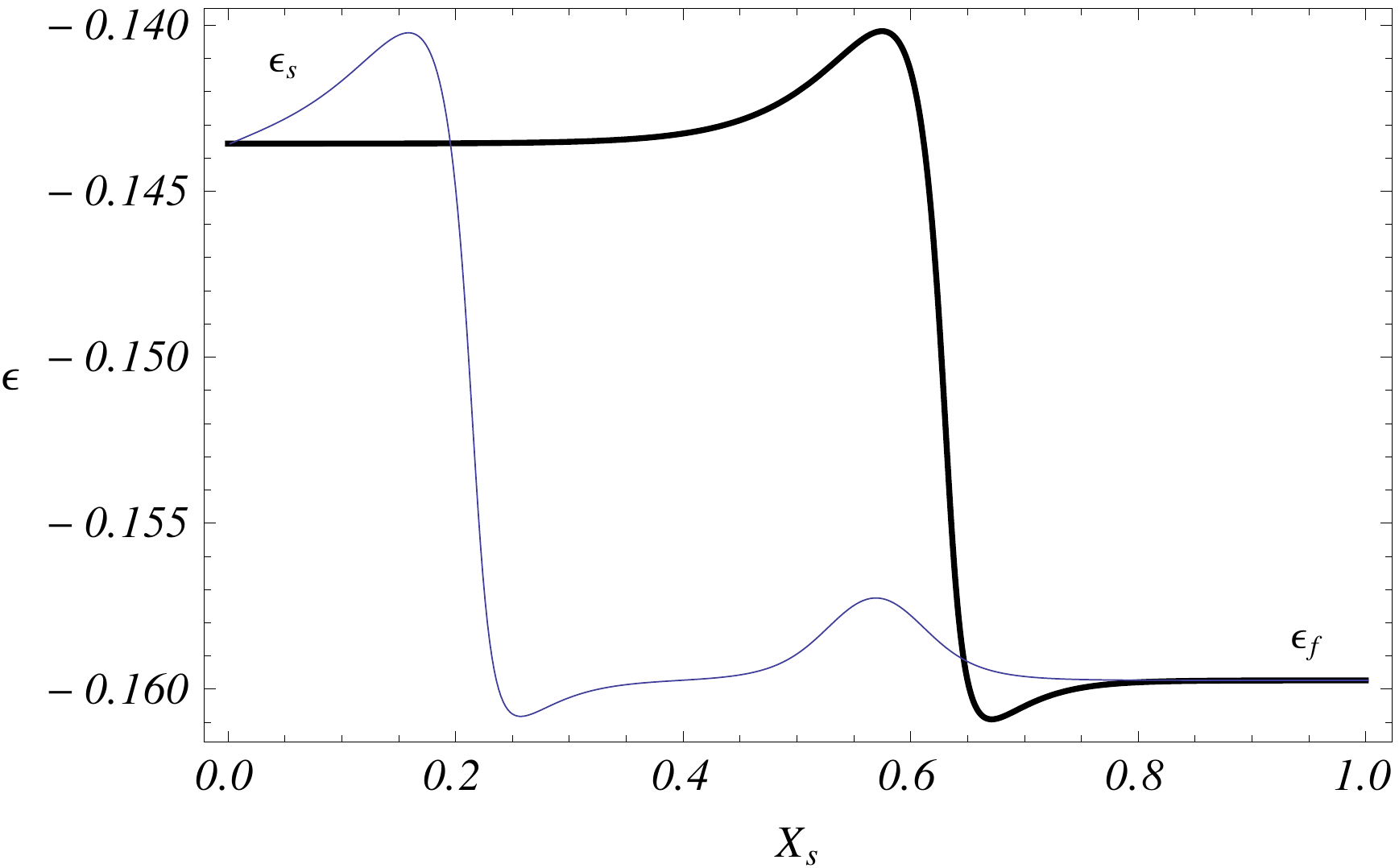}}} 
}
\put(0,320)
{
\resizebox{7.8cm}{!}{\rotatebox{0}{\includegraphics{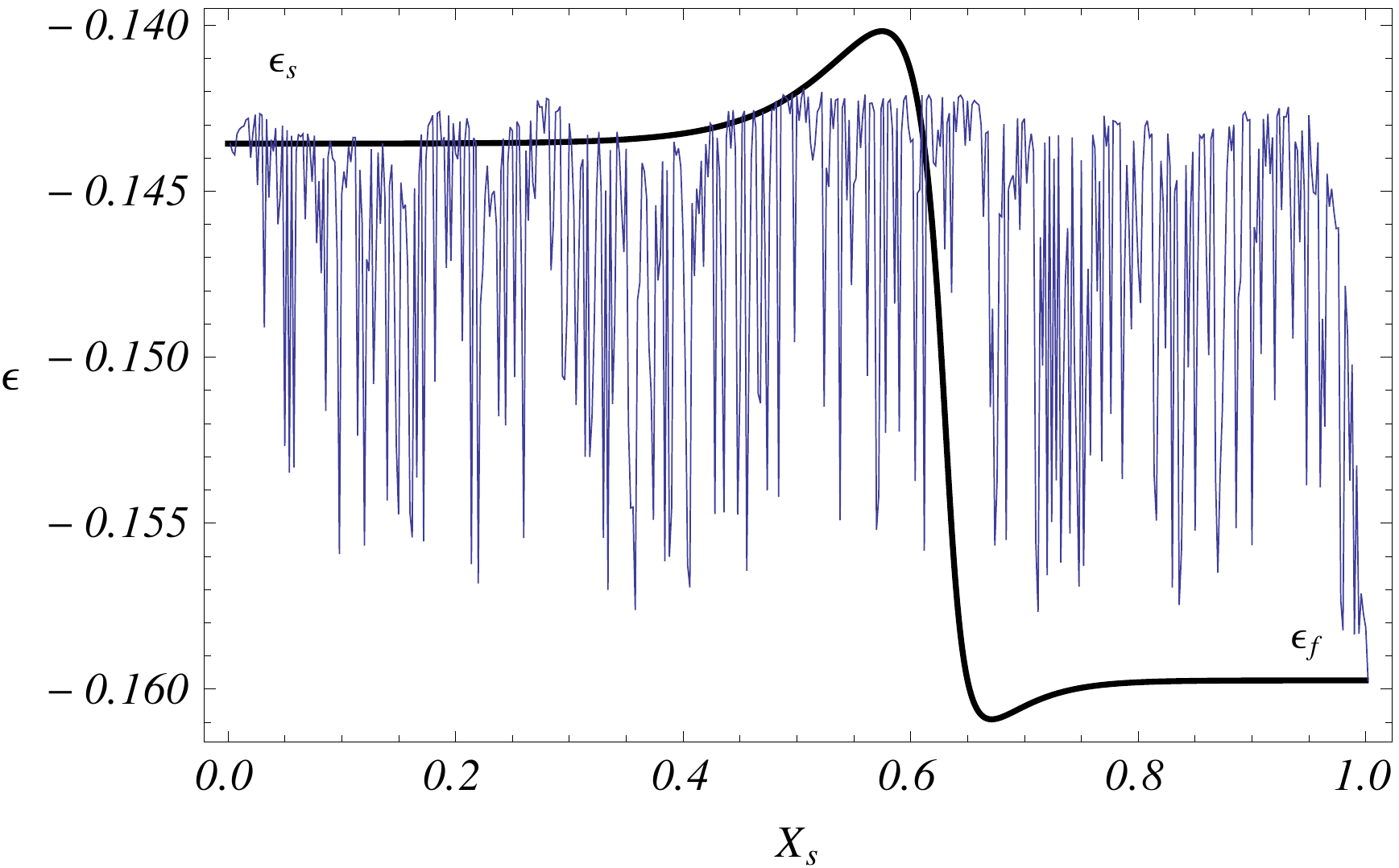}}} 
}
\put(235,320)
{
\resizebox{7.8cm}{!}{\rotatebox{0}{\includegraphics{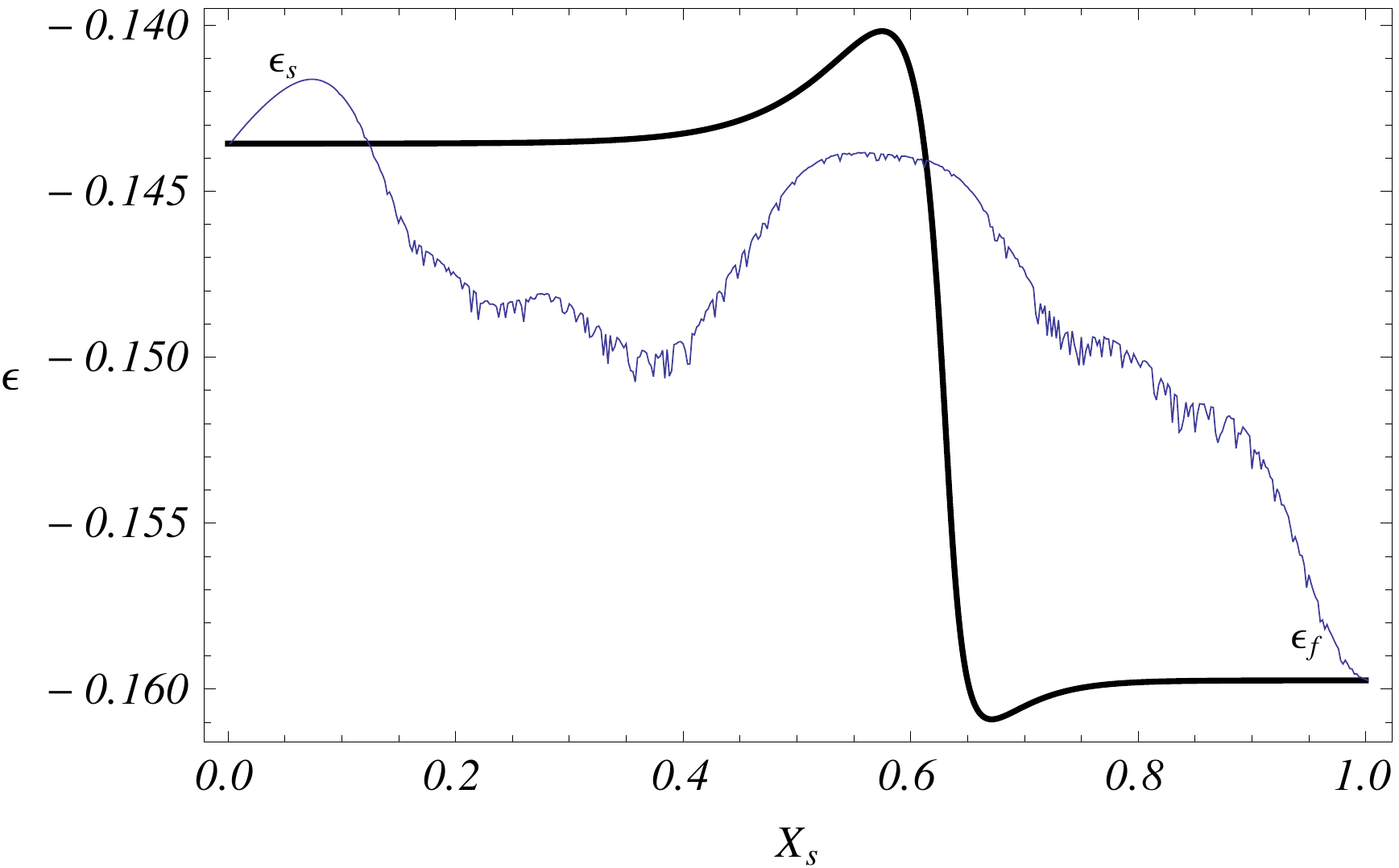}}} 
}
\end{picture}  
%\vskip 2 cm 
\caption{Profiles (solid thin lines)
$\varepsilon(X_\rr{s},t)$ 
obtained by solving the Allen--Cahn--like system (\ref{num06}) 
with a random initial condition and 
Dirichelet boundary conditions 
$m(0)=m_\rr{s}$, $\varepsilon(0)=\varepsilon_\rr{s}$,
$m(1)=m_\rr{f}$, and $\varepsilon(1)=\varepsilon_\rr{f}$
on the finite interval $[0,1]$, 
at the coexistence pressure
for 
$a=0.5$, $b=1$, $\alpha=100$,
$k_1=k_2=k_3=10^{-3}$. 
The solid thick line is the corresponding stationary profile.
Profiles at times
$t=2,5.4,7,15,200,50000$
are depicted in lexicographic order.
}
\label{f:acdin01} 
\end{figure}

\begin{figure}[t]
\begin{picture}(200,460)(50,0)
\put(0,0)
{
\resizebox{7.8cm}{!}{\rotatebox{0}{\includegraphics{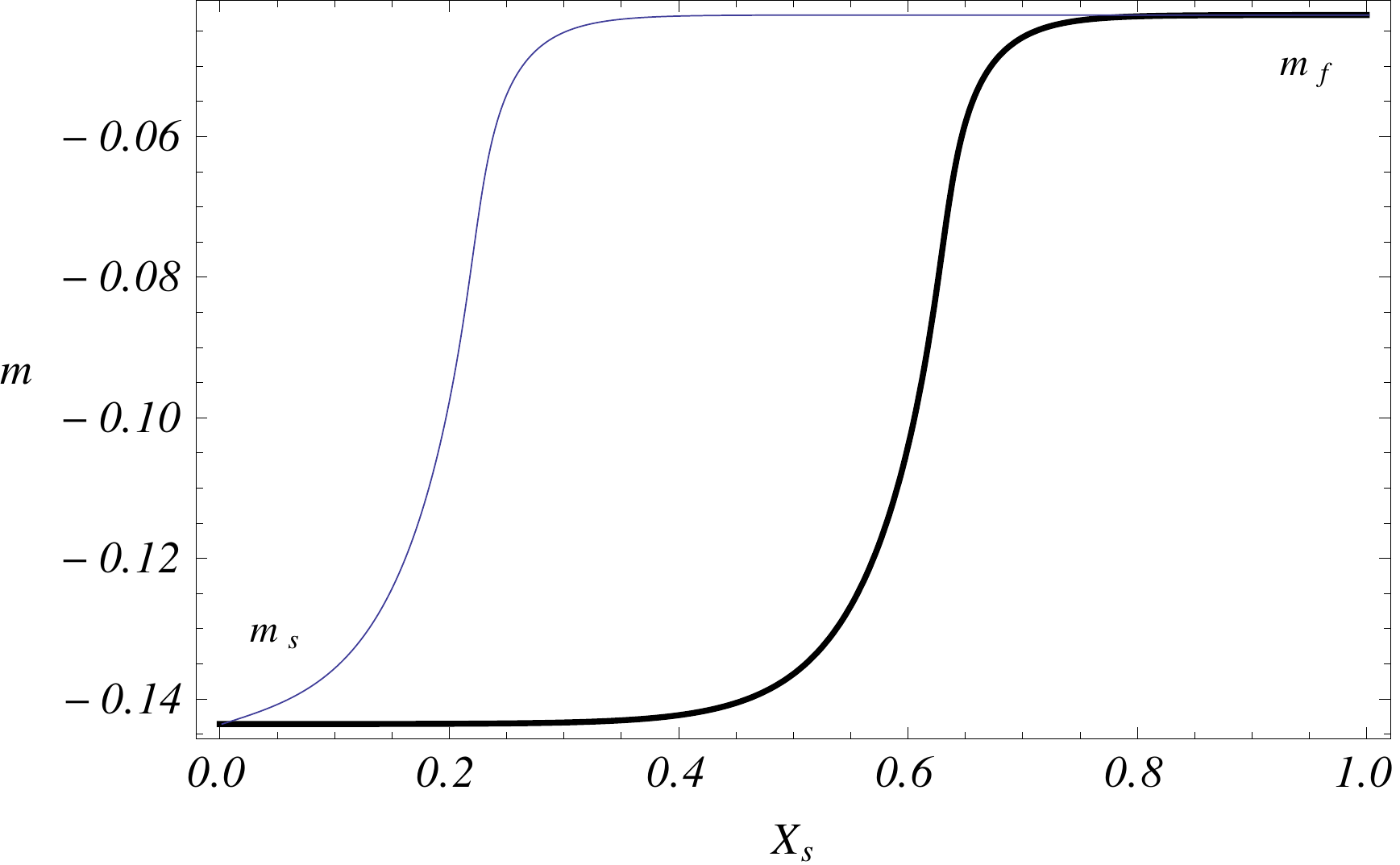}}} 
}
\put(235,0)
{
\resizebox{7.8cm}{!}{\rotatebox{0}{\includegraphics{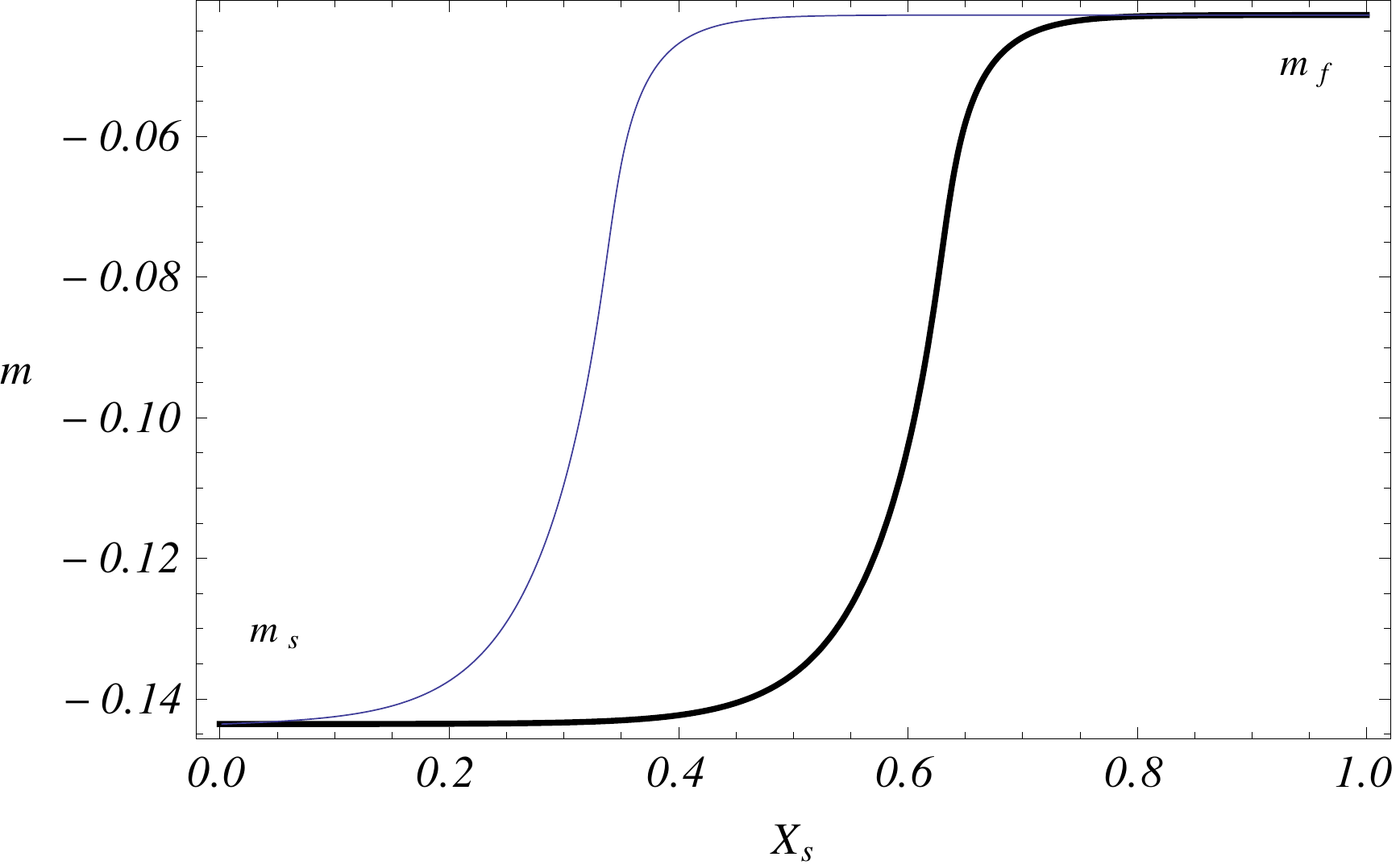}}} 
}
\put(0,160)
{
\resizebox{7.8cm}{!}{\rotatebox{0}{\includegraphics{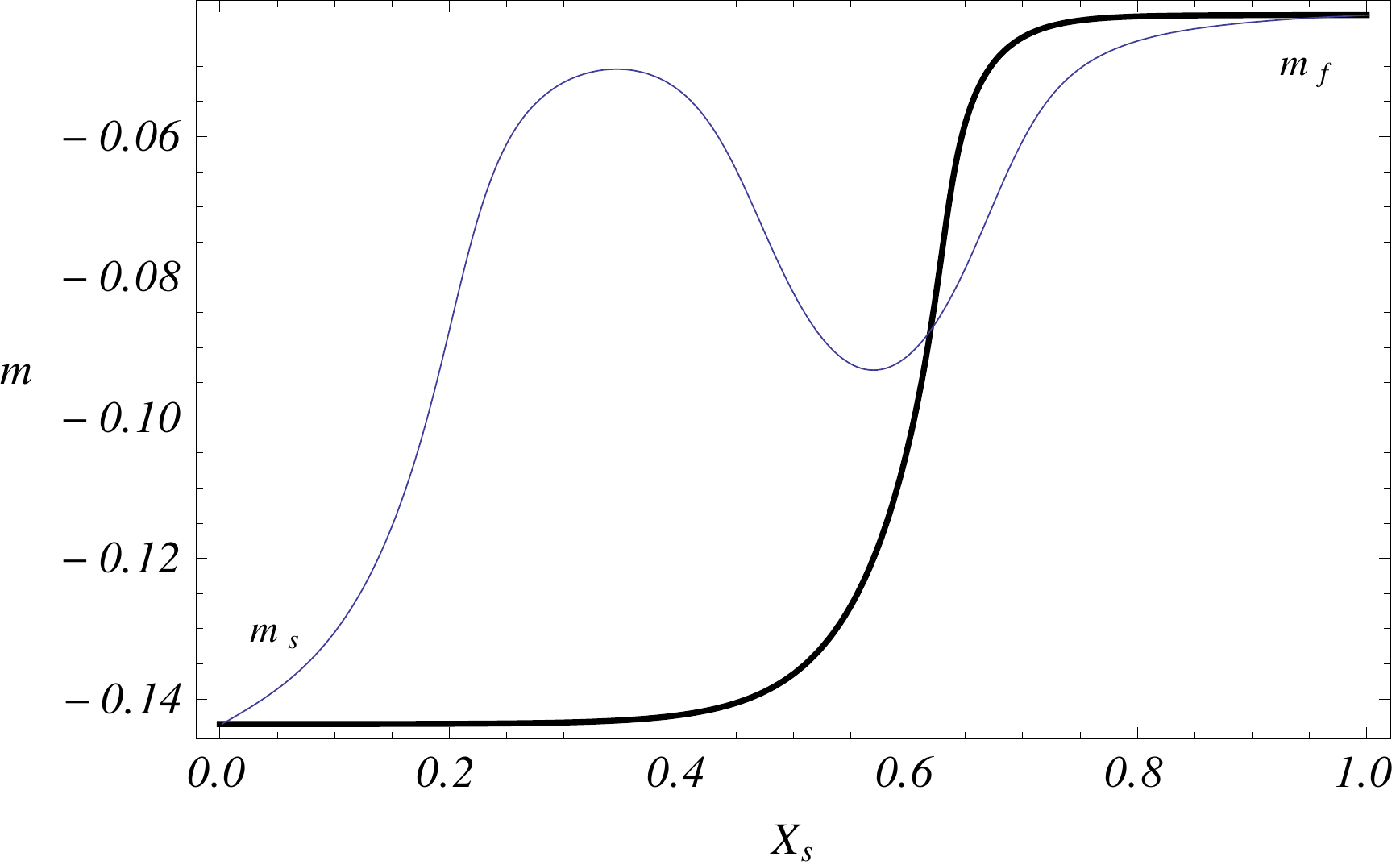}}} 
}
\put(235,160)
{
\resizebox{7.8cm}{!}{\rotatebox{0}{\includegraphics{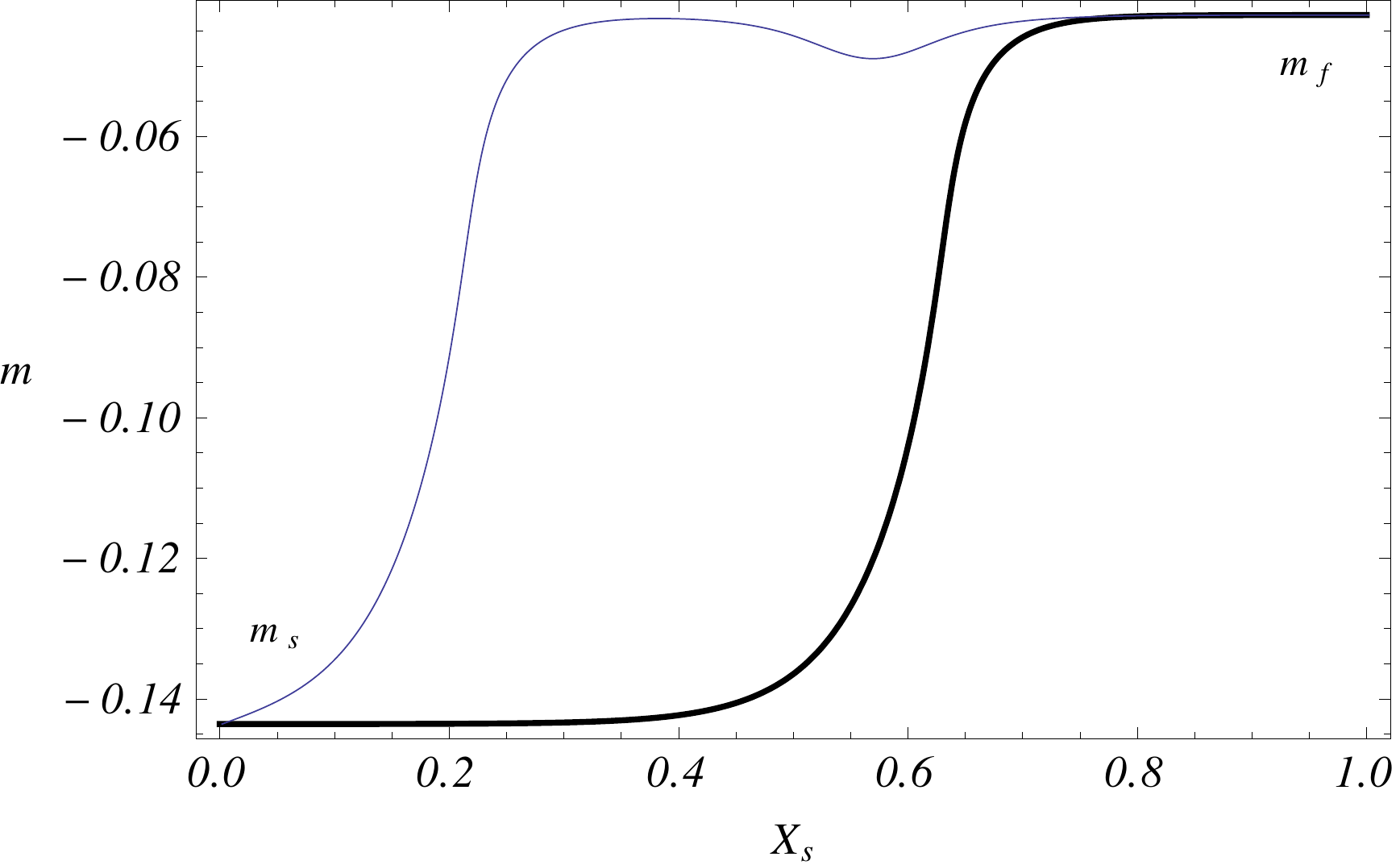}}} 
}
\put(0,320)
{
\resizebox{7.8cm}{!}{\rotatebox{0}{\includegraphics{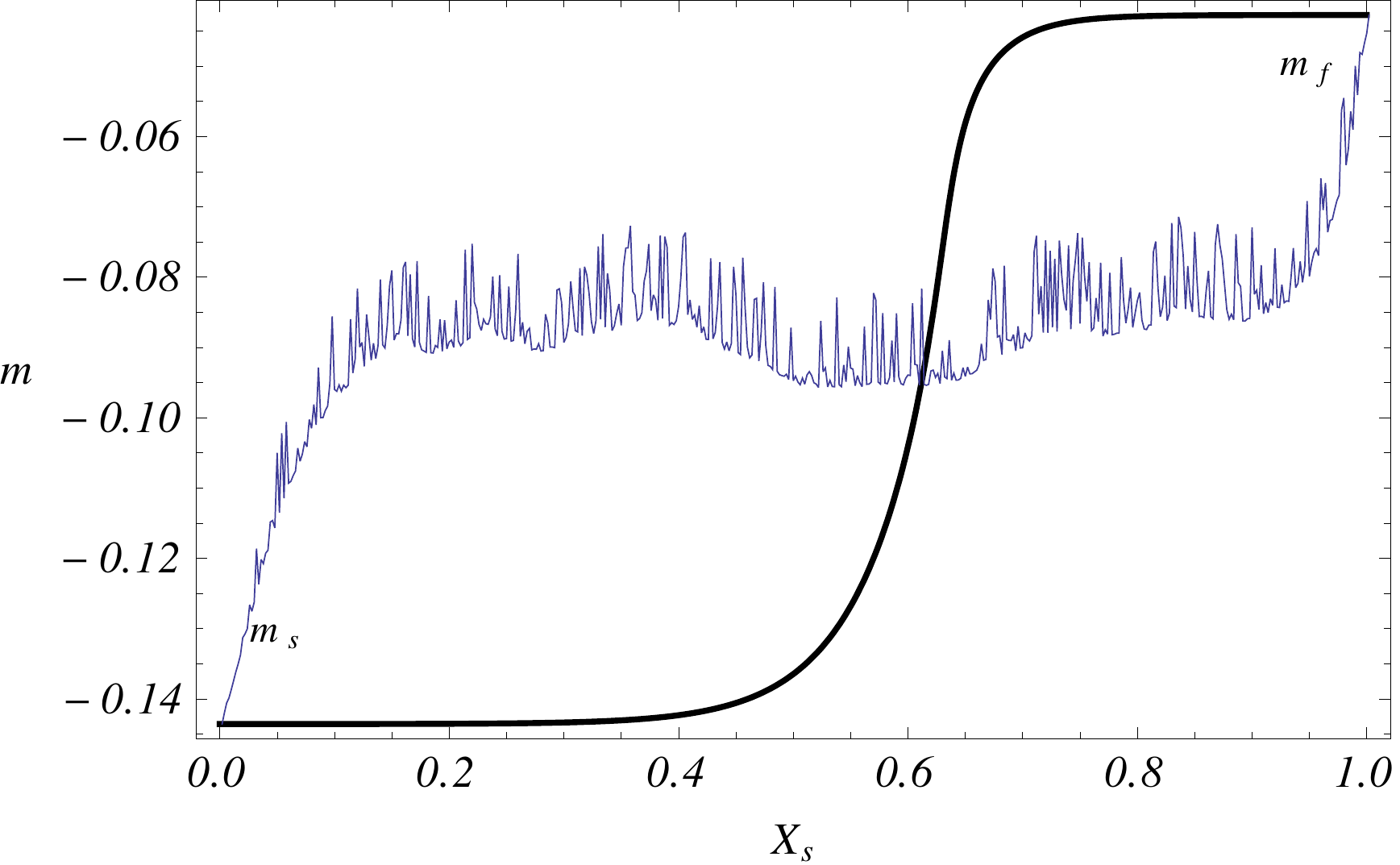}}} 
}
\put(235,320)
{
\resizebox{7.8cm}{!}{\rotatebox{0}{\includegraphics{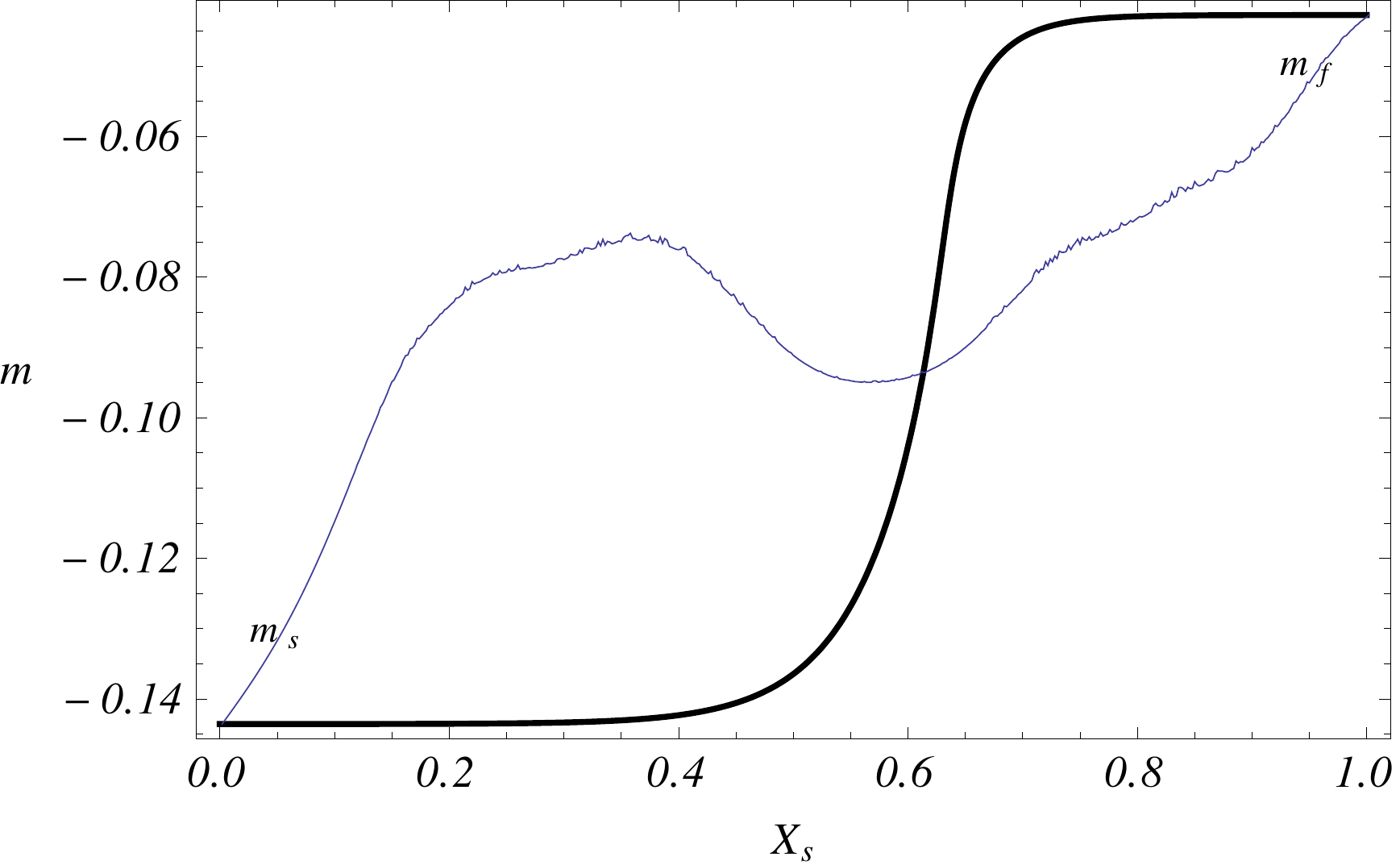}}} 
}
\end{picture}  
%\vskip 2 cm 
\caption{The same as in figure~\ref{f:acdin01} for the 
$m(X_\rr{s},t)$ profiles.
}
\label{f:acdin02} 
\end{figure}

\begin{figure}[t]
\begin{picture}(200,150)(50,0)
\put(0,0)
{
\resizebox{5cm}{!}{\rotatebox{0}{\includegraphics{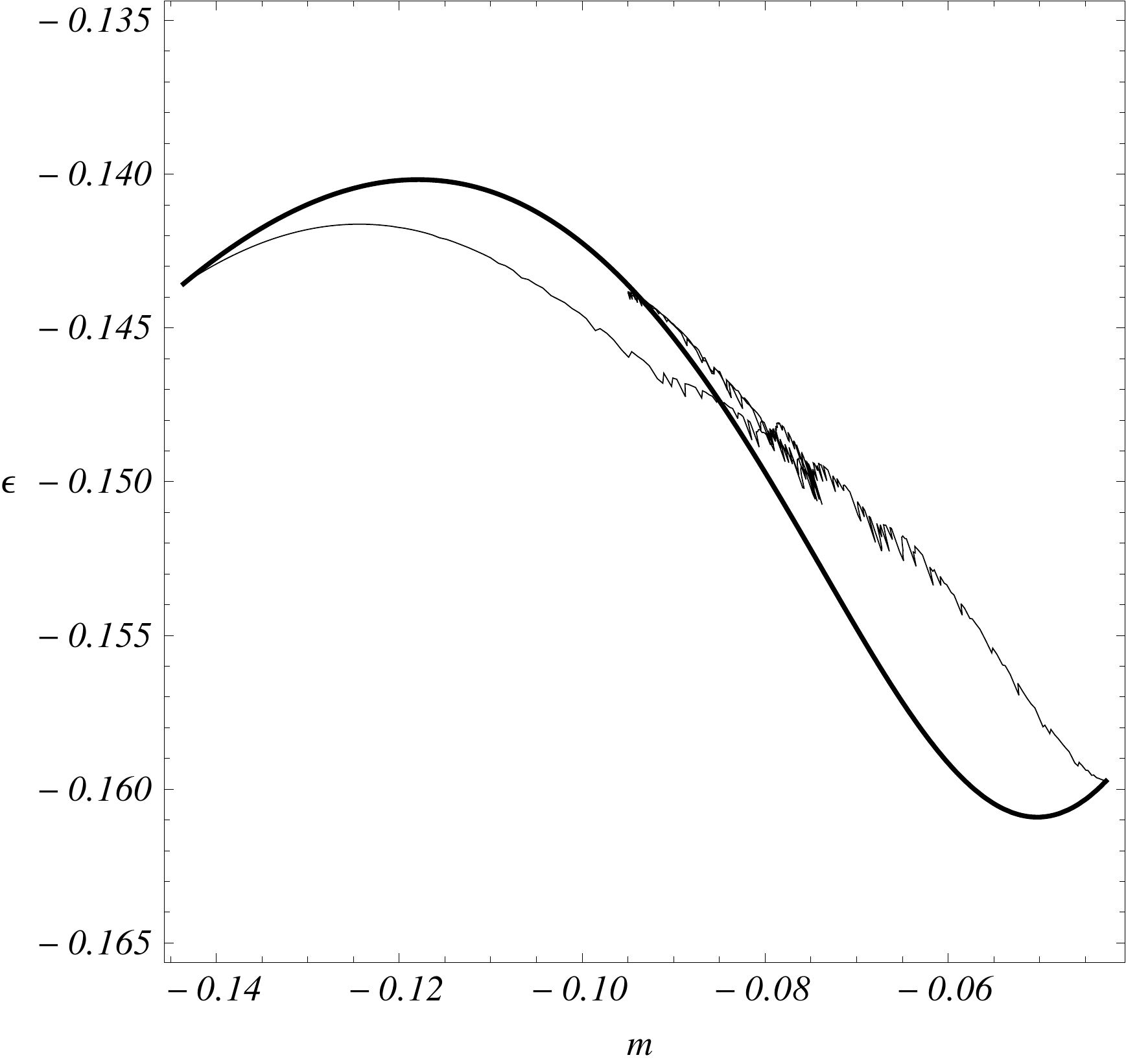}}} 
}
\put(160,0)
{
\resizebox{5cm}{!}{\rotatebox{0}{\includegraphics{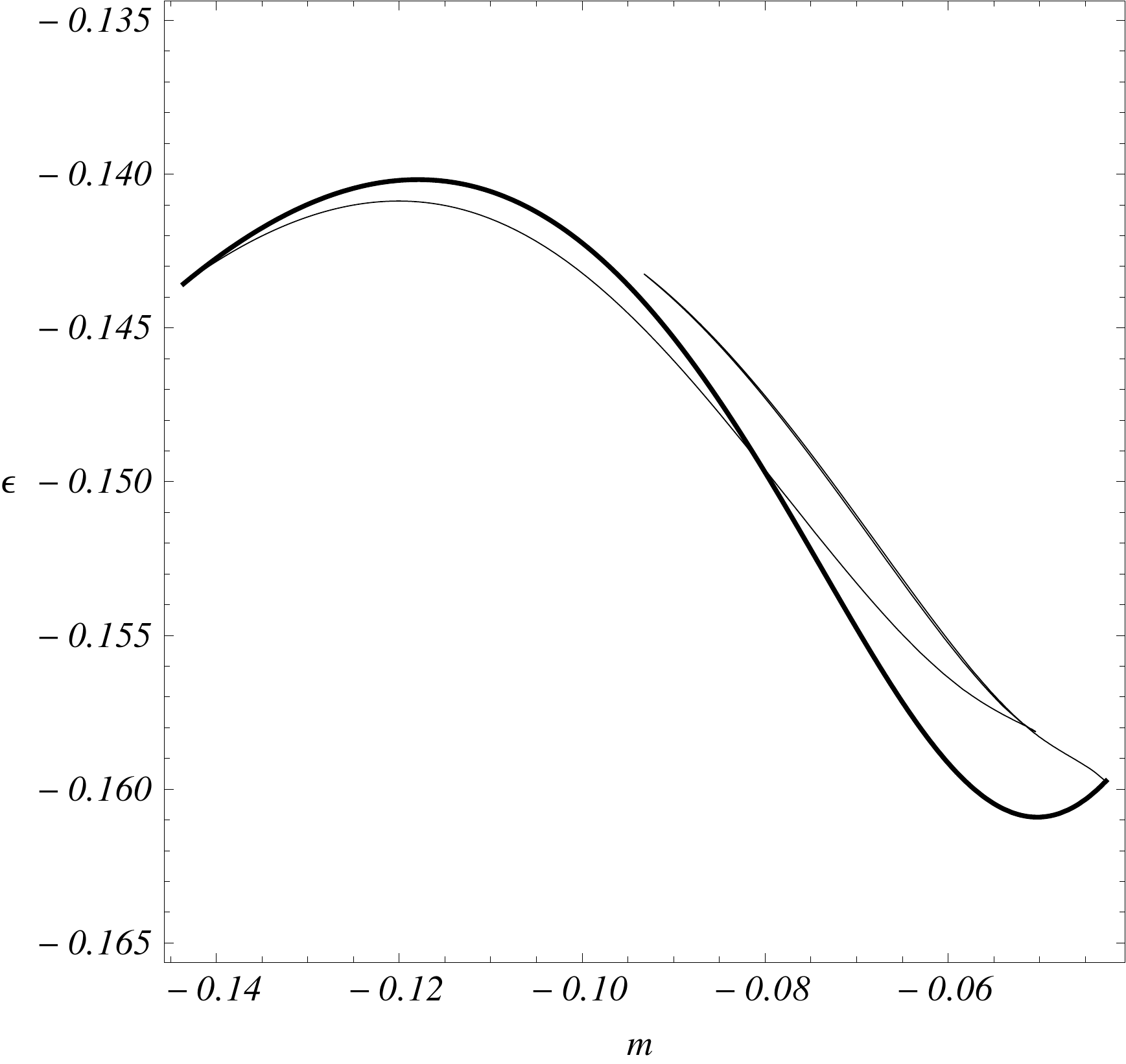}}} 
}
\put(320,0)
{
\resizebox{5cm}{!}{\rotatebox{0}{\includegraphics{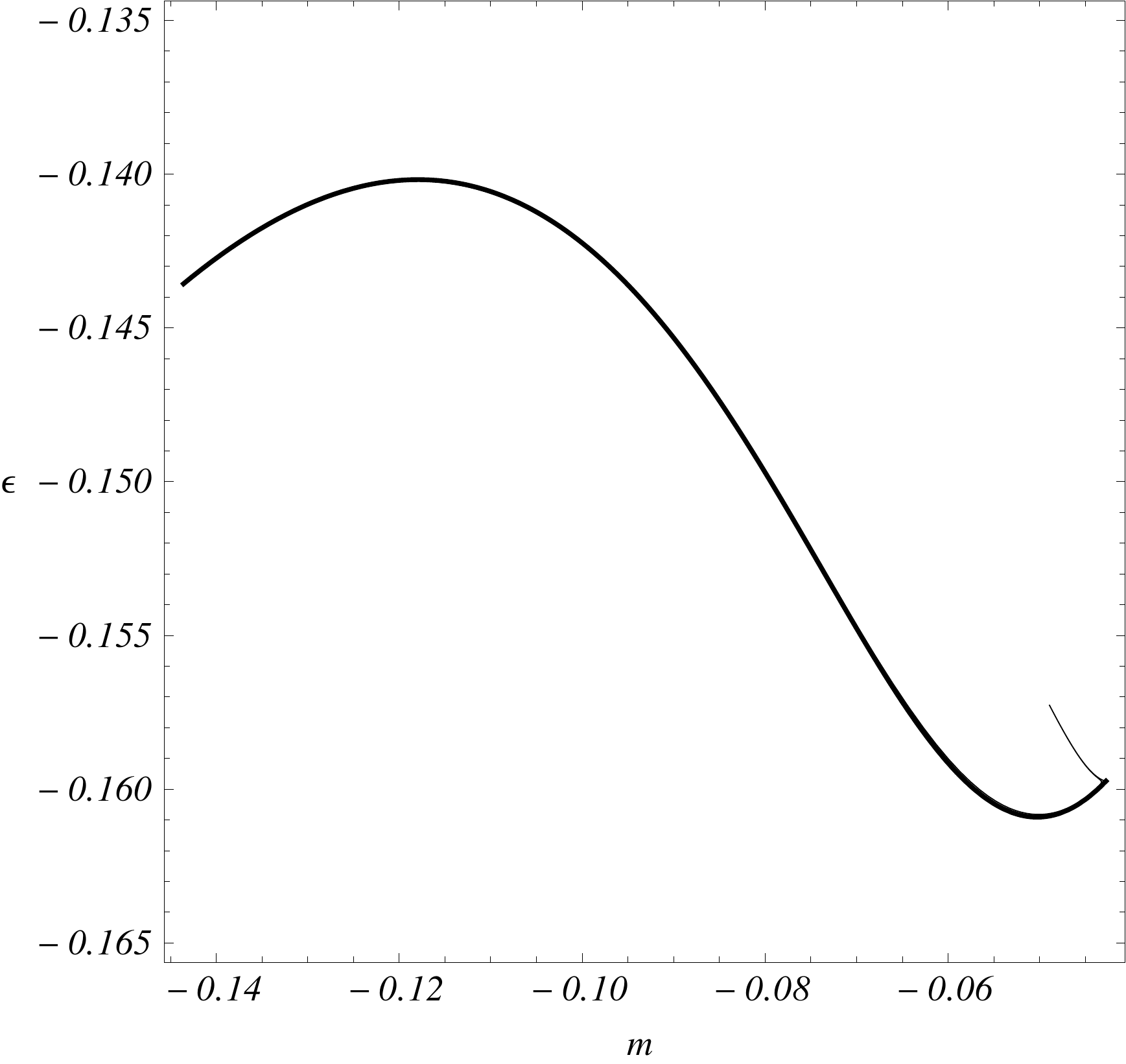}}} 
}
\end{picture}  
%\vskip 2 cm 
\caption{The solution of the same problem as in figure~\ref{f:acdin01} 
is depicted on the plane $m$--$\varepsilon$
at times $t=5.4,7,15$ from the left to the right.
}
\label{f:acdin03} 
\end{figure}

\begin{figure}
\begin{picture}(200,460)(50,0)
\put(0,0)
{
\resizebox{7.8cm}{!}{\rotatebox{0}{\includegraphics{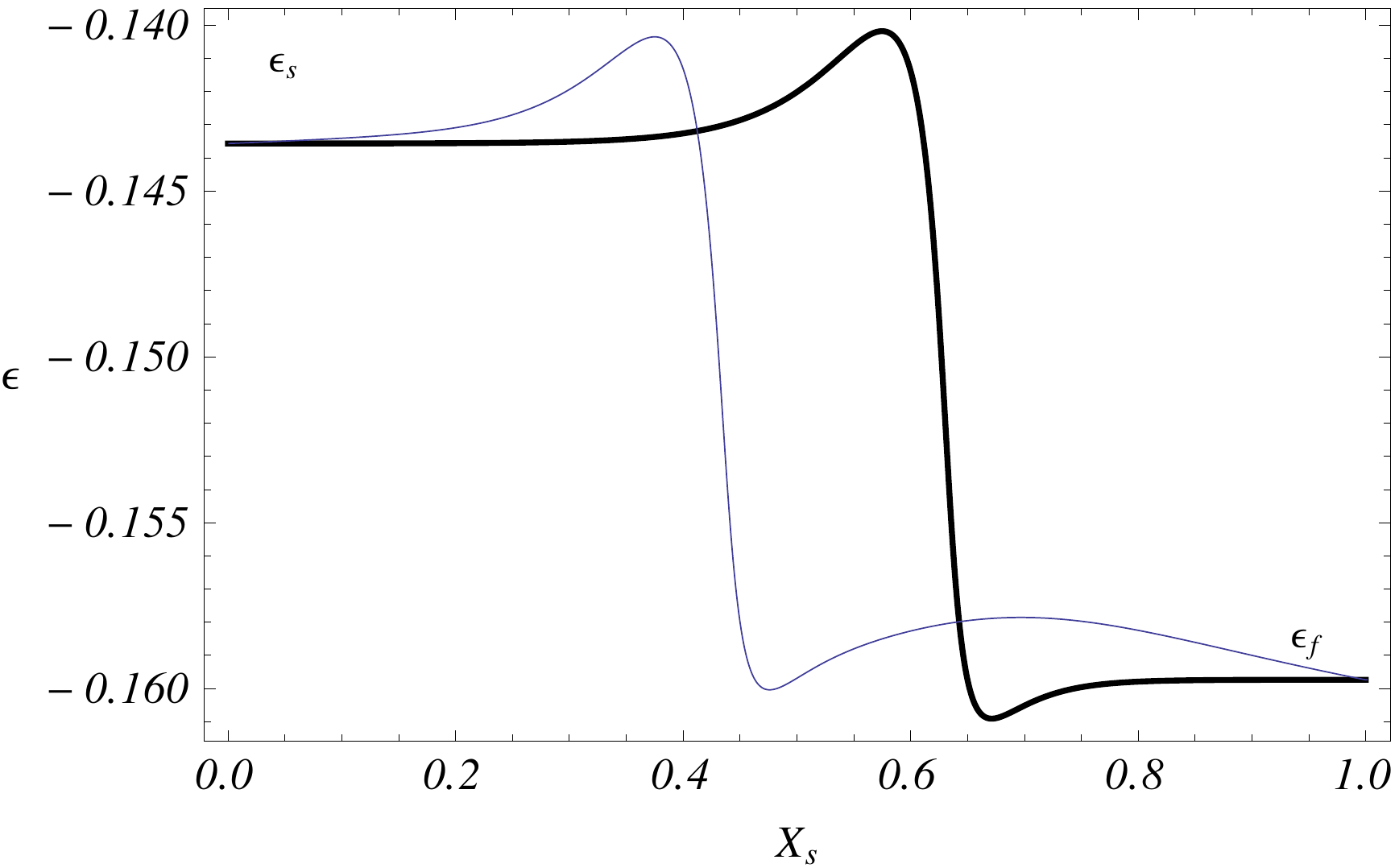}}} 
}
\put(235,0)
{
\resizebox{7.8cm}{!}{\rotatebox{0}{\includegraphics{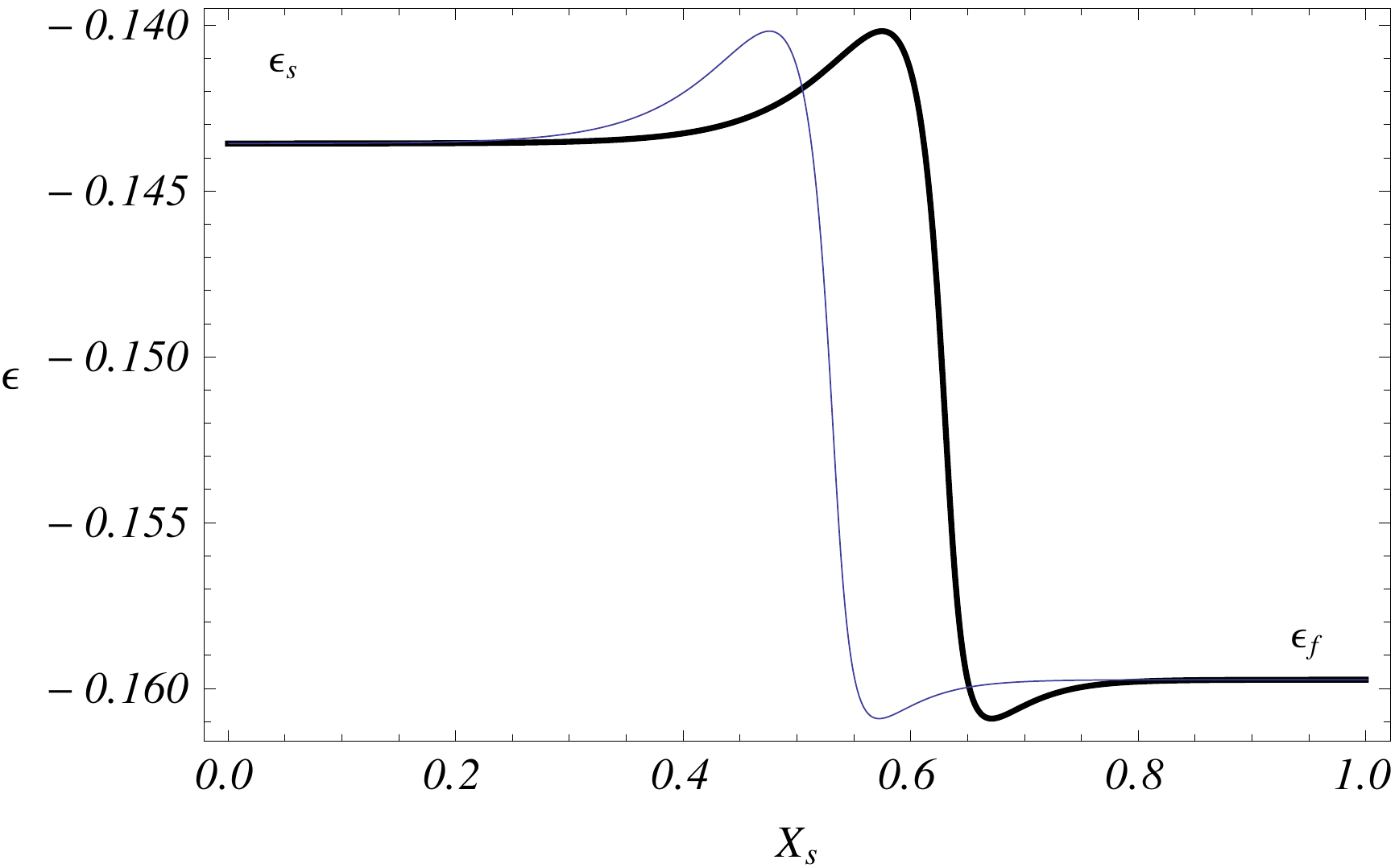}}} 
}
\put(0,160)
{
\resizebox{7.8cm}{!}{\rotatebox{0}{\includegraphics{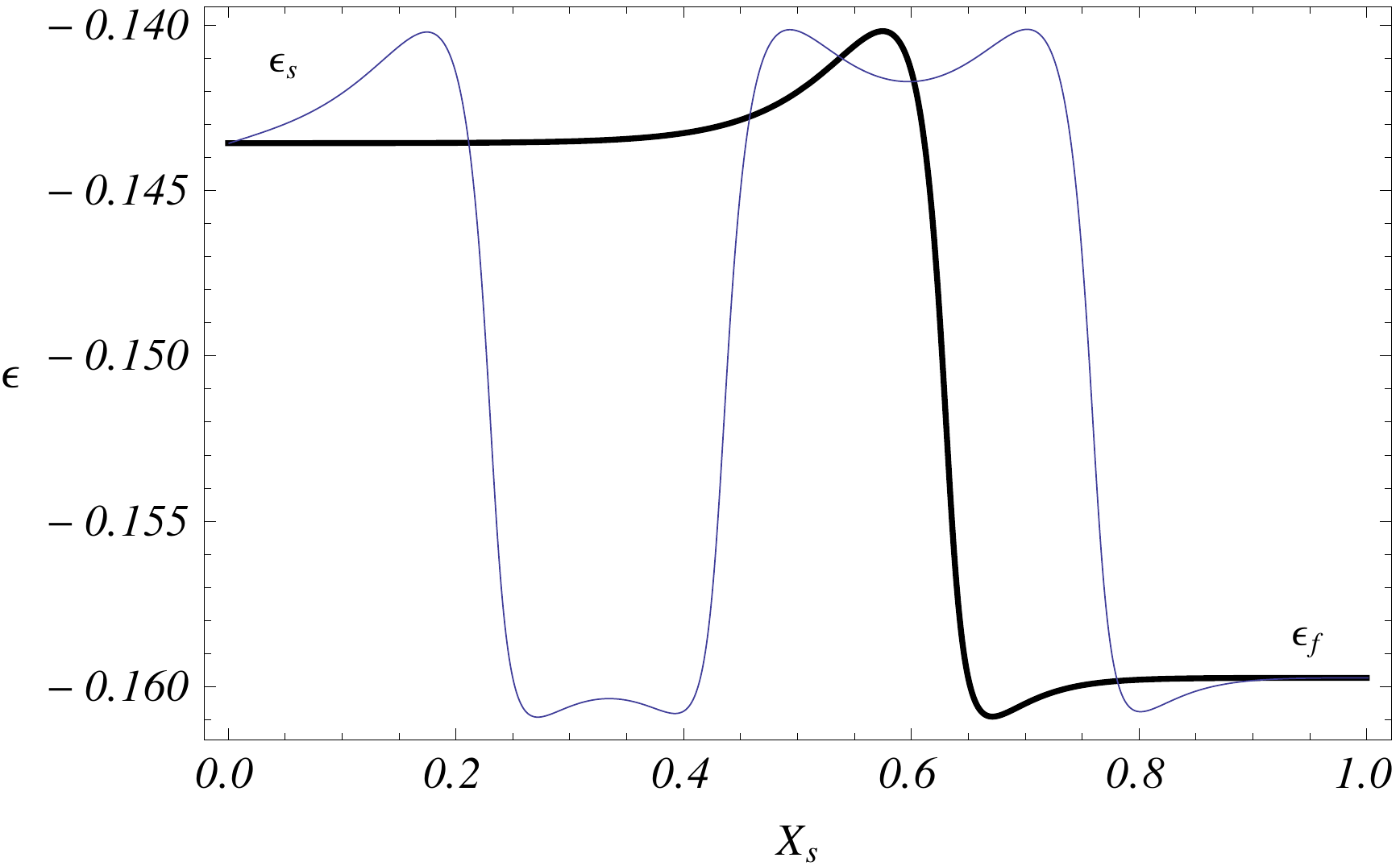}}} 
}
\put(235,160)
{
\resizebox{7.8cm}{!}{\rotatebox{0}{\includegraphics{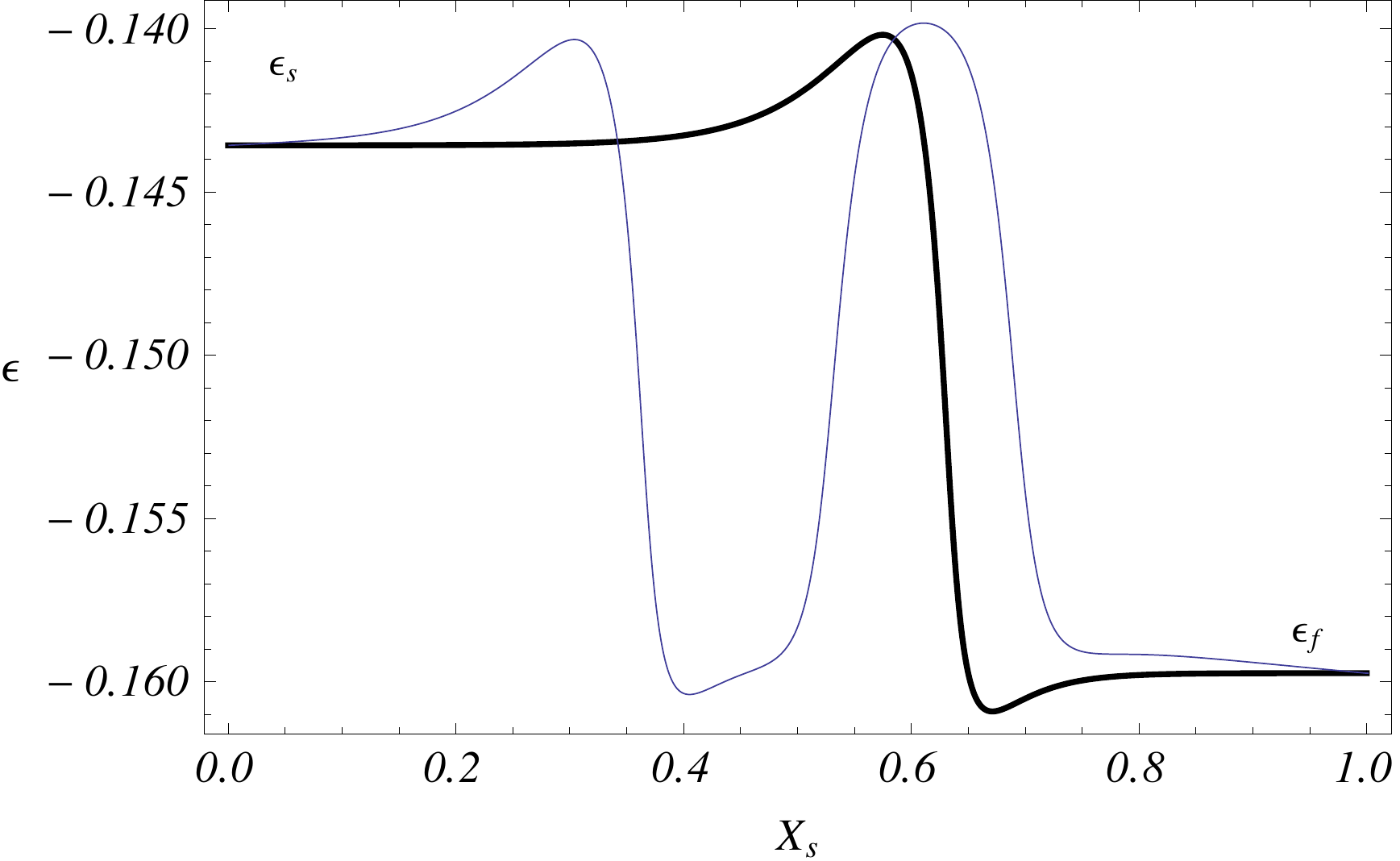}}} 
}
\put(0,320)
{
\resizebox{7.8cm}{!}{\rotatebox{0}{\includegraphics{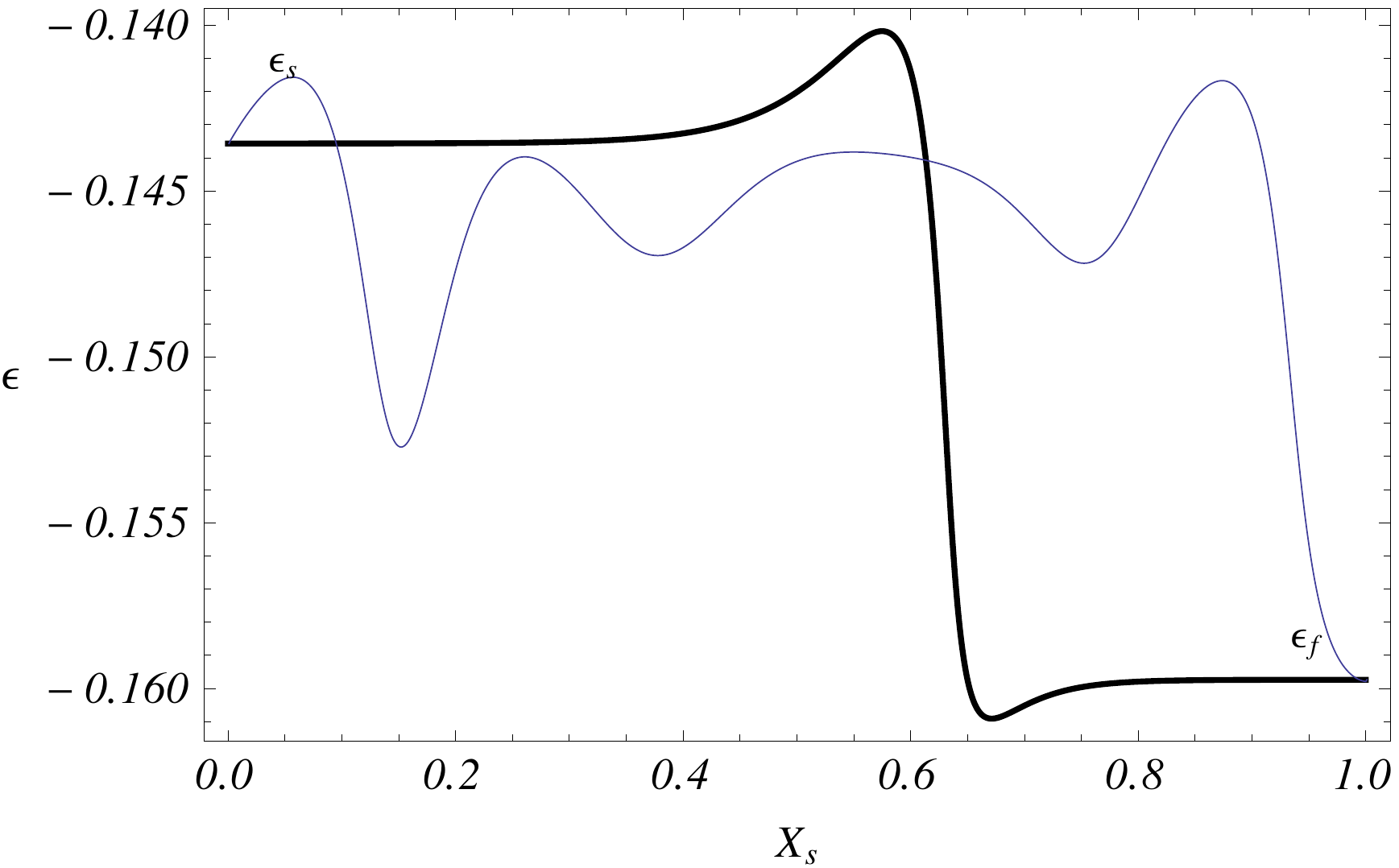}}} 
}
\put(235,320)
{
\resizebox{7.8cm}{!}{\rotatebox{0}{\includegraphics{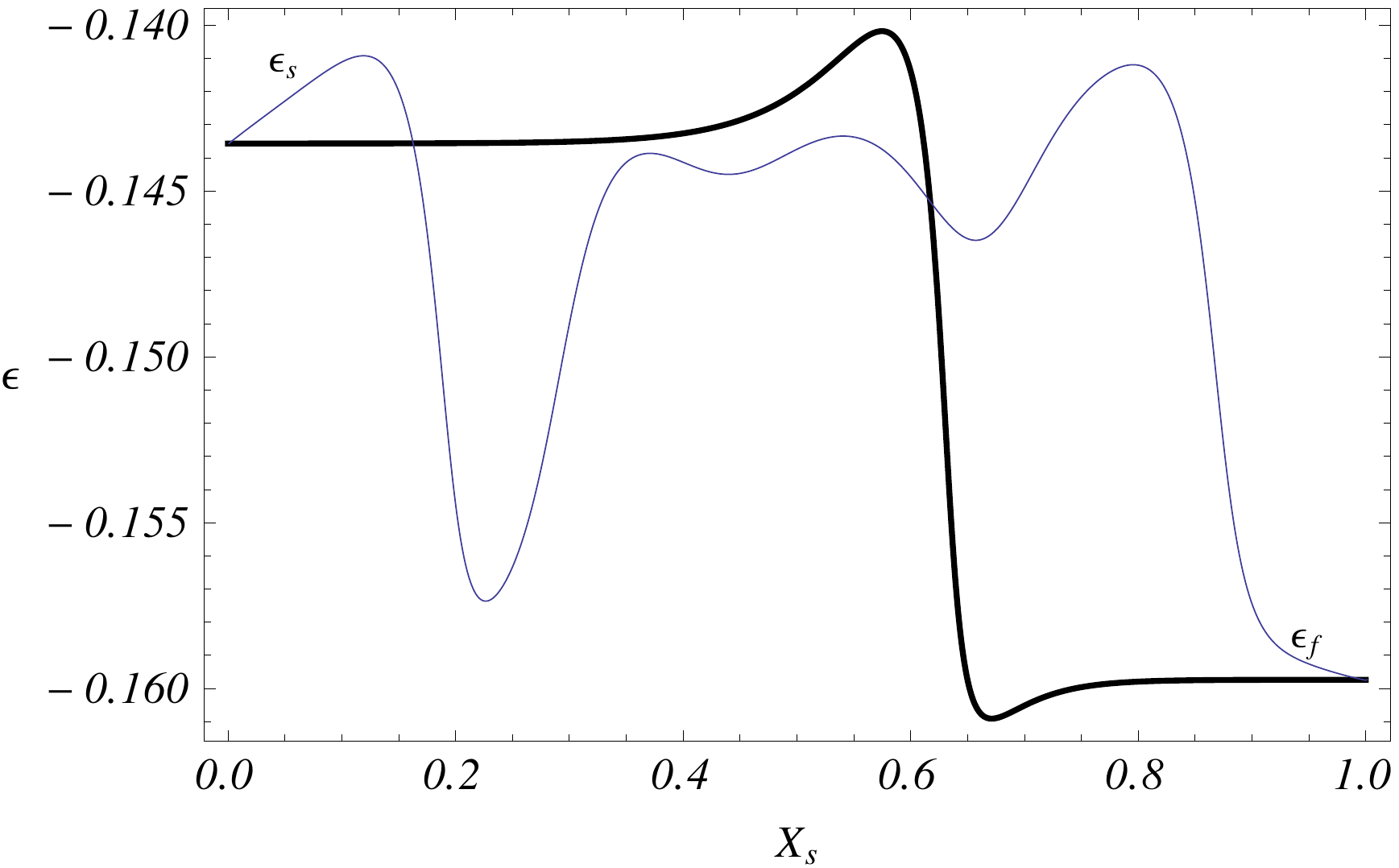}}} 
}
\end{picture}  
%\vskip 2 cm 
\caption{Profiles (solid thin lines)
$\varepsilon(X_\rr{s},t)$ 
obtained by solving the Cahn--Hilliard--like system (\ref{num05}) 
with the same random initial state as the one used 
in figure~\ref{f:acdin01}.
We used Dirichelet boundary conditions 
$m(0)=m_\rr{s}$, $\varepsilon(0)=\varepsilon_\rr{s}$,
$m(1)=m_\rr{f}$, and $\varepsilon(1)=\varepsilon_\rr{f}$
on the finite interval $[0,1]$, 
at the coexistence pressure
for 
$a=0.5$, $b=1$, $\alpha=100$,
$k_1=k_2=k_3=10^{-3}$. 
The solid thick line is the corresponding stationary profile.
Profiles at times
$t=0.009,0.04,0.2,3.7,3.9,89$
are depicted in lexicographic order.
}
\label{f:chdin01} 
\end{figure}

\begin{figure}
\begin{picture}(200,460)(50,0)
\put(0,0)
{
\resizebox{7.8cm}{!}{\rotatebox{0}{\includegraphics{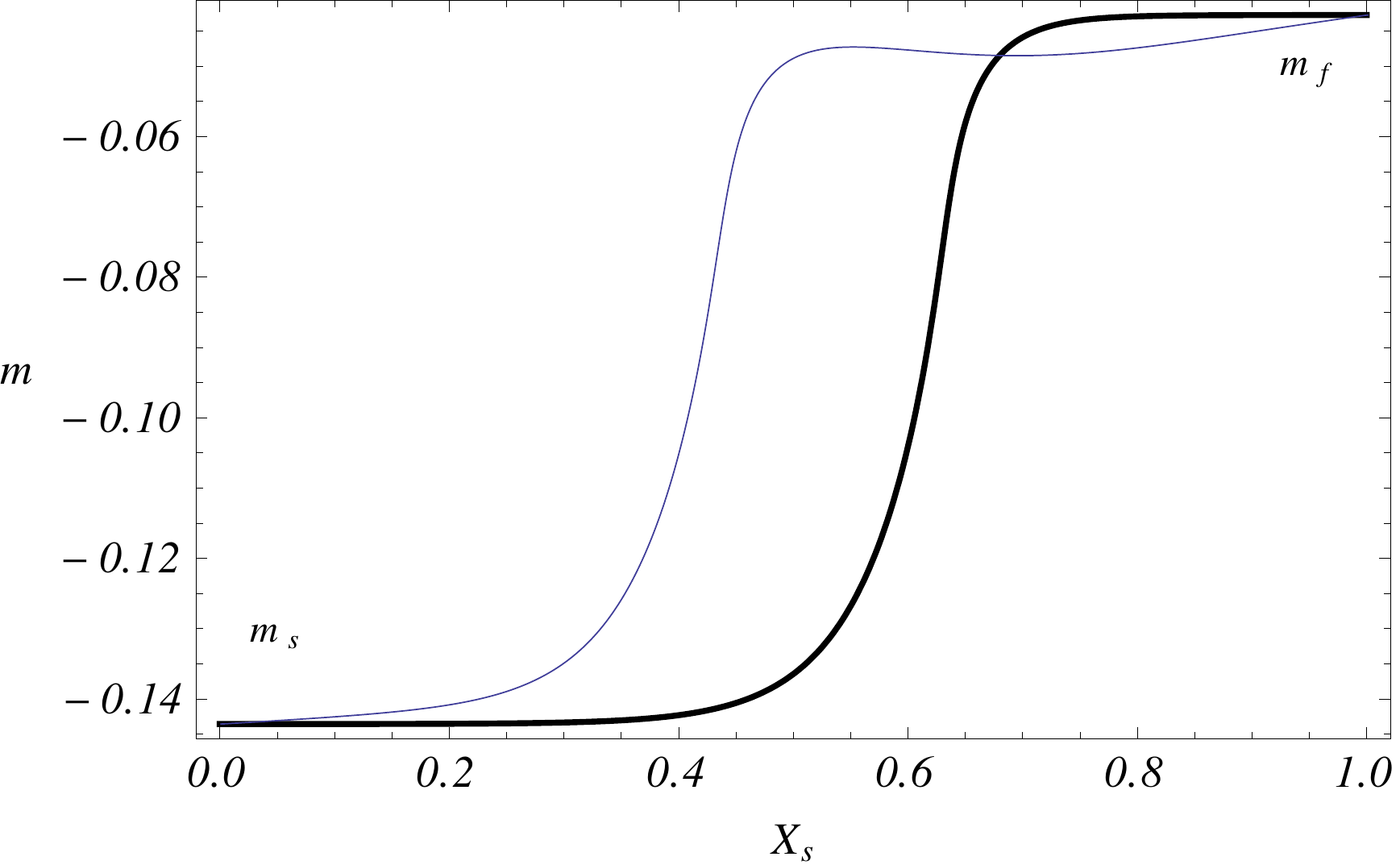}}} 
}
\put(235,0)
{
\resizebox{7.8cm}{!}{\rotatebox{0}{\includegraphics{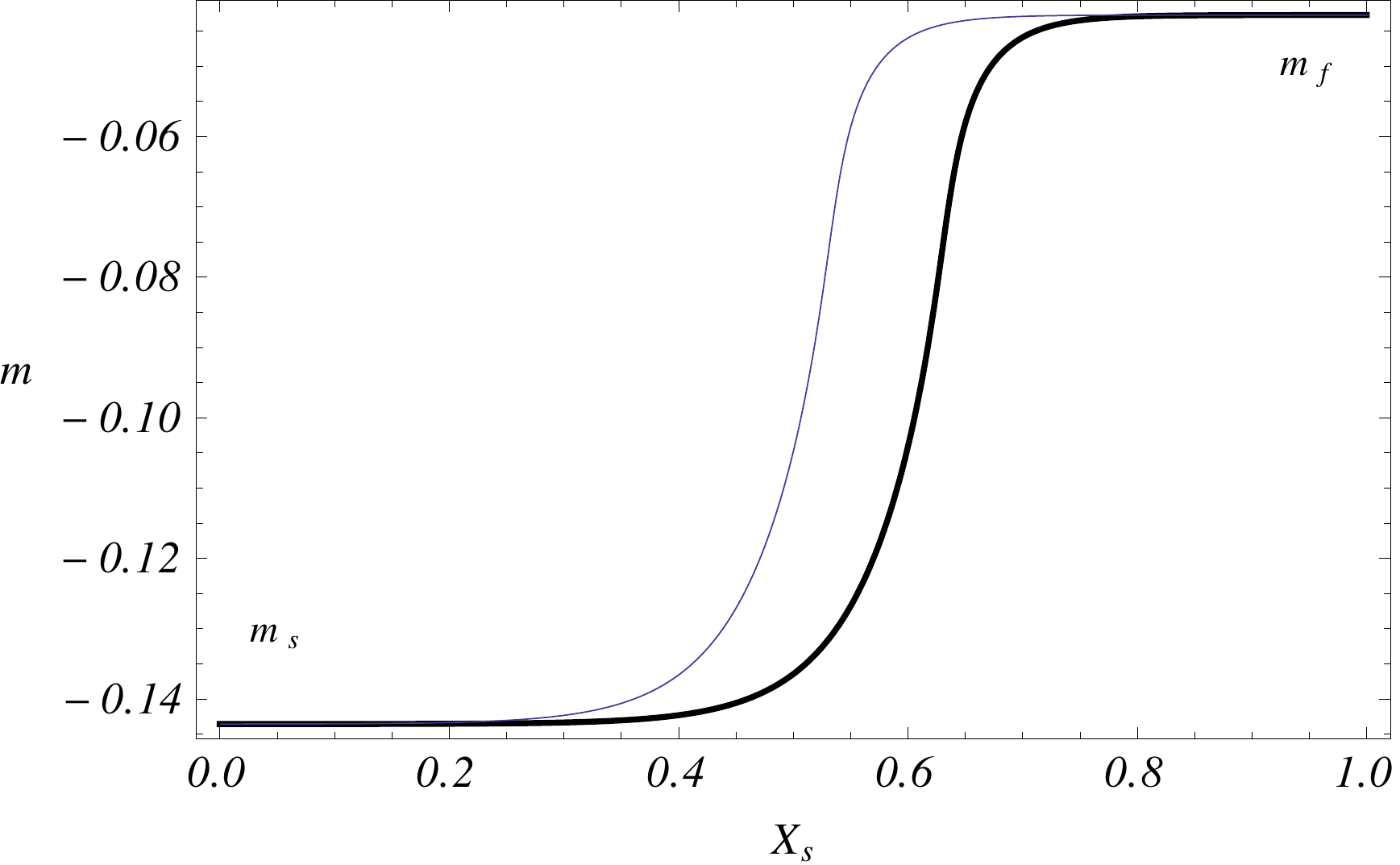}}} 
}
\put(0,160)
{
\resizebox{7.8cm}{!}{\rotatebox{0}{\includegraphics{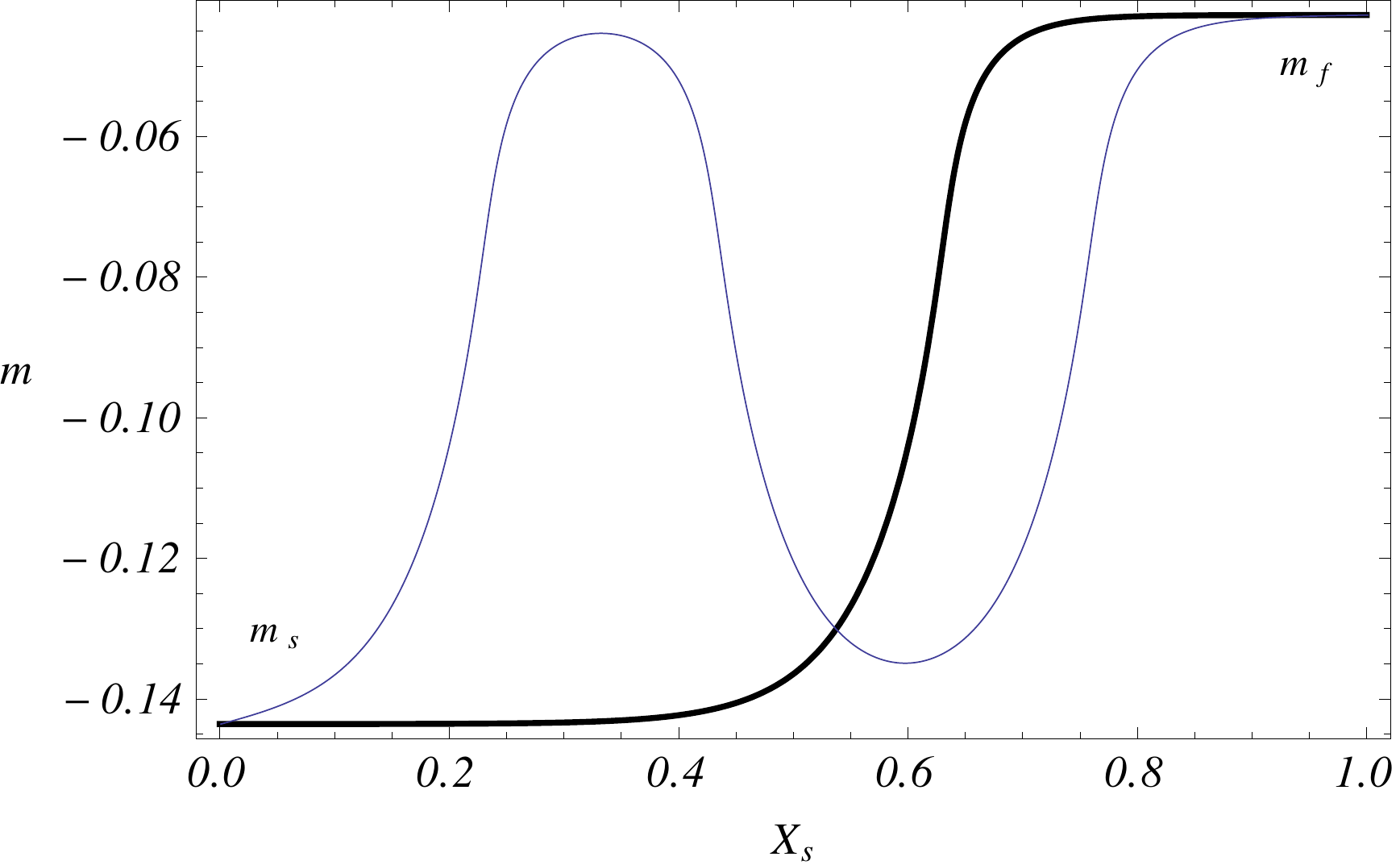}}} 
}
\put(235,160)
{
\resizebox{7.8cm}{!}{\rotatebox{0}{\includegraphics{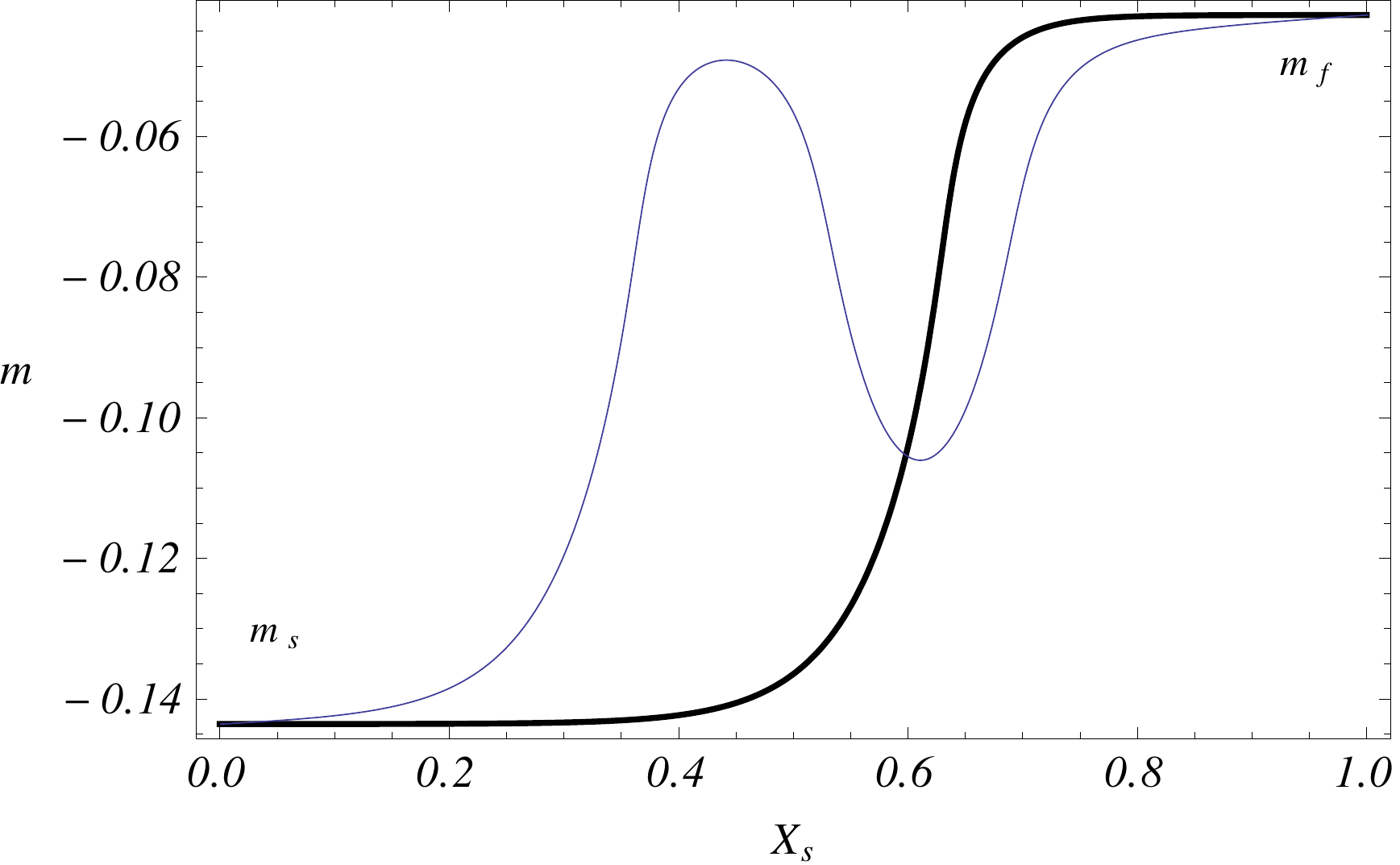}}} 
}
\put(0,320)
{
\resizebox{7.8cm}{!}{\rotatebox{0}{\includegraphics{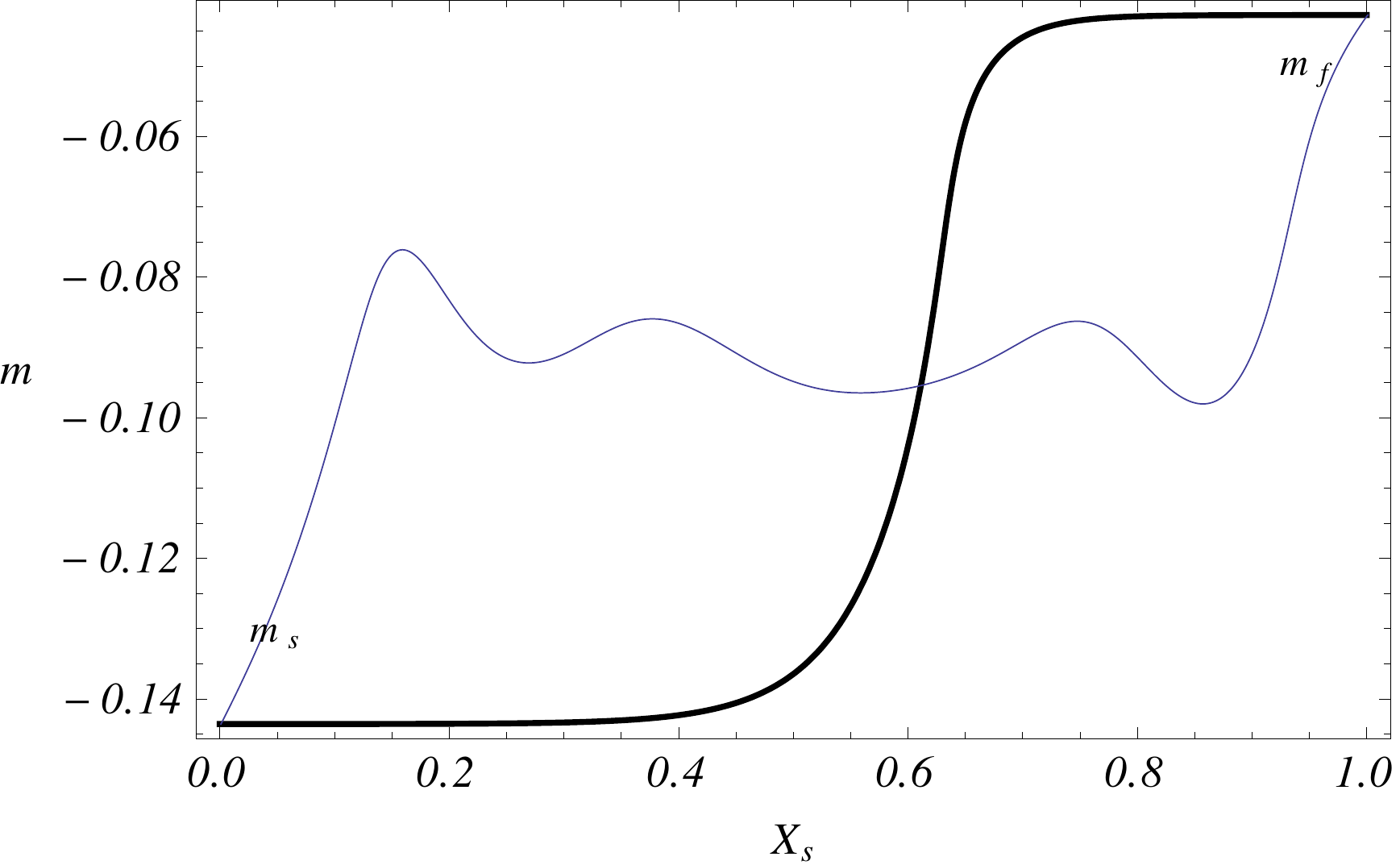}}} 
}
\put(235,320)
{
\resizebox{7.8cm}{!}{\rotatebox{0}{\includegraphics{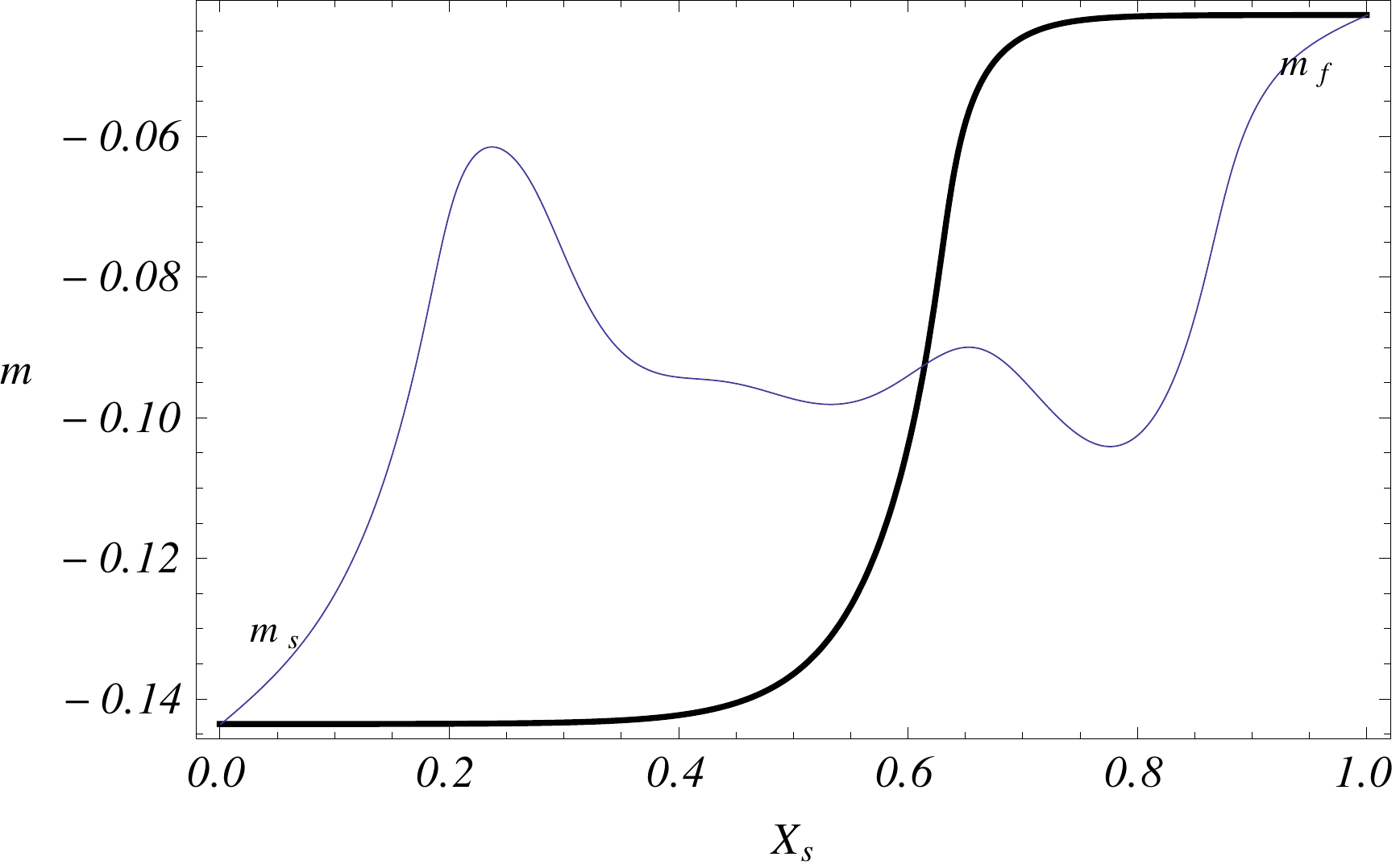}}} 
}
\end{picture}  
%\vskip 2 cm 
\caption{The same as in figure~\ref{f:chdin01} for the 
profiles $m(X_\rr{s},t)$.
}
\label{f:chdin02} 
\end{figure}

\begin{figure}
\begin{picture}(200,150)(50,0)
\put(0,0)
{
\resizebox{5cm}{!}{\rotatebox{0}{\includegraphics{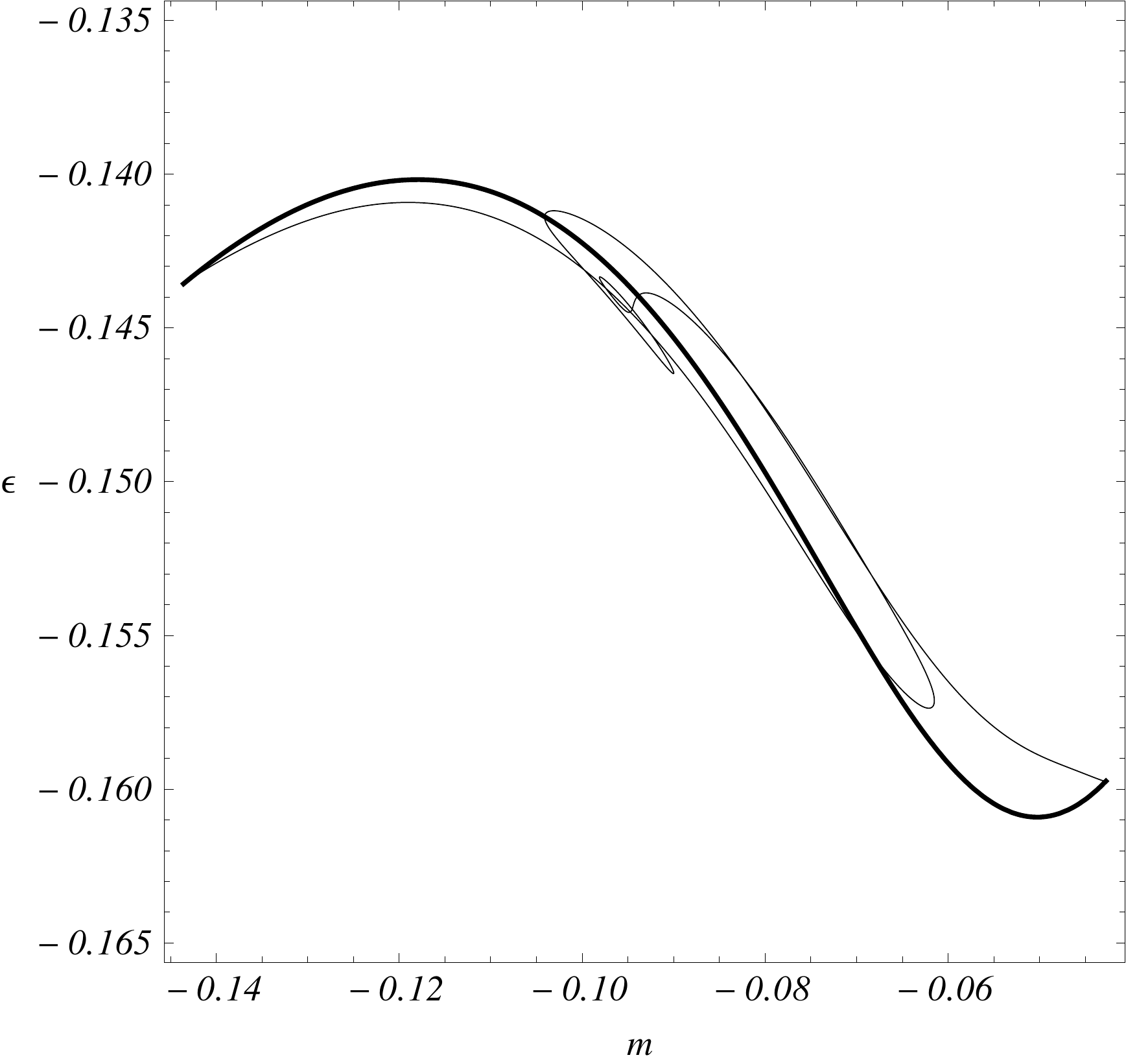}}} 
}
\put(160,0)
{
\resizebox{5cm}{!}{\rotatebox{0}{\includegraphics{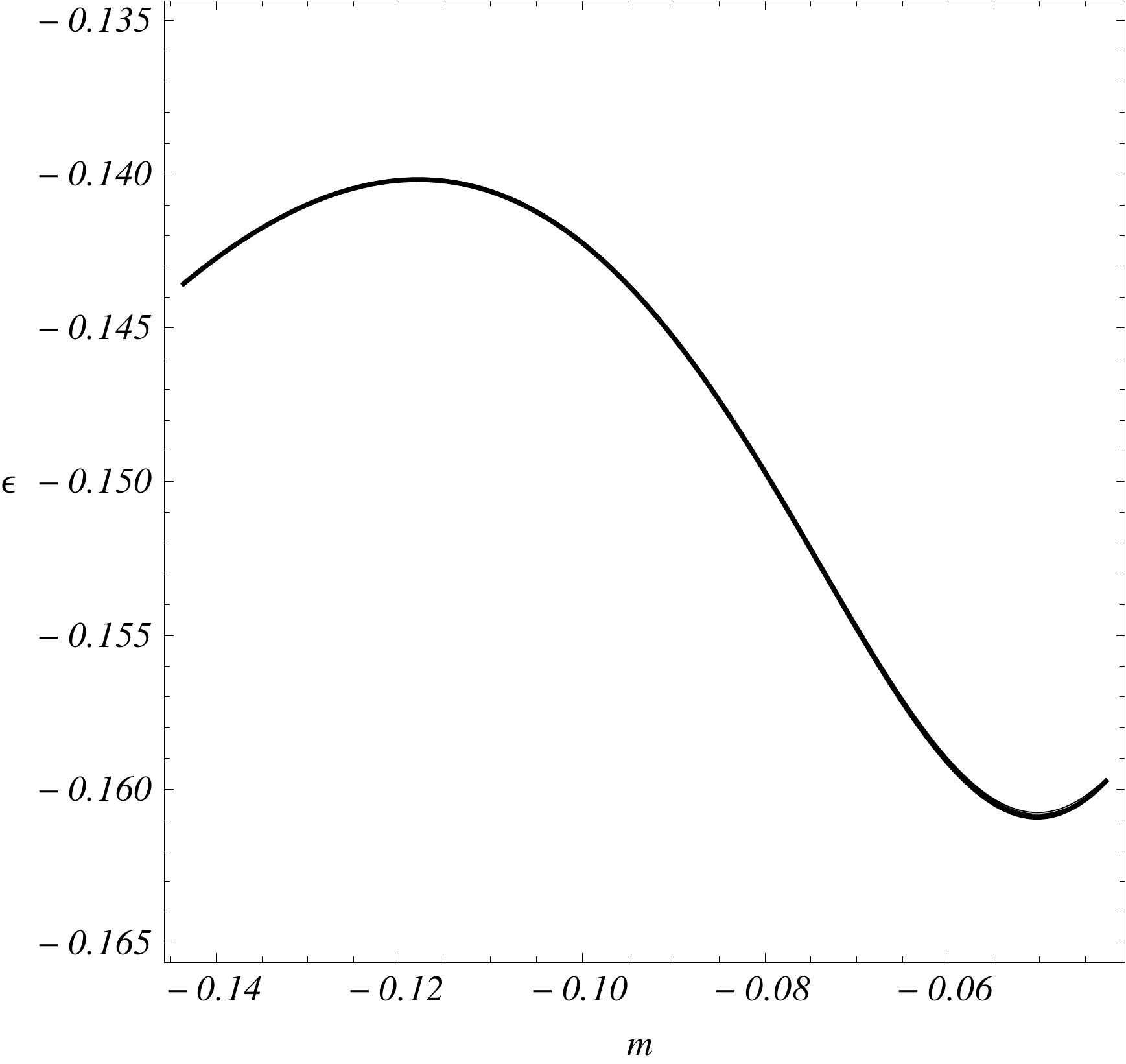}}} 
}
\put(320,0)
{
\resizebox{5cm}{!}{\rotatebox{0}{\includegraphics{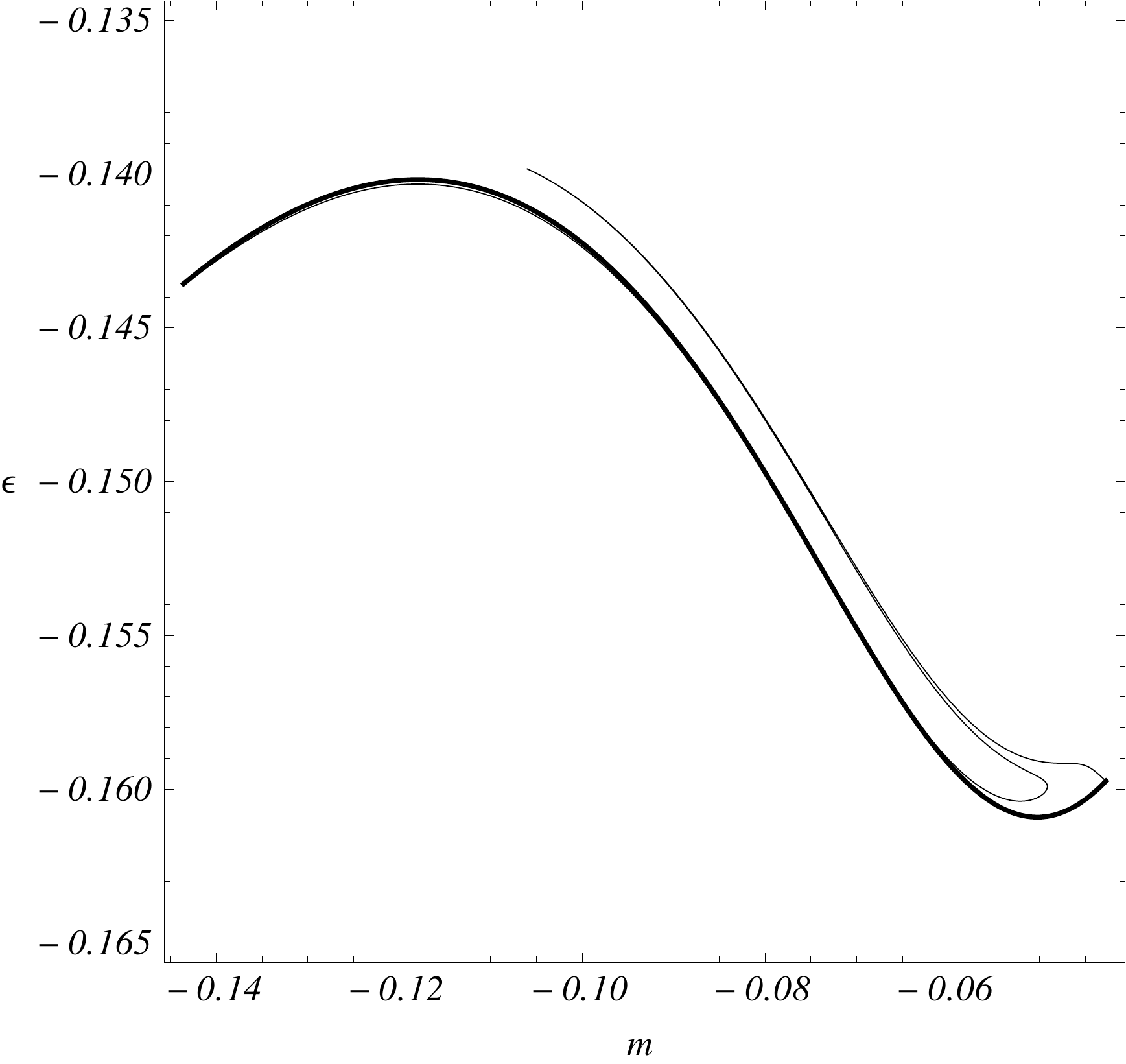}}} 
}
\end{picture}  
%\vskip 2 cm 
\caption{The solution of the same problem as in figure~\ref{f:chdin01} 
is depicted on the plane $m$--$\varepsilon$
at times $t=0.04,0.2,3.7$
from the left to the right.
}
\label{f:chdin03} 
\end{figure}

The same difference is appreciated when considering stratified initial data;
again the short--time behavior provides for both evolutions a 
deformation of the initial profile into a drop--like new one
at different time scales. Looking at the profiles related to
different characteristic times one can assess that 
the evolution process governed by the Allen--Cahn--like set of partial
differential equations is affected by initial conditions during a 
characteristic time interval much larger than the corresponding one
typical of the Cahn--Hilliard--like equations, see figures~\ref{f:acdin01}$c$ and
\ref{f:chdin01}$c$, which correspond to the formation of the drop-like
shape of the profile.

It is worth to notice that both the Allen--Cahn and the Cahn--Hilliard evolution
processes separately exhibit a consistent behavior when the initial data
is changed.
In particular, starting from spatially random or 
stratified initial conditions, 
both equations provide, after a transitory regime, 
a quasi--static evolution of an 
interface--like profile which nucleates and propagates at the same time 
and with
the same velocity. 
On the other hand
the numerical simulations suggest that
the quasi--static evolution process starts at different characteristic 
times and develops with different velocities when the 
Darcy and Stokes dissipation mechanisms are considered.

\begin{figure}[h]
\begin{picture}(200,150)(50,0)
\put(120,0)
{
\resizebox{8cm}{!}{\rotatebox{0}{\includegraphics{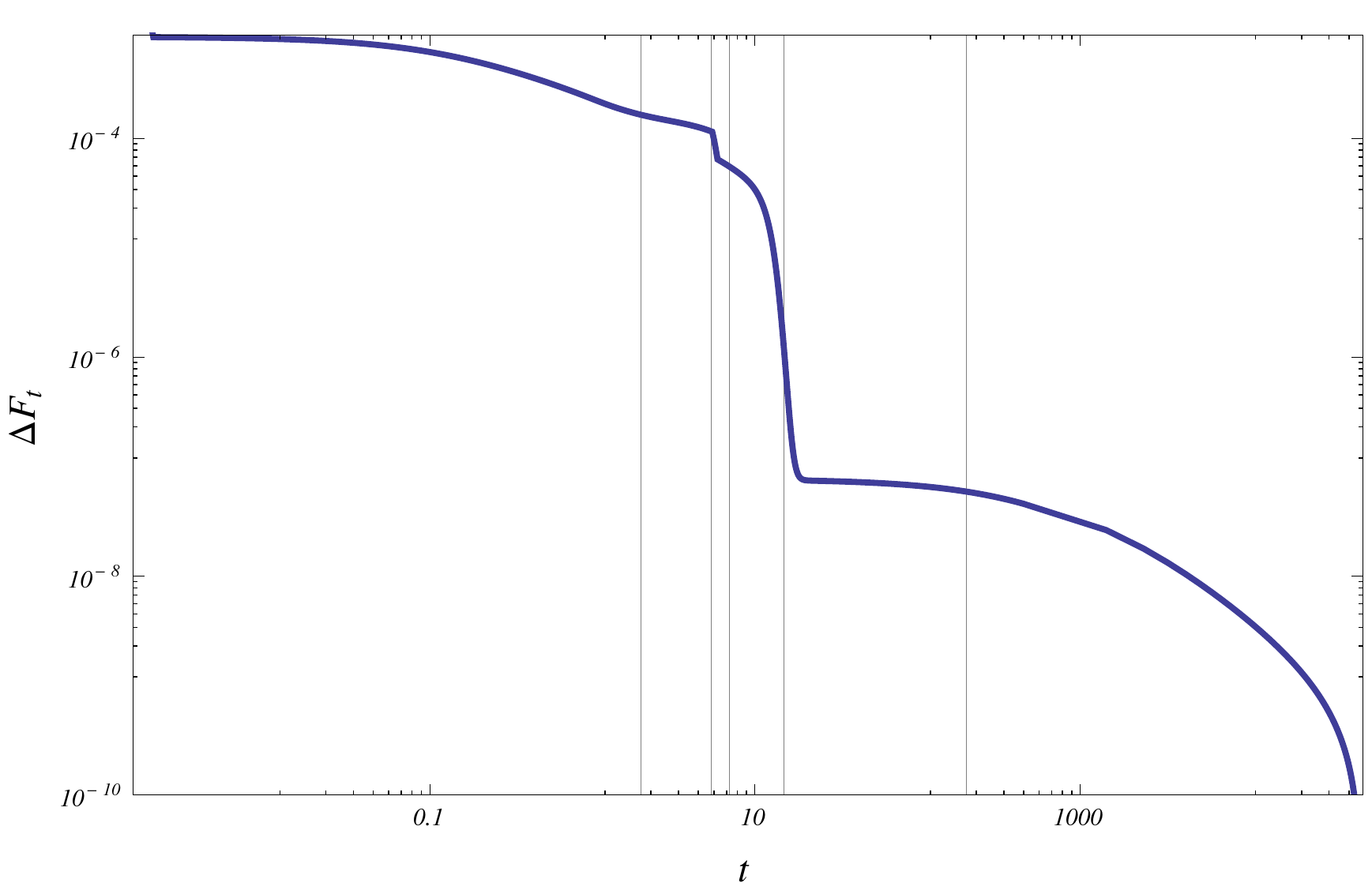}}} 
}
% \put(235,0)
% {
% \resizebox{8cm}{!}{\rotatebox{0}{\includegraphics{fig06b-porosi004.pdf}}} 
% }
\end{picture}  
%\vskip 2 cm 
\caption{For the same problem as in figure~\ref{f:acdin01} 
the difference of the energy (\ref{liapunov}) at time $t$ and that 
corresponding to the stationary profile is reported as function of time. 
The vertical thin lines denote the times $t=2,5.4,7,15,200$ 
at which the $\varepsilon$ and $m$--profiles are depicted in 
figures~\ref{f:acdin01}--\ref{f:acdin03}.
}
\label{f:acdin04} 
\end{figure}

\begin{figure}[h]
\begin{picture}(200,150)(50,0)
\put(130,0)
{
\resizebox{8cm}{!}{\rotatebox{0}{\includegraphics{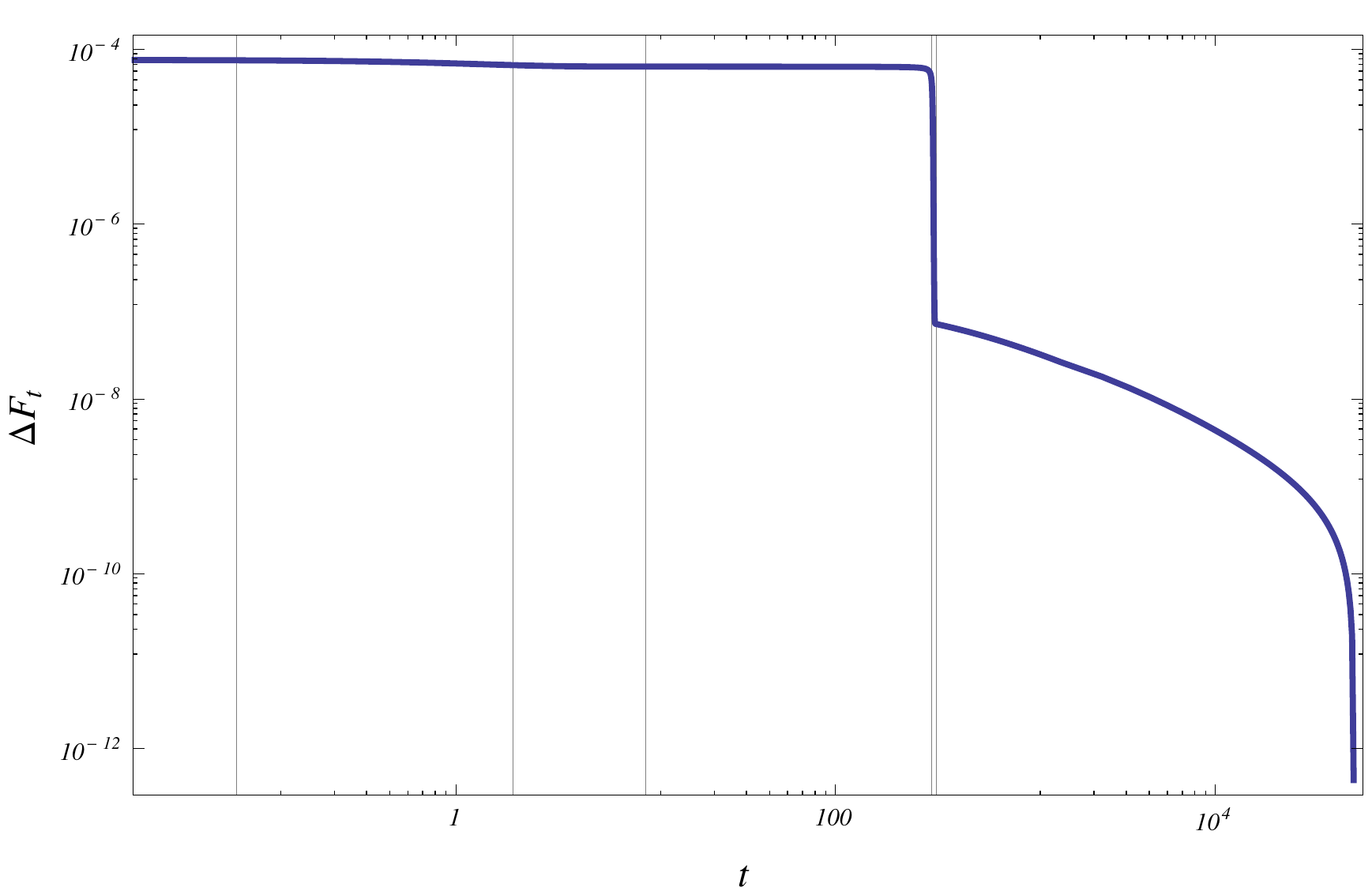}}} 
}
% \put(235,0)
% {
% \resizebox{8cm}{!}{\rotatebox{0}{\includegraphics{fig08b-porosi004.pdf}}} 
% }
\end{picture}  
%\vskip 2 cm 
\caption{For the same problem as in figure~\ref{f:acdin05} 
the difference of the energy (\ref{liapunov}) at time $t$ and that 
corresponding to the stationary profile is reported as function of time. 
The vertical thin lines denote the times 
$t=0.07,2,10,320,340$
at which the $\varepsilon$ profiles are depicted in 
figures~\ref{f:acdin05}.
}
\label{f:acdin06} 
\end{figure}

\begin{figure}[h]
\begin{picture}(200,150)(50,0)
\put(125,0)
{
\resizebox{8cm}{!}{\rotatebox{0}{\includegraphics{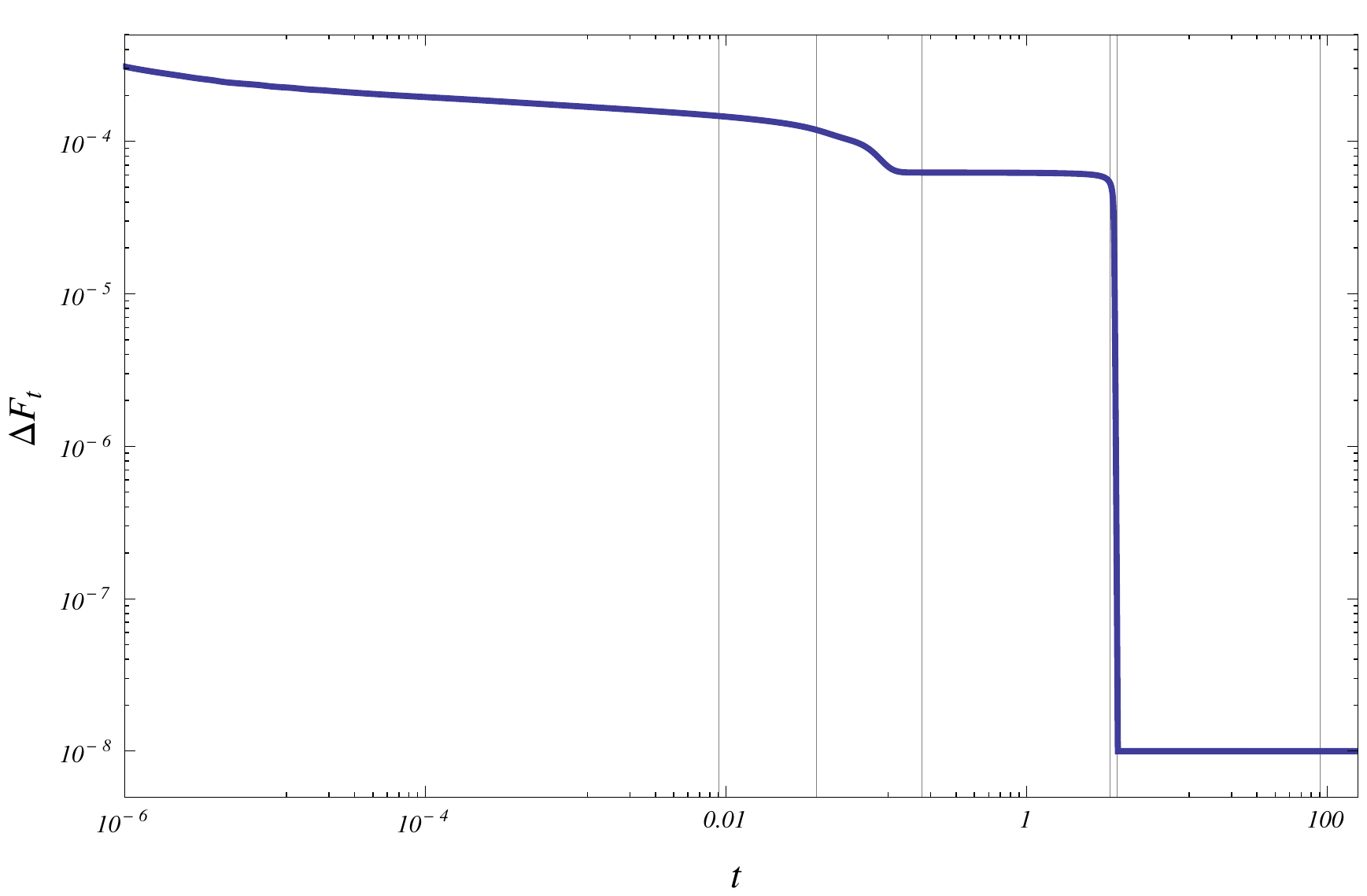}}} 
}
% \put(235,0)
% {
% \resizebox{8cm}{!}{\rotatebox{0}{\includegraphics{fig12b-porosi004.pdf}}} 
% }
\end{picture}  
%\vskip 2 cm 
\caption{For the same problem as in figure~\ref{f:chdin01} 
the difference of the energy (\ref{liapunov}) at time $t$ and that 
corresponding to the stationary profile is reported as function of time. 
The vertical thin lines denote the times $t=0.009,0.04,0.2,3.7,3.9,89$ 
at which the $\varepsilon$ and $m$--profiles are depicted in 
figures~\ref{f:chdin01}--\ref{f:chdin03}.
}
\label{f:chdin04} 
\end{figure}

\begin{figure}[h]
\begin{picture}(200,150)(50,0)
\put(125,0)
{
\resizebox{8cm}{!}{\rotatebox{0}{\includegraphics{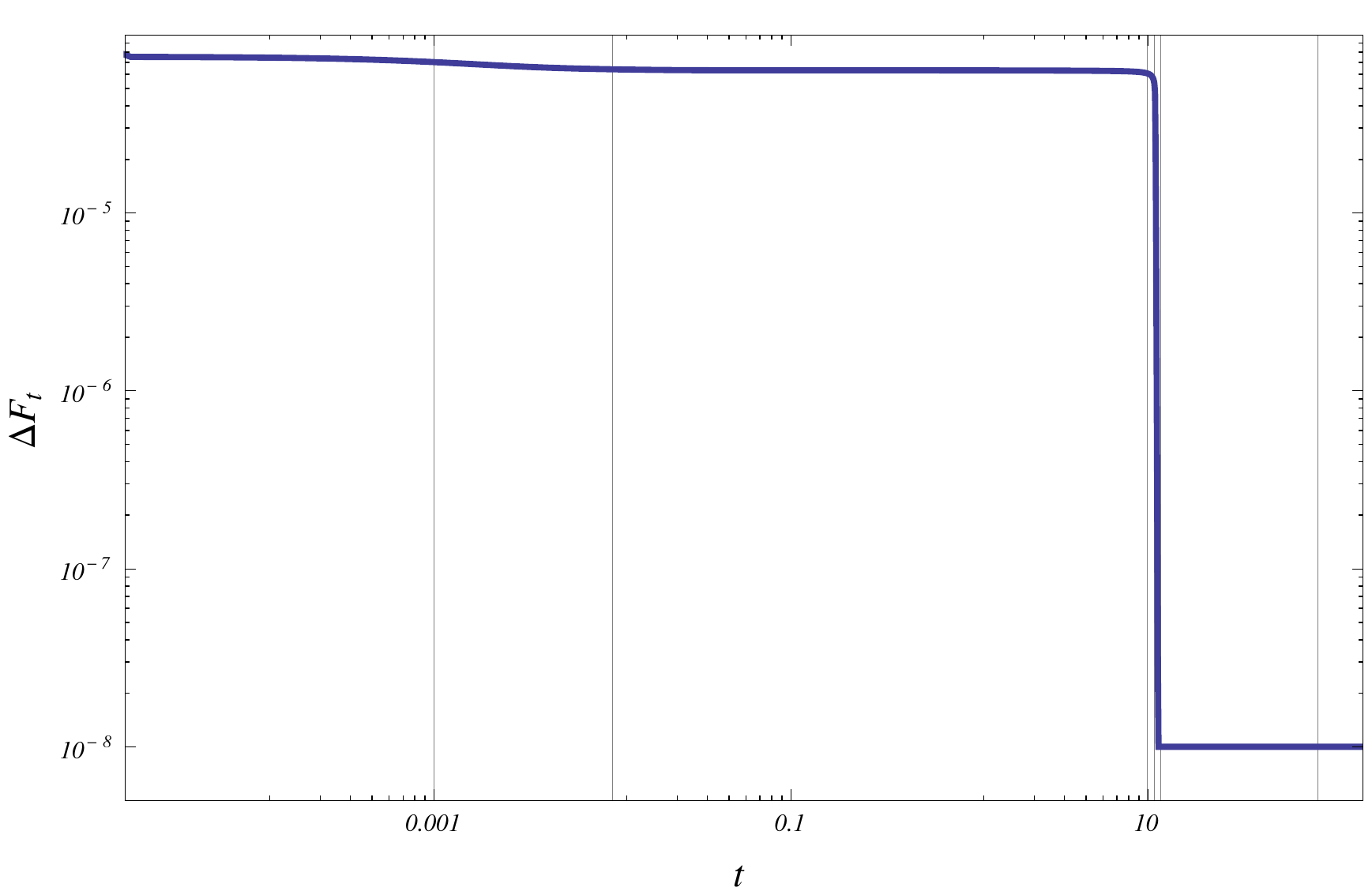}}} 
}
% \put(235,0)
% {
% \resizebox{8cm}{!}{\rotatebox{0}{\includegraphics{fig14b-porosi004.pdf}}} 
% }
\end{picture}  
%\vskip 2 cm 
\caption{For the same problem as in figure~\ref{f:chdin05} 
the difference of the energy (\ref{liapunov}) at time $t$ and that 
corresponding to the stationary profile is reported as function of time. 
The vertical thin lines denote the times 
$t=0.001,0.01,10.3,11,11.1,89$
at which the $\varepsilon$ profiles are depicted in 
figures~\ref{f:chdin05}.
}
\label{f:chdin06} 
\end{figure}

As already mentioned these two mechanisms correspond respectively to 
the presence (Darcy \& Cahn--Hilliard regime)
or absence (Stokes \& Allen--Cahn regime) of scale separation and, thus, 
to the cases in which 
the fluid dissipative flow
is parametrised either by the permeability of the solid or by the viscosity
of the fluid. 
In these two situations the micro--structure of the corresponding porous 
media is quite different: 
materials with scale separation are constituted by solid
grains which do not hinder the fluid flow strongly; the grains can be regarded 
as undeformable and impermeable, suffering an equivalent drag force due to
the average fluid velocity. 
On the other hand materials with no scale separation
are micro--porous and the fluid flow is much more 
hindered by the micro--structure
of the skeleton. This could provide a possible explanation for the very high 
discrepancy between the characteristic times of evolution for the Cahn--Hilliard 
and the Allen--Cahn--like system of equation: when dealing with a Cahn--Hilliard
dissipative process the time necessary to approach the 
stationary state is lower than that necessary to reach the same state 
through an 
Allen--Cahn evolution, see figures~ \ref{f:acdin04}, \ref{f:acdin06}, 
\ref{f:chdin04}, and~\ref{f:chdin06}. 

Comparing the distance between the current and 
the stationary strain energy for the Cahn--Hilliard and Allen--Cahn 
evolution process, three or two different regimes can be 
appreciated. These indeed correspond to the 
smearing out of the random oscillation (if any), to the formation of 
the intermediate drop--like solution, and 
tothe nucleation of the interface--like
profile.

Deeper investigations will be developed in the future in order to better 
explain such phenomenon, also accounting for different boundary conditions. 
Particular attention should be devoted to the study of initial boundary 
value problems involving Neumann boundary conditions, which naturally allow 
to describe the behavior of granular media under consolidation.

%\section{Conclusions}
%\label{s:conclusioni}
%\par\noindent
%%% Figura

%%% Fine figura
%
%%% Figura
%%% Fine figura

%%% Figura
%%% Fine figura

%%% Figura
%%% Fine figura

%%% Figura
%%% Fine figura

%%% Figura
%%% Fine figura

%%% Figura
%%% Fine figura

%%% Figura
%%% Fine figura

%%% Figura
%%% Fine figura

%%% Figura
%%% Fine figura
 
%%% Figura
%%% Fine figura

%%% Figura
%%% Fine figura

%%% Figura
%%% Fine figura

%%% Figura
%%% Fine figura

\appendix

\section{Stokes virtual dissipation in terms of the basic fields}
\label{s:bold}
\par\noindent
Derivation of the equation (\ref{lv16}) for the virtual work of the dissipation
due to the Stokes effect.
We first note that from the equality (\ref{lv05})
we have that 
\begin{equation}
\label{bd00}
[v_\rr{f}(x,t)-v_\rr{s}(x,t)]'
=
-
\Big(
     %\frac{\partial}{\partial X_\rr{s}}
     \Big(
          \frac{\dot\phi}{\phi'}
          \chi_\rr{s}'
     \Big)'
\Big)_{\chi^{-1}_\rr{s}(x,t)}
\,
(\chi^{-1})'_\rr{s}(x,t)
=
\Big(
     \frac{1}{\chi'_\rr{s}}
     \Big(
          \frac{\dot\phi}{\phi'}
          \chi_\rr{s}'
     \Big)'
\Big)_{\chi^{-1}_\rr{s}(x,t)}
\end{equation}
where in the first step we used 
the chain rule 
for the derivative of a composed function and in the second 
step the expression of 
the derivative of the inverse function.
In the second term of the equation above the notation means that 
the derivative of $\dot\phi\chi'_\rr{s}/\phi'$ is computed with respect to
$X_\rr{s}$
and the result is then evaluated in $X_\rr{s}=\chi^{-1}_\rr{s}(x,t)$.

Now, we note that by evaluating the expression (\ref{sv02}) for the variation 
of the field $\chi_\rr{f}$ in $X_\rr{s}=\chi^{-1}_\rr{s}(x,t)$, for 
any $x\in B_t$ and $t\in\bb{R}$, we get 
\begin{equation}
\label{bd01}
\delta\chi_\rr{f}(\chi^{-1}_\rr{f}(x,t),t)
-
\delta\chi_\rr{s}(\chi^{-1}_\rr{s}(x,t),t)
=
-\frac{\chi'_\rr{s}(\chi^{-1}_\rr{s}(x,t),t)}
      {\phi'(\chi^{-1}_\rr{s}(x,t),t)}
\,\delta\phi(\chi^{-1}_\rr{s}(x,t),t)
\end{equation}
As before, from (\ref{bd01}) we get
\begin{equation}
\label{bd02}
[\delta\chi_\rr{f}(\chi^{-1}_\rr{f}(x,t),t)
-
\delta\chi_\rr{s}(\chi^{-1}_\rr{s}(x,t),t)]'
%=
%-
%\Big(
%     \Big(
%          \frac{\chi'_\rr{s}}{\phi'}
%          \delta\phi
%     \Big)'
%\Big)_{\chi^{-1}_\rr{s}(x,t)}
%\,
%\chi^{-1}'_\rr{s}(x,t)
=
\Big(
     \frac{1}{\chi'_\rr{s}}
     \Big(
          \frac{\chi'_\rr{s}}{\phi'}
          \delta\phi
     \Big)'
\Big)_{\chi^{-1}_\rr{s}(x,t)}
\end{equation}
Finally,
by inserting (\ref{bd00}) and (\ref{bd02}) in (\ref{lv10}) and 
by performing the change of variables $x=\chi_\rr{s}(X_\rr{s},t)$
we get (\ref{lv16}).

\section{Finite difference approximations}
\label{s:numerico}
\par\noindent
In this appendix we collect the finite difference substitution rules
that we have adopted in our numerical computations. 
Let $n$ a positive integer number. 
Let $\sigma=1/n$ and $\tau\in\bb{R}$ be respectively the space and the
time increments. The space interval $[0,1]$ is subdivided into 
$n$ small intervals
of length $\sigma$.
Given a field $h(X_\rr{s},t)$, for any $i=1,\dots,n-1$ and $j\in\bb{N}$, 
we set
\begin{displaymath}
h'(i\sigma,j\tau)
\approx
\frac{1}{2\sigma}[h((i+1)\sigma,j\tau)
                    -h((i-1)\sigma,j\tau)]
\end{displaymath}
For the second space derivative we set 
\begin{displaymath}
h''(i\sigma,j\tau)
\approx
\frac{1}{\sigma^2}[h((i+1)\sigma,j\tau)-2h(i\sigma,j\tau)
                    +h((i-1)\sigma,j\tau)]
\end{displaymath}
for $i=1,\dots,n-1$ and $j\in\bb{N}$.
For the third space derivative we set 
\begin{displaymath}
\begin{array}{l}
{\displaystyle
\!\!\!\!\!
h'''(i\sigma,j\tau)
\approx
\frac{1}{2\sigma^3}[h((i+2)\sigma,j\tau)
                    -2h((i+1)\sigma,j\tau)
}
\\
{\displaystyle
 \phantom{aaaaaaaaaaaaaaaa}
                    +2h((i-1)\sigma,j\tau)
                    -h((i-2)\sigma,j\tau)]
}
\end{array}
\end{displaymath}
for $i=2,\dots,n-2$ and $j\in\bb{N}$.
For the fourth space derivative we set 
\begin{displaymath}
\begin{array}{l}
{\displaystyle
\!\!\!\!\!
\!\!\!\!\!
h''''(i\sigma,j\tau)
\!
\approx
\!
\!
\frac{1}{\sigma^4}[h((i+2)\sigma,j\tau)
                    -4h((i+1)\sigma,j\tau)
                    +6h(i\sigma,j\tau)
}
\\
{\displaystyle
 \phantom{aaaaaaaaaaaaaaaaaaaaa}
                    -4h((i-1)\sigma,j\tau)
                    +h((i-2)\sigma,j\tau)]
}
\end{array}
\end{displaymath}
for $i=2,\dots,n-2$ and $j\in\bb{N}$.
For the time derivative we finally set 
\begin{displaymath}
\dot{h}(i\sigma,j\tau)
\approx
\frac{1}{\tau}[h(i\sigma,j\tau)
                    -h(i\sigma,(j-1)\tau)]
\end{displaymath}
for $i=0,\dots,n$ and $j=1,\dots$.

%\begin{acknowledgements}
%The authors are grateful to G.\ Fusco for helpful discussions and hints.
%\end{acknowledgements}

\end{document}